\documentclass[10pt,journal,compsoc]{IEEEtran}

\newcommand{\hooklongrightarrow}{\lhook\joinrel\longrightarrow}
\newcommand{\hooklongleftarrow}{\longleftarrow\joinrel\rhook}
\RequirePackage{float} 
\usepackage{mathtools} 

\usepackage{graphicx}

\usepackage{microtype}                 
\PassOptionsToPackage{warn}{textcomp}  
\usepackage{textcomp}                  
\usepackage{mathptmx}                  
\usepackage{times}                     
\usepackage{cite}                      
\usepackage{tabu}                      
\usepackage{booktabs}                  

\newcommand{\blue}[1]{\textcolor{black}{#1}}

\usepackage{graphicx}
\usepackage{array,multirow}
\usepackage[table,xcdraw]{xcolor}
\usepackage{textalpha}
\newcommand*{\img}[1]{%
    \raisebox{-.2\baselineskip}{%
        \includegraphics[
        height=0.8\baselineskip,
        width=0.8\baselineskip,
        keepaspectratio,
        ]{#1}%
    }%
}
\usepackage{xspace}  
\newcommand{\myParagrapho}[1]{\smallskip\noindent\textbf{#1}\xspace}
\usepackage{bbding}
\usepackage{pifont}
\usepackage{amssymb}
\usepackage[hidelinks]{hyperref}
\usepackage{enumitem}

\newcommand{\myparagraph}[1]{\smallskip\noindent\textbf{#1}\xspace}

\newcommand{\squishlist}{
	\begin{list}{$\bullet$}
		{ \setlength{\itemsep}{0pt}      \setlength{\parsep}{3pt}
			\setlength{\topsep}{3pt}       \setlength{\partopsep}{0pt}
			\setlength{\leftmargin}{1.5em} \setlength{\labelwidth}{1em}
			\setlength{\labelsep}{0.5em} } }
	
\newcommand{\squishlisttwo}{
  \begin{list}{$\bullet$}
    { \setlength{\itemsep}{0pt}    \setlength{\parsep}{0pt}
      \setlength{\topsep}{0pt}     \setlength{\partopsep}{0pt}
      \setlength{\leftmargin}{2em} \setlength{\labelwidth}{1.5em}
      \setlength{\labelsep}{0.5em} } }
  
\newcommand{\squishend}{	\end{list}  }

\usepackage{amsmath}

\begin{document}

\title{ZigzagNetVis: Suggesting temporal resolutions for graph visualization using zigzag persistence}



%
%
%
%

\author{Rapha\"el Tinarrage, Jean R. Ponciano, Claudio D. G. Linhares, Agma J. M. Traina, and Jorge Poco
\IEEEcompsocitemizethanks{
\IEEEcompsocthanksitem R. Tinarrage and J. Poco are with the School of Applied Mathematics, Funda\c{c}\~ao Getulio Vargas, Rio de Janeiro, Brazil.
\protect\\
E-mails: \{raphael.tinarrage, jorge.poco\}@fgv.br.
\IEEEcompsocthanksitem J. Ponciano and A. Traina are with the Institute of Mathematics and Computer Sciences, University of S\~ao Paulo, S\~ao Carlos, Brazil.
\protect\\
E-mails: jeanponciano@usp.br, agma@icmc.usp.br.
\IEEEcompsocthanksitem C. Linhares is with the Department of Computer Science and Media Technology,
Linnaeus University, Växjö, Sweden.
\protect\\
E-mail: claudio.linhares@lnu.se
}
\thanks{Manuscript received April 19, 2021; revised August 16, 2021.}}

%
%

\markboth{Journal of \LaTeX\ Class Files,~Vol.~14, No.~8, August~2015}%
{Tinarrage \MakeLowercase{\textit{et al.}}: ZigzagNetVis: Suggesting temporal resolutions for
graph visualization using zigzag persistence}

\IEEEtitleabstractindextext{%
\begin{abstract}
Temporal graphs are commonly used to represent complex systems and track the evolution of their constituents over time. 
Visualizing these graphs is crucial as it allows one to quickly identify anomalies, trends, patterns, and other properties that facilitate better decision-making. 
In this context, selecting an appropriate \textit{temporal resolution} is essential for constructing and visually analyzing the layout.
The choice of resolution is particularly important, especially when dealing with temporally sparse graphs. 
In such cases, changing the temporal resolution by grouping events (i.e., edges) from consecutive timestamps --- a technique known as timeslicing --- can aid in the analysis and reveal patterns that might not be discernible otherwise. 
However, selecting an appropriate temporal resolution is a challenging task.
In this paper, we propose ZigzagNetVis, a methodology that suggests temporal resolutions potentially relevant for analyzing a given graph, i.e., resolutions that lead to substantial topological changes in the graph structure. 
ZigzagNetVis achieves this by leveraging zigzag persistent homology, a well-established technique from Topological Data Analysis (TDA).
To improve visual graph analysis, ZigzagNetVis incorporates the colored barcode, a novel timeline-based visualization inspired by persistence barcodes commonly used in TDA.
We also contribute with a web-based system prototype that implements suggestion methodology and visualization tools. 
Finally, we demonstrate the usefulness and effectiveness of ZigzagNetVis through a usage scenario, a user study with 27 participants, and a detailed quantitative evaluation.

%
\end{abstract}

}

\maketitle

\section{Introduction}

\IEEEPARstart{T}{emporal} graphs (or temporal networks) constitute a powerful framework for modeling dynamic and complex systems from a variety of domains, including computer science, social sciences, and biology~\cite{holme2012temporal}. The visual representation of temporal graph data provides an intuitive and interactive way to explore complex relationships and dynamic changes over time. By using appropriate visualization techniques, researchers and practitioners are able to gain insights concerning the temporal evolution of the graph structure, to identify trends and anomalies, and detect important events that impact the system being studied.

Many studies have proposed graph drawing methods and visualizations to enhance the analysis of real-world temporal graphs. Examples include animated and timeline-based visualizations~\cite{SurveyDynamicVisualization} (e.g., animated node-link diagrams and \textit{Massive Sequence View} layout~\cite{Elzen2014}), optimization of node positioning~\cite{8807379,CNO}, edge data sampling~\cite{EOD}, summarization of visual representations~\cite{Stanley2018,8249874}.

Another important type of strategy concerns graph timeslicing, i.e., the choice of a timeslice length that defines the temporal granularity at which the graph will be studied (e.g., daily or weekly). 
In this context, although non-uniform timeslicing methods have been proposed in recent years~\cite{wang2019nonuniform, PONCIANO2021170,MultiPiles}, the most adopted strategy is uniform timeslicing, where timeslices of equal length represent the graph over time~\cite{7192717,CNO,samplingLuis,EOD,8365984, LargeNetVis,3399922}. 

\blue{
Once the timeslice length has been chosen, one divides the time interval into windows, and builds in each of them a graph, called a \textit{snapshot}, enabling the use of standard graph analysis techniques.
In this paper, in order to present a more general point of view, we will use the term \textit{temporal resolution}, which corresponds to timeslice length, but expressed in terms of the graph's initial resolution rather than an arbitrary unit of time (both quantities are proportional).
}

Different temporal resolutions reveal different patterns, making the choice of resolution crucial for effective analysis.
This is particularly relevant when dealing with temporally sparse graphs; in this case, global pattern identification might not be easy (or even possible) with too-fine resolutions due to the elevated number of timestamps. However, choosing a suitable temporal resolution is not a trivial task. 
In most cases, it requires exploratory analyses leading to empirical choices or input from a domain expert with prior knowledge of suitable resolutions.

\blue{
To date, a handful of studies have tackled the problem of \textit{automatic} resolution selection.
Some are based on features computed on each snapshot (e.g., mean degree or clustering coefficient) and between consecutive snapshots (e.g., Jaccard similarity between nodes or edges); optimal resolutions are then obtained via maximization or peak detection \cite{monetexplorer,soundarajan2016generating}.
However, by considering only the consecutive snapshots, these strategies miss information regarding the graph's global behavior.
Incorporating larger-scale dynamics has been explored, notably by finding the largest intervals over which features ``persist'', through minimization of a trade-off information/variance 
\cite{sulo2010meaningful,orman2021finding,fish2017supervised,soundarajan2016generating,darst2016detection,uddin2017optimal}.
However, these methods require additional hyperparameters or only study snapshots through specific features, thereby losing the structural information of the underlying graphs.
}

In order to study a temporal graph as a whole, and not fragmented into isolated snapshots, we will use tools from Topological Data Analysis (TDA), and more particularly persistent homology (PH) and zigzag persistent homology (zigzag PH).
This theory, which aims to capture relevant topological and geometric features from datasets, has already been applied to a wide range of problems concerning the analysis and visualization of graphs \cite{Aktas2019}.
\blue{
Although its application to dynamic graphs is still in its early stages, a common methodology is emerging: gathering the snapshots into a zigzag module, and analyze its persistence barcode
\cite{gamble2012applied,kim2018formigrams,kim2020analysis,myers2023temporal,myers2023topological}.}
We emphasize that, in this context, one of the main benefits of employing zigzag PH instead of ordinary PH is that the former allows tracking the appearance, disappearance, merge, and split of connected components, while the latter only allows appearance and merge.
To the best of our knowledge, no study has applied PH to the problem of temporal resolution selection.


\myparagraph{Our contributions.}
This paper introduces ZigzagNetVis, a methodology that employs zigzag PH to suggest potentially relevant temporal resolutions for visualizing temporal graphs. 
These resolutions are identified based on the degree of topological change they induce.
As we will discuss throughout the article, leveraging ideas from TDA yields new valuable insights for this problem.

\blue{
First of all, the structure of zigzag module, by including not only pointwise information (snapshots) but also dynamic information (their relationship), allows one to study a temporal graph as a whole.
We propose a topological interpretation of the effect of changing resolution, classified as \textit{timestamps shift} or \textit{structural change}.
}

\blue{
Second, compared to certain features used in the literature, PH can be clearly interpreted and visualized through the \textit{persistence barcodes}, a structure that, in the same vein as a tracking graph, captures the dynamics of a temporal graph.
An important feature of the barcodes is that we can compare them via the \textit{bottleneck distance}.
Based on this idea, we devise an explainability pipeline that spots the most important differences between resolutions.
}

In addition, to enhance the visual analysis, ZigzagNetVis incorporates a novel timeline-based visualization inspired by the persistence barcodes. It was specifically designed to enhance the analysis of connected components' structure and evolution.

\blue{
Last, we address an important related issue: the question of selecting an “optimal” resolution is ill-posed.
Indeed, different resolutions may be relevant for uncovering different patterns.
Furthermore, no reference benchmark is available.
We contribute to this problem by bringing together various results scattered in the literature, and by comparing our approach with other traditional methods through an empirical study of two real-world datasets.
}

In summary, our main contributions are: 
(i) A layout-agnostic method that leverages zigzag PH to suggest potentially relevant temporal resolutions for graph visualization; 
\blue{
(ii)
An explainability method for identifying the major topological differences caused by two different resolutions;
}
(iii) A timeline layout inspired by the barcodes from TDA and which depicts the evolutionary behavior of the graph's connected components;
(iv) The prototype of a web-based system with interactive linked views to assist in the graph analysis;
(v) Evaluation using a usage scenario, a user study (27 participants), and a quantitative comparison with existing features. 
%

\section{Background and Related Work}
\label{sec:related_work}

\subsection{Temporal graphs and timeslicing}
\label{subsec:timeslicing}

\myParagrapho{Timeslicing.}
Let $N$ be an integer representing the maximal time value.
A \emph{temporal graph} is a graph $G$ and a collection of pairs $(e,t)$, where $e$ is an edge of $G$ and $t$ is an integer in $[0, N]$.
In practice, $e$ represents an interaction between its nodes, occurring at time $t$.
This formalism underpins many models of dynamic phenomena, ranging from communication networks to biological mechanisms \cite{holme2012temporal}.
The value $r_0 = 1$ is called the \emph{initial resolution} and the integers $t \in [0,N]$ are referred to as the \emph{initial timestamps}. As in~\cite{DyNetVis}, the initial resolution represents the time interval in which the graph data was originally recorded, e.g., timestamps in $r_0 = 1$ span a 1-day interval in the Enron network~\cite{enron} and 20 seconds in the Primary School network~\cite{primarySchool}.

\begin{figure}[t]
\centering
\includegraphics[width=\linewidth]{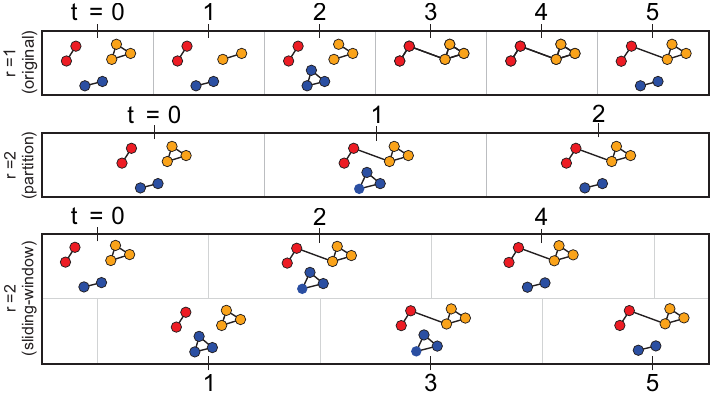}
\caption{A temporal graph of initial resolution $1$ (first row) and its partition and sliding-window timeslicing at resolution $r=2$ (second and third rows).}
\label{fig:timeslicing}
\end{figure}

In the context of temporal graph analysis, one is interested in the graphical representation and analysis of temporal graphs.
To this end, the usual approach (used, e.g., in~\cite{CNO,samplingLuis,3399922}) consists in choosing an integer $r > 1$, regularly cutting the interval $[0,N]$ into $M = \lfloor N/r\rfloor$ sub-intervals $[k r, (k+1)r]$, where $k \in[0,M]$ is an integer, and building $M+1$ graphs $\{G_k\}_{k = 0}^{M}$. Each graph $G_k$ contains the edges $(e,t)$ where $t \in [k r, (k+1)r]$, and and the nodes of these edges.
In other words, we build the graphs by collecting the edges active during the corresponding intervals and discarding the isolated nodes.
The parameter $r$ is called the \emph{resolution}, and the integers $k\in[0,M]$ are the corresponding \emph{timestamps}.
In what follows, we will refer to this process as \emph{partition timeslicing}.
The first and second rows of Fig.~\ref{fig:timeslicing} represent the collection of graphs obtained via this process for resolutions $1$ and $2$, respectively.
One observes that, for the initial resolution, there are two timestamps where the blue nodes are not present. This phenomenon disappears at resolution $2$. In general, as the resolution increases, both the number of edges and nodes present at each timestamp may grow.

We will also consider another cutting process, called \emph{sliding-window timeslicing}~\cite{7192717,myers2023temporal}.
As before, let $r$ be a resolution parameter.
For each initial timestamp $k$, we build the graph $G_k$ whose edges are those with the activation time $t$ contained in $[k-r/2, k+r/2]$.
Unlike partition timeslicing, which separates edges into disjoint intervals, sliding-window timeslicing allows activation intervals to overlap (see Fig.~\ref{fig:timeslicing}).
Note that the graphs obtained for an even resolution $r = 2s$ are identical to those obtained for the next odd resolution $r = 2s+1$, since the edges' activation times are integers. Thus, in the rest of this article, we will consider sliding-window timeslicing for even values of resolution only.

A characteristic shared by these two approaches is that all timeslices have the same length, known as \textit{uniform timeslicing}. Although not as popular, the idea of \textit{non-uniform timesciling} has also been considered in recent years. This type of timeslicing allows timeslices with different lengths over time. In graph visualization, we may find timeslices whose lengths depend on how many consecutive timestamps have similar graph structure~\cite{MultiPiles} and how active (in terms of bursts of events) the graph is over time. As an example of this last case, while Ponciano et al.~\cite{PONCIANO2021170} use long timeslices to represent intervals with bursts of events, Wang et al.~\cite{wang2019nonuniform} adopt short timeslices to analyze such intervals. 

In this paper, we focus on uniform timeslicing, the most commonly adopted approach~\cite{7192717,EOD,8365984, LargeNetVis,3399922,monetexplorer}.
In this context, the choice of the resolution $r$ can strongly impact the analysis: an overly coarse cut erases short-duration phenomena, while an overly fine cut disrupts continuous phenomena~\cite{PONCIANO2021170}.
This impact has been studied in \cite{Krings2012,samplingLuis}, by comparing features of the resulting temporal graphs (e.g., the average degree of their nodes or the size of their connected components).
It is worth noting that the problem persists in the context of non-uniform timeslicing, since such methods often rely on selecting an initial resolution, that has to be chosen wisely \cite{orman2021finding}.
Although it is a crucial parameter, the resolutions are often chosen heuristically, and the question of their selection is barely raised.
Keeping in mind the potential applications of temporal graph analysis, our work aims to describe and implement an automatic method of resolution selection.

\myParagrapho{Automatic resolution detection.}
Among the works that tackle this problem, two strategies are found.
The first is to define the parameter via the maximization of features' values or the minimization of trade-offs between features~\cite{monetexplorer,sulo2010meaningful,uddin2017optimal}. 
For example, MoNetExplorer~\cite{monetexplorer} is a visual analytic system that ranks candidate timeslice lengths (i.e., window sizes) based on three motif-based features: motif stability, motif fidelity, and motif clusterness, which are computed for each candidate.
Not every possible length becomes a candidate; only those following a predefined base time unit. For example, if the base is a month, there are 12 candidates (lengths ranging from one to twelve months).
Other examples include \cite{sulo2010meaningful}, where the resolution is found by minimizing a trade-off between the compression ratio and the variance of a sequence of features, as well as \cite{uddin2017optimal}, which seeks to minimize the variance of several features (positional dynamicity, degree, closeness, and betweenness centrality).

The second strategy involves abrupt change detection in time series, for instance, from the Jaccard distance between snapshots~\cite{orman2021finding}, from compression ratios \cite{fish2017supervised} or from coefficients of convergence \cite{soundarajan2016generating}.  
In the same vein, \cite{clauset2012persistence} detects peaks from the autocorrelation of the time series of features.
Our method employs this strategy since it can be naturally combined with features from Topological Data Analysis. 
To highlight the value of our method, we provide a precise theoretical justification and an extensive empirical analysis.


As pointed out by the studies above \cite{Krings2012}, several distinct resolutions may be relevant for analyzing a temporal graph. 
Therefore, the question of a ``correct'' interval length is ill-posed.
In this work, we circumvent this issue by suggesting various values --- without relying on predefined user-selected candidates or other parameters --- and explaining their relevance.




\subsection{Persistent Homology applied to Graphs}
\label{subsec:PH_applied_to_graph}


The mathematical tools used in this article are drawn from Topological Data Analysis (TDA), a field at the intersection of computational geometry, algebraic topology, and data analysis \cite{chazal2021introduction}.
Persistent homology (PH), one of its most popular techniques, allows us to infer homology groups of a dataset \cite{niyogi2008finding}.
It has been applied to a wide range of problems, including medicine, physics, computer vision, and machine learning, among others.
However, its application to the study of temporal graphs is relatively new.

\myParagrapho{Analysis of graphs.} 
PH is mainly used when the dataset is a point cloud, an image, a scalar function, or a graph.
We refer the reader to the survey \cite{Aktas2019} for an extensive presentation of how TDA has been applied to graph analysis.
As an intermediary construction between the input data and PH, the user must choose a \emph{filtration}, i.e., a non-decreasing family of subspaces that covers the data.
To this end, several popular filtrations exist, such as the Rips filtration.

However, in our context, the input data is not a single graph but a sequence of graphs, and PH cannot be used directly.
This is due to the fact that the sequence may not be non-decreasing: as time progresses, nodes or edges may disappear.
As a consequence, the temporal graph may not form a filtration.
To get around this problem, one strategy involves applying PH to each graph in the sequence and analyzing the results, as \cite{8365984} does in the context of temporal graph exploration.
Although it allows exhibiting global properties of the data, this method does not use the full potential of PH, since persistence is computed only at the level of each graph, and not throughout the sequence.
In particular, no temporal information is contained in the persistence diagram.
Moreover, this method lacks the theoretical guarantees of TDA, such as stability.

As an alternative, one can use \textit{zigzag persistent homology} (zigzag PH), which we will describe in Sec.~\ref{subsec:intro_zigzag}.
This variation of PH has already been used in the context of topological bootstrapping, thresholding, and parameter selection.
Unlike ordinary PH, it is based on the notion of \textit{zigzag filtrations}, which do not have to be non-decreasing.
In particular, it can be applied to a temporal graph, allowing one to compute the persistence of the sequence of graphs all at once \cite{gamble2012applied,kim2018formigrams,kim2020analysis,myers2023temporal,myers2023topological}.
By computing the persistence barcode, the main object of TDA and described in the next section, one can detect the global behavior of the graph (e.g., the evolution of its connected components, periodic or chaotic patterns).
\blue{Our work brings these ideas to the problem of resolution selection by investigating the link between the stability of zigzag persistence modules and the choice of a resolution. In addition, we also devised a new visualization layout based on PH.}

We point out that, in this article, the topology of the graphs will be studied through the lenses of the homology $H_0$, that is, the connected components.
Ordinary PH enables us to track these components over time, limited to the case of appearance and merge. 
In addition, employing zigzag PH allows one to study the disappearance and splitting of connected components, phenomena that occur in temporal graphs.
As exemplified by numerous articles in the TDA literature, $H_0$ contains sufficient information to solve certain problems \cite{paris2007topological,plonka2016relation,bendich2016topological,chazal2013persistence}.
Furthermore, in the particular context of temporal graph visualization, it has been reported that the analysis of connected components allows for a rich exploration of the data \cite{tracking_graph1,tracking_graph2,nested_graphs2,nested_graphs1}.
Since the purpose of this article is to visualize the formation of groups within networks, i.e., of connected components, we will focus on $H_0$.
The higher homology groups $H_i$, $i>0$, although they could capture additional information (e.g., tunnels, voids), are beyond the scope of the paper.

\myParagrapho{Visualization.}
TDA has also seen applications in the context of (non-temporal) graph visualization.
By quantifying the strength of connections between the nodes of the graph, TDA can improve force-directed layouts and facilitate interaction with them \cite{doppalapudi2022untangling,8807379}.
One may also consider the connectivity between communities, resulting in new representations, such as those in \cite{tracking_graph2,8017588,fdata2021}.

In contrast, applications of TDA to the visualization of \textit{temporal} graphs are few.
The first work is found in \cite{nested_graphs1}, where the persistence diagram is used as a means to visualize the connected components generated by a scalar field.
However, in this case, PH is computed at the level of each snapshot, and therefore does not capture information about the dynamics of the data.To our knowledge, only \cite{8365984,myers2023temporal} propose visual layouts incorporating temporal information.
The former consists of a curve, exhibiting patterns and changes in behavior over time.
However, it does not provide information concerning the topology of the graphs at each snapshot. 
The latter layout uses the persistence barcodes given by the zigzag PH.
It displays the topological properties of the graphs at each timestamp and shows how they evolve over time.
Nevertheless, in some contexts, focusing only on graphs' topological properties, such as their number of connected components, can be too coarse and make analysis and visualization difficult for the user.
An important contribution of our work is to enhance this representation by incorporating information about the size and composition of the connected components.
These enhanced barcodes, that we call ``colored barcodes'', show promising results for graph visualization.

We draw the reader's attention to the fact that a close connection can be established between the persistence barcodes offered by TDA and certain popular visualization techniques.
In particular, the persistence barcodes of (ordinary) PH can be deduced from the merge tree of the data, and that of zigzag PH from its tracking graph \cite{tracking_graph1,tracking_graph2}.
In particular, the barcode graph \cite{dey2021computing} or formigram \cite{kim2018formigrams,kim2020analysis}, a handy tool of TDA, can be understood as a tracking graph.
This connection is studied in further detail in Sec.~\ref{subsec:ColoredBarcodeDesign}.
Similarly, the visualization proposed in this paper (Fig.~\ref{fig:teaser}(A)), , which incorporates additional information into the persistence barcodes, is related to the idea of nested tracking visualization \cite{nested_graphs2,nested_graphs1}. Both approaches draw flows between adjacent timestamps to represent events like merges and splits (in our case, triggered by user interaction). 
This last connection, however, only concerns visual representation, since these tools are designed to handle different information (nested components vs. disjoint components). Sec.~\ref{subsec:ColoredBarcodeDesign} discusses our design decisions and explains in more detail why nested tracking visualizations are not applicable in our case.

\subsection{Zigzag persistent homology}
\label{subsec:intro_zigzag}

We now succinctly introduce the topological tools used in this paper, and refer the reader to \cite{chazal2021introduction} for a thorough presentation.

\myParagrapho{Persistence modules.}
Zigzag persistent homology, introduced in \cite{carlsson2010zigzag}, is based on the notion of \emph{simplicial homology}.
Given an integer $i \geq 0$, the $i^\mathrm{th}$ homology functor $H_i$ is an operator that takes as input a graph $G$, and returns a vector space, denoted $H_i(G)$, which contains topological information about $G$.
\blue{As already discussed, we will only consider $H_0(G)$, the group of connected components, since it already enables a rich analysis of the graph's structure.}
It is a vector space whose dimension is equal to the number of connected components of $G$.

To define a zigzag PH, one has to first build a \emph{zigzag filtration}, that is, a sequence of graphs, such that for each pair of consecutive graphs, one of them is included in the other. In order to build such a filtration, consider the sequence of graphs $\{G_k\}_{k = 0}^{M}$ defined in the previous section, using the partition or sliding-window timeslicing.
By considering the union graph $G_k \cup G_{k+1}$ for all the pairs of consecutive graphs, one obtains a zigzag filtration
\begin{equation*}    
G_0 \hooklongrightarrow G_0 \cup G_1 \hooklongleftarrow G_1 \hooklongrightarrow G_1 \cup G_2 \hooklongleftarrow G_2 \hooklongrightarrow \dots
\end{equation*}

In this filtration, one is able to \emph{track the evolution} of the connected components: how they merge, split, appear or disappear. 

By applying the $H_0$-homology to this filtration, the graphs are transformed into vector spaces, and the inclusions into linear maps:
\begin{equation*}    
H_0(G_0) \rightarrow H_0(G_0 \cup G_1) \leftarrow H_0(G_1) \rightarrow H_0(G_1 \cup G_2) \leftarrow H_0(G_2) \rightarrow \dots
\end{equation*}

This sequence forms a \emph{zigzag persistence module}, an algebraic structure that condenses all the information concerning the evolution of the connected components.
For instance, one reads directly from these maps whether a connected component splits or is preserved; similarly, one reads whether two connected components merge.

\myParagrapho{Barcodes.}
To each persistence module is attached a \emph{persistence barcode}, denoted $\mathcal{B}$. It is a collection of intervals $[b,d]$, called \emph{bars}.
They are interpreted as follows.
For each timestamp $k$, the number of bars present at this time is equal to the number of connected components in the graph $G_k$.
Moreover, we can see how these connected components evolve:
To a bar $[b,d]$ corresponds a connected component of the graph born at time $b$ (either because new points appeared in the graph, or because an existing component split in two) and died at time $d$ (either because the points that compose it disappeared, or because it merged with another component).
The barcode is the main object of TDA, and can be understood as a visual representation of persistence modules.

\begin{figure}[t]
\centering
\includegraphics[width=\linewidth]{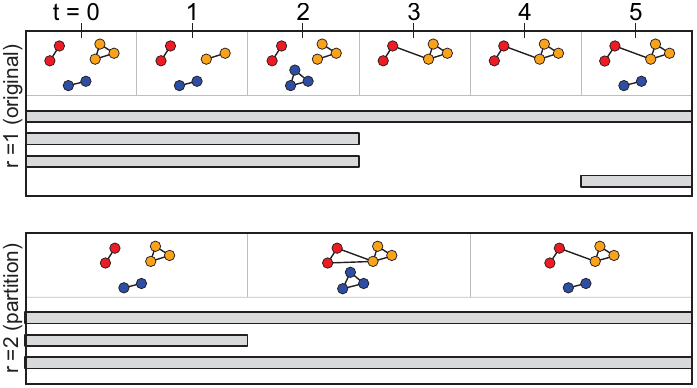}
\caption{Barcodes associated with a temporal graph at resolution $1$ and $2$. \blue{Each horizontal bar refers to a connected component throughout time.}}
\label{fig:barcodes}
\end{figure}

As an example, we give in Fig.~\ref{fig:barcodes} the persistence barcodes associated with a temporal graph at resolutions $1$ and $2$, as in Fig.~\ref{fig:timeslicing}.
Let us analyze the first barcode.
It contains a long bar $[0,5]$, indicating that there is a connected component that persists all along the filtration. 
We may think of it as representing the nodes colored in orange, or red.
Moreover, there are three smaller bars, depicting connected components that survive for a shorter time: one bar $[0,2]$ represents a component that merges with another (the red nodes with the orange nodes), another bar $[0,2]$ represents a component that disappears (the blue nodes) and reappears at $t=5$. 
Besides, the second barcode of Fig.~\ref{fig:barcodes} contains only three bars. 
Indeed, in the corresponding filtration, the blue nodes are always present, resulting in a long bar $[0,5]$ in the barcode.

It should be pointed out that the barcode does not allow one to identify directly which connected components the bars represent.
In certain cases, in the presence of many bars, for instance, this task can be difficult to perform visually. 
One part of our work consisted in defining an improved version of the barcode, called the colored barcode, which allows us to fix this problem (see Sec.~\ref{subsec:ColoredBarcode}).

\begin{figure}[t]
\centering
\includegraphics[width=\linewidth]{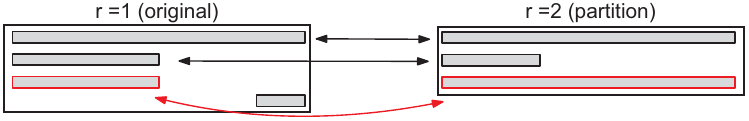}
\caption{A pairing between the barcodes of Fig.~\ref{fig:barcodes}. We outline in red the most distant paired bars (distance 3), causing the bottleneck distance.}
\label{fig:bottleneck}
\end{figure}

\myParagrapho{Bottleneck distance.}
Another fundamental feature of TDA is the possibility of comparing two persistence barcodes, through the notion of \textit{bottleneck distance}. 
In a few words, this distance seeks a pairing between the bars of the barcodes and computes the largest distance between a bar in the first barcode and one in the second. 
The exact definition is given in our supp. material (Sec.~A).
With respect to the bottleneck distance, two barcodes are close if the large bars of one can be matched with the large bars of the other, the short bars being forgotten.
Fig.~\ref{fig:bottleneck} shows a pairing between the barcodes in Fig.~\ref{fig:barcodes}.
The most distant bars in this pairing are the two bottom ones, $[0,2]$ and $[0,5]$. The distance between these bars is $3$, which is also the bottleneck distance, denoted $d_\mathrm{B}(\mathcal{B},\mathcal{B}')$.

The bottleneck distance lies at the core of our method and will be used as a means to select resolutions in Sec.~\ref{subsec:resolution_suggestion_method}.
Namely, we will compare the temporal graphs coming from two different resolutions via the bottleneck distance between the persistence barcodes coming from their zigzag filtrations.
This distance computes the global topological agreement between these temporal graphs, allowing us to determine whether they are similar or not, just as in the context of abrupt change detection.
In addition, the bottleneck distance offers two advantages.
First, it allows for a theoretical treatment: we will study in Sec.~\ref{subsubsec:renormalization} what values of distance are to be expected, and when they indicate a relevant change.
\blue{
Secondly, and as a consequence of its definition, the bottleneck distance is always caused by a pair of bars or a bar alone.
From a practical point of view, one can identify which nodes of the graph are responsible for the topological difference between two barcodes.
Based on this observation, we will describe an explainability pipeline in Sec.~\ref{sec:analysis_suggested_resolutions}.
}

\begin{figure*}[!t]
\centering
\includegraphics[width=\linewidth]{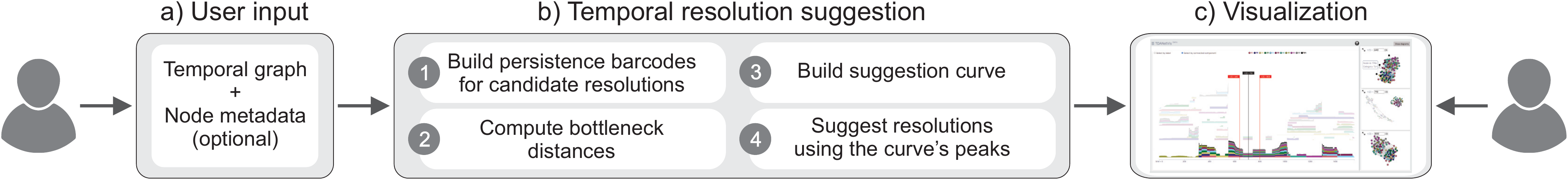}
\caption{ZigzagNetVis workflow. (a) Users input a temporal graph and node metadata (optional). (b) We suggest resolutions using a four-step procedure. (c) Users visualize the graph using any resolution through the colored barcode and node-link diagrams, visualizations that compose our prototype.}
\label{fig:workflow}
\end{figure*}

\section{Design Tasks and Workflow}

\myParagrapho{Design tasks.} 
Besides suggesting temporal resolutions, we seek to effectively explore the graph, and identify global and local behaviors and patterns, under a given temporal resolution.
In that sense, we designed our visual components and interaction to meet high-level tasks derived from low-level tasks and dimensions proposed in Bach et al.'s taxonomy for temporal graph exploration~\cite{mainTaskTaxonomy1}.

Specifically, we combine the three task dimensions described in this taxonomy: \textit{temporal/when} (easy identification and reaching of specific time steps); \textit{topological/where} (easy identification, situation, and tracking of elements with properties of interest); and \textit{behavioral/what} (easy understanding of the behaviors and changes that affect elements of interest). These dimensions help generate the following tasks, which should be satisfied during the graph analysis under any temporal resolution.

\myparagraph{T1:} Analyze particular groups of elements (entire network, connected components, or nodes) in terms of identification, situation, and inspection at a given time of interest. 

\myparagraph{T2:} Analyze the temporal evolution of particular groups of elements, identifying, e.g., the addition or deletion of elements and abrupt increases or decreases of an element property (referred to as \textit{peak} or \textit{valley} events in Ahn et al.'s taxonomy~\cite{mainTaskTaxonomy2}). 

\myparagraph{T3:} Identify and compare structural changes that occur at particular times of interest.


In addition to the \textit{when}, \textit{where}, and \textit{what} dimensions from Bach et al.'s taxonomy, we further consider \textit{why} and \textit{how} task descriptions from Brehmer \& Munzner’s multi-level typology of visualization tasks~\cite{taxonomiaTamara}. From the \textit{why} point of view, our tasks enable \textit{discoveries}, which include the generation and verification of hypotheses. To achieve that, users first \textit{locate} groups of elements of interest (tasks T1, T2) or at particular times (task T3). Alternatively, they can freely \textit{explore} the visualization to find elements/times of interest (e.g., based on global patterns or anomalies). Once these are found, users may \textit{identify}, \textit{compare}, and \textit{summarize} elements or patterns (T1-T3). From the \textit{how} perspective, our views will meet the tasks by \textit{encoding} the network data and by providing \textit{manipulation} methods such as \textit{selection}, \textit{navigation}, and \textit{filtering}. They will also \textit{introduce} new elements to the visualization by \textit{importing} network data on demand.





\myParagrapho{Workflow.} As illustrated in Fig.~\ref{fig:workflow}(a), users first input a temporal graph and its node categorical metadata (optional). 
The resolution suggestion then proceeds as follows (Fig.~\ref{fig:workflow}(b), details in Sec.~\ref{subsec:resolution_suggestion_method}): we build persistence barcodes for every candidate resolution (predefined range of values, e.g., [1,100]); we compute the bottleneck distance between pairs of barcodes, and build a suggestion curve using the distances.
Resolutions are then suggested based on the curve’s peaks.
Finally, users visualize the graph under any resolution by using our proposed layout --- the \textit{colored barcode} (Sec.~\ref{subsec:ColoredBarcode}) --- and associated node-link diagrams, visualizations that compose our system prototype (Fig.~\ref{fig:workflow}(c), details in Sec.~\ref{sec:systemPrototype}).

\section{Temporal resolution suggestion}
\label{subsec:resolution_suggestion_method}

\subsection{Description of the method}
\label{subsec:description_method}
As discussed above, the choice of a resolution significantly impacts the analysis of a temporal graph. 
In practice, one wishes to select an ``optimal'' resolution.
However, the problem is ill-posed: various resolutions may be relevant, leading to different analyses.
To circumvent this issue, our strategy selects a collection of resolutions, each of which reveals different behaviors of the temporal graph.

Let us consider an initial set of resolutions $\{r_0, \dots, r_n\}$, to be tested, and a parameter $m$, the number of requested resolutions. Our method consists in partitioning this set into $m$ subsets of consecutive resolutions,
\begin{equation}
\{r_{i_0} = r_0, \dots, ~r_{i_1}\}, 
~~\{r_{i_1}, \dots, r_{i_2}\}, 
~~\dots, 
~~\{r_{i_{m-1}}, \dots, r_{i_{m}}=r_{n}\},
\label{eq:partition_resolutions}
\end{equation}

where each subset consists of resolutions for which the temporal graphs exhibit similar behavior. We will quantify this similarity using zigzag PH, as explained in the next paragraph.
As a last step, we will choose a resolution in each of these subsets --- for instance, the first ones, $r_{i_0}, \dots, r_{i_{m-1}}$ --- therefore yielding an exhaustive sample of all possible behaviors exhibited by the temporal graph.

\begin{figure}[!t]
\center
\includegraphics[width=\linewidth]{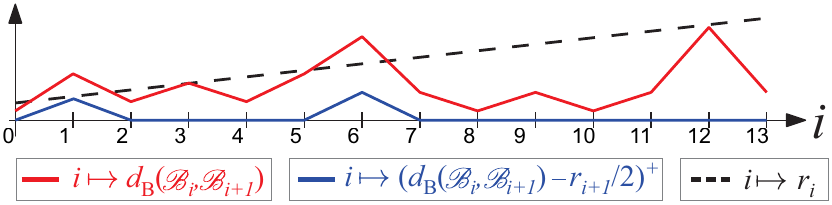}
\caption{
A suggestion curve in red, its corresponding normalized suggestion curve in blue (for partition timeslicing), and the curve $i\mapsto r_{i}$ in black.
}
\label{fig:suggestion curve}
\end{figure}

Our method for obtaining a partition as in Eq. \eqref{eq:partition_resolutions} consists in comparing each pair of consecutive resolutions $r_{i}$ and $r_{i+1}$, and in detecting abrupt changes in the corresponding temporal graphs. This detection is performed using zigzag PH, as follows.
First, we perform timeslicing on the temporal graph $G$ for both resolutions, using partition or sliding-window, as described in Sec.~\ref{subsec:timeslicing}.
Second, we compute the corresponding persistence barcodes $\mathcal{B}_{i}$ and $\mathcal{B}_{i+1}$, as well as their bottleneck distance $d_\mathrm{B}(\mathcal{B}_{i},\mathcal{B}_{i+1})$, as described in Sec.~\ref{subsec:intro_zigzag}.
Gathering all the bottleneck distances yields a sequence
\begin{equation*}
d_\mathrm{B}(\mathcal{B}_{0},\mathcal{B}_{1}),
~~d_\mathrm{B}(\mathcal{B}_{1},\mathcal{B}_{2}),
~~\dots,
~~d_\mathrm{B}(\mathcal{B}_{n-1},\mathcal{B}_{n}),
\end{equation*}

which we represent as a curve, drawn in red in Fig.~\ref{fig:suggestion curve}. We refer to it as the \emph{suggestion curve}.

On the suggestion curve, peaks correspond to consecutive resolutions for which the associated barcodes are significantly different, which we interpret as structural topological changes in the temporal graphs. 
Finally, we identify the peaks of this curve and use them as separators to obtain the partition of Eq.~\eqref{eq:partition_resolutions}.
We give further explanations in the next section.

In a nutshell, our methodology employs the bottleneck distance between consecutive resolutions as a feature to perform abrupt change detection.
While change detection based on features is common in temporal graph analysis (see, e.g.,~\cite{MultiPiles}), incorporating PH offers several advantages.
First, thanks to the high interpretability of PH, we can give a heuristic analysis in Sec.~\ref{subsubsec:renormalization}, already yielding important insights.
Moreover, as studied further in the supp. material, the bottleneck distance appears to be a stable and relevant quantity, gathering information from various other features of the literature.

\subsection{Timestamps shifts and structural changes}
\label{subsubsec:renormalization}

In the previous section, we have built the suggestion curve $i \mapsto d_\mathrm{B}(\mathcal{B}_{i},\mathcal{B}_{i+1})$.
In order to identify relevant peaks of this curve, we need to give some comments regarding the values it can take.

\myParagrapho{Partition timeslicing.}
Let us first consider that we have chosen the partition timeslicing.
By going from resolution $r_i$ to $r_{i+1}$, one alters the timestamps: a timestamp for the first resolution will be at a distance at most $r_{i+1}/2$ of a timestamp for the second one. 
Consequently, we expect that the bars of the persistence barcode will be displaced by a distance of at most $r_{i+1}/2$.
This interpretation leads us to distinguish two values of the bottleneck distance.
\squishlist
    \item If $d_\mathrm{B}(\mathcal{B}_{i},\mathcal{B}_{i+1})\leq r_{i+1}/2$, the distance is merely caused by artificial changes coming from the modification of the timestamps' values. We call it \emph{timestamps shift}.
    \item If $d_\mathrm{B}(\mathcal{B}_{i},\mathcal{B}_{i+1})> r_{i+1}/2$, the distance is no longer just caused by the displacement of the timestamps: we consider that the temporal graph has undergone a \emph{structural change}.
\squishend
In order to estimate the structural changes only, we must detect the values of the suggestion curve that exceed $r_{i+1}/2$.
In other words, we seek the positive values of the \emph{normalized suggestion curve}:
\begin{equation*}
    i \longmapsto (d_\mathrm{B}(\mathcal{B}_{i},\mathcal{B}_{i+1})-r_{i+1}/2)^+
\end{equation*}

where $(\cdot)^+$ denotes the positive part of a real number.
This curve is represented in Fig.~\ref{fig:suggestion curve}.
In this example, we would detect the resolutions $r_2$ and $r_7$ as values that cause structural changes since they are the first resolutions after the peaks occurring at $r_1$ and $r_6$.

Fig.~\ref{fig:barcodes} provides another example.
Going from resolution $r_i = 1$ to $r_{i+1} = 2$, we have seen previously that the bottleneck distance is equal to $3$, greater than $r_{i+1}/2 = 1$, hence we observe a structural change.
It is caused by the two blue bars merging together.
If we had considered only the red and yellow bars, we would have observed a bottleneck distance of $2$, i.e., a timestamps shift.

\myParagrapho{Sliding-window timescling.}
We now turn to the case of sliding-window timeslicing.
By going from resolution $r_i$ to $r_{i+1}$, the activation windows of the edges are only altered by a value $(r_{i+1}-r_i)/2$.
Consequently, we expect that the bars of the barcode will be displaced by a distance of at most $(r_{i+1}-r_i)/2$.
This leads us to define a \emph{timestamps shift} if $d_\mathrm{B}(\mathcal{B}_{i},\mathcal{B}_{i+1})\leq (r_{i+1}-r_i)/2$, and \emph{structural change} if $d_\mathrm{B}(\mathcal{B}_{i},\mathcal{B}_{i+1})> (r_{i+1}-r_i)/2$.
Accordingly, we define the \emph{normalized suggestion curve} as
\begin{equation*}
  i \longmapsto (d_\mathrm{B}(\mathcal{B}_{i},\mathcal{B}_{i+1})-(r_{i+1}-r_i)/2)^+.  
\end{equation*}

As before, we identify structural changes through its positive values.

In practice, users can select the preferred timeslicing method prior to applying the resolution suggestion technique.
However, the results obtained for partition or sliding-window may be different.
In the case of partition, a particularly inconvenient phenomenon occurs. 
Two bars of the barcode might merge between $r_i$ and $r_{i+1}$, provoking a structural change, and then split between $r_{i+1}$ and $r_{i+2}$, again provoking a structural change.
We call this phenomenon \emph{instability}, and we explain the situation in more detail in our supp. material (Sec.~B.1).
Consequently, we recommend that users use sliding-window timeslicing, and we make this choice in the rest of this article, except when stated otherwise.

\myParagrapho{Peak detection.}
In real-life examples, the normalized suggestion curve may contain many positive values.
However, returning all the corresponding resolutions to the user would not be relevant.
Instead, we choose to return only the most prominent peaks of the curve.
In practice, prominence is computed using the package \texttt{signal} of \texttt{scipy}.
We return only $m=5$ maxima, five being an arbitrary value that we found suitable.
We will give concrete outputs of our algorithm on eight temporal graphs in Section~\ref{sec:analysis_suggested_resolutions}.

\myParagrapho{Other distances.}
\blue{
In TDA, one chooses a distance according to the context: while the bottleneck distance calculates the maximal discrepancy between two barcodes, the Wasserstein distance incorporates all perturbations.
The latter option is of interest, for instance, when low-persisting features matter \cite{chazal2021introduction}.
In contrast, our work aims to detect structural changes, which are evidenced by the perturbation of a single bar of the barcode. Therefore, the bottleneck distance appears as a natural choice (see, for instance, Fig.~\ref{fig:barcodes}). 
This observation is supported by Sec.~C.2.4 of our supp. material, where it is shown that the Wasserstein distance leads to less interpretable results.
In the same vein, one could use, instead of the bottleneck distance, any feature that quantifies the proximity between two temporal graphs.
To this end, many quantities exist, such as those presented in Section C.2.3 of our supp. material (e.g., mean degree, density, or burstiness).
However, they all appear to either lack stability or provide limited information.
Our experimental study shows that the bottleneck distance acts as a relevant trade-off between stability and information, ``incorporating'' several popular features.
}

\section{Visualization}

\subsection{Colored barcode layout}
\label{subsec:ColoredBarcode}

In practice, the barcodes of TDA may not contain enough information: one is not able to identify which nodes are part of which bar.
Indeed, the barcode is built from the homology groups $H_0(G_k)$ of the graphs, where the information about the nodes has been lost.
A contribution of our work is to adapt and implement an algorithm that identifies the nodes that compose each bar.

\myParagrapho{Nodes identification.}
Consider a temporal graph, the sequence of graphs $\{G_k\}_{k = 0}^{M}$ obtained by timeslicing, and the $H_0$-barcode $\mathcal{B}$ of its zigzag filtration.
We wish to find, for each bar $I\in\mathcal{B}$ and each timestamp $k \in I$, a connected component $C^I_k$ such that
\squishlist
    \item for each timestamp $k$, if $\mathcal{B}^k$ denotes the set of bars living at time $k$, then the set $\{C^I_k \mid I \in \mathcal{B}^k\}$ is a partition of the set of nodes of $G_k$,
    \item for each bar $I\in\mathcal{B}$ and each $k \in I$ such that $k+1 \in I$, we have $C^I_k\cap C^I_{k+1}\neq\emptyset$. 
\squishend
The first point guarantees that we do not attribute the same node to two bars at the same timestamp, and the second point that, within a bar, we choose a sequence of connected components that are connected one to another.
Such a choice is possible as a consequence of previous work, which is detailed below.

Once the node identification has been done, this information can be incorporated into the persistence barcode. 
By attributing to each node or cluster of nodes a color (representing, e.g., node metadata information), we paint the bars in accordance with the nodes it contains.
We also vary the height of the bars to indicate the number of nodes.
We call this representation the \emph{colored barcode}. In case it is not possible to assign different colors to nodes (e.g., when there are no node metadata), we use a single color and only consider the variation of the heights of the bars.

\begin{figure}[t]
\centering
\includegraphics[width=\linewidth]{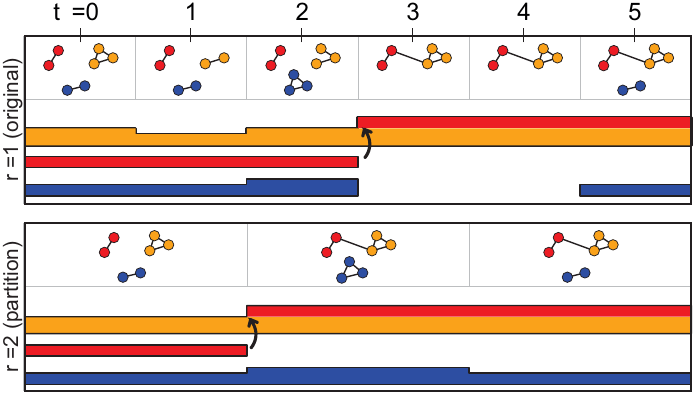}
\caption{Colored barcodes corresponding to the barcodes of Fig.~\ref{fig:barcodes}. Vertical arrows depict component merging.}
\label{fig:colored_barcode}
\end{figure}

We give in Fig.~\ref{fig:colored_barcode} two examples of colored barcodes, where the nodes are divided into three clusters: red, orange, and blue.
They correspond to the (non-colored) barcodes of Fig.~\ref{fig:barcodes}.
On the first colored barcode (Fig.~\ref{fig:colored_barcode}(top)), one reads that a connected component persists throughout the filtration, initially composed of orange nodes and later receiving the participation of red nodes. One can also visualize the connected component formed by the blue nodes, which disappears and reappears at $t=5$.

The choice of nodes composing each bar is not unique.
For instance, on the first barcode of Fig.~\ref{fig:colored_barcode}, the long bar starts with only orange nodes, until $t=3$, where red nodes connect.
In this example, one could have chosen to start this long bar with red nodes instead.
The analysis of the colored barcode, however, is independent of this choice. 
The user must keep in mind that, when two connected components merge, only one of the two has been arbitrarily chosen to appear at the beginning of the corresponding bar.

\begin{figure}[t]
\centering
\includegraphics[width=\linewidth]{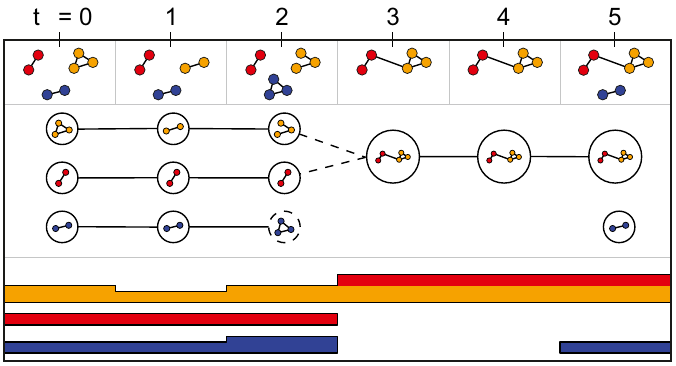}
\caption{\blue{On top of a zigzag filtration (top) is built the barcode graph (middle). By keeping the nodes' information, an adaptation of \cite[Algorithm 1]{dey2021computing} enables us to compute the colored barcode (bottom).}}
\label{fig:algo_zigzag}
\end{figure}

\myParagrapho{Algorithm.}
We now turn to the implementation of nodes identification, based on the work of Dey and Hou \cite{dey2021computing}. 
As described in the article, there exists an intermediate construction between the zigzag filtration and the persistence barcode, called the \emph{barcode graph}.
It is built recursively by studying how the temporal graph $\{G_k\}_{k=0}^{M}$ evolves: creating, removing, merging, or splitting connecting components.
Formally, each node of the barcode graph is associated with a connected component of $G_k$ at a certain time $k$.
Moreover, an edge is added between two components at times $k$ and $k+1$ if they share a node (see Fig.~\ref{fig:algo_zigzag}).
We draw the reader's attention to the fact that the barcode graph is a tracking graph (see Sec.~\ref{subsec:PH_applied_to_graph}).

Algorithm 1 of the aforementioned article allows one to deduce, from the barcode graph, the persistence barcode of the zigzag filtration.
To do so, the authors recursively build the \textit{barcode forest}, a complementary construction.
For the most part of the algorithm, when iterating through the filtration, five events may happen: \textsc{Entrance}, \textsc{Departure}, \textsc{No-Event}, \textsc{Merge} and \textsc{Split}. They respectively represent that a node entered the filtration, that a node left the filtration, that an edge entered or left without changing the topology of the graph, that an edge entered the filtration provoking two connected components to merge, or that an edge left the filtration provoking a connected component to split in two.
Only \textsc{Departure} and \textsc{Merge} provoke the appearance of a new bar.

In our context, since we wish to identify the nodes that compose the bars, we incorporate a further step in this procedure.
During the event \textsc{Departure} (as the dashed node of Fig.~\ref{fig:algo_zigzag}), we simply collect the connected components written on the path, and add this information to the bar of the barcode.
For \textsc{Merge} (as the dashed edges of Fig.~\ref{fig:algo_zigzag}), there are two potential paths; we choose one arbitrarily, associate the new bar with the connected components written on it, and remove these components from the graph.

Note that the latter choice is not unique \blue{as the ``elder rule'' holds for ordinary PH, but not in the zigzag
context. For instance, if the filtration consists of one connected component, that splits in two
(at time $t_1$) and then merges (at time $t_2$), then the barcode will consist of two bars, one bar being [$t_1$, $t_2$]. However, two choices of identification are possible for this bar, and none is canonical.}
In practice, when choosing a path to remove, we remove the one that starts with the least number of nodes. 
This allows for maintaining homogeneity within the bars; that is, bars containing many nodes will continue to have many nodes.

One final detail should be noted.
The original algorithm takes as input a zigzag filtration such that two consecutive graphs are obtained one from the other by adding or removing a single node or edge.
However, partition or sliding-window timeslicings may yield filtrations where several nodes or edges are added or removed at the same time.
Consequently, we must apply a pre-processing step to assign each event a unique time.
Once the algorithm has been performed, we go back to the initial timestamps.

\subsection{Design decisions}
\label{subsec:ColoredBarcodeDesign}
In practice, our colored barcode is a timeline visualization that
can be thought of as a series of stacked area charts, each referring to a connected component and its node members throughout time (Fig.~\ref{fig:teaser}(A)). As mentioned above, the color and height indicate the label (node metadata) and number of nodes, respectively.


\myParagrapho{Alternative designs.}
We have considered \textit{alternative design choices} before proposing this layout. We decided to use a timeline visualization instead of animations to better meet analyses that rely on multiple and often distant timestamps (tasks T2, T3), complying with~\cite{SurveyDynamicVisualization}. We then studied the suitability of existing timeline visualizations for our context. An option would be a visualization based on tracking graphs~\cite{LargeNetVis, tracking_graph1,tracking_graph2}. In particular, LargeNetVis's Global View~\cite{LargeNetVis} is a grid-based layout where rows and columns represent respectively network communities and timeslices. In this view, communities are encoded as circles with varying sizes, and their temporal evolution is depicted through links connecting communities from consecutive timeslices. Although we could adapt it to encode connected components, it would still not provide immediate information about components' node members. The identification and tracking of the members would also not be immediate with other visualizations, for instance, MSV~\cite{Elzen2014}, PAOH~\cite{PaohVis}, and TAM~\cite{LargeNetVis}. 

\blue{We have also considered nested tracking graph visualizations~\cite{nested_graphs1, nested_graphs2}, but we opted for a different approach due to their inherent limitations in our specific context, particularly those concerning \textit{visual scalability}, in terms of the number of timestamps, and the intrinsic \textit{components' hierarchy} they consider. First, nested tracking graphs depict the graph evolution by representing events (i.e., merges and splits) through horizontal flows drawn between uniformly spaced timestamps. While our visualization also enables the analysis of these events upon interaction, our primary focus lies on depicting the graph structure at each timestamp and the most salient connected components, namely the longest bars throughout time. We, therefore, provide a more scalable solution for our case by employing vertical flows alongside timestamps that are positioned one right after the other. As an example, while Figs.~\ref{fig:teaser}~and~\ref{fig:colored_barcodes_PrimarySchool_HighSchool} illustrate our layout for a graph with 1,555 timestamps, all visual analyses from~\cite{nested_graphs1, nested_graphs2} consider a maximum of 100 timestamps.}

Secondly, nested tracking graphs leverage hierarchical relationships coming from superlevel sets to derive component visibility, which can potentially lead to occlusions and impair the identification of patterns that would be crucial to our context (e.g., the number of nodes in a given component or its life-cycle). Conversely, we consider graphs with components without a hierarchical relationship (disjoint). Stacking them instead of nesting allows for immediate and effortless recognition of individual components and their attributes. 
Overall, we consider our colored barcode layout to be a better solution for our tasks and goals. It is suitable for graphs with many timestamps, emphasizes structural clarity, and enables efficient identification and exploration of key connected components and timestamps.

\myParagrapho{Bars' positioning.}
We use two representations for the bars' positioning in our colored barcode.
The first consists of fixing the bars' bottom ordinate and distributing them upwards only (see Fig.~\ref{fig:teaser}) --- we call this approach \textit{``bottom-based ordering''}.
Inspired by well-established cluster positioning on Sankey- and nested graph visualizations (e.g.,~\cite{nested_graphs1, nested_graphs2}), our second representation considers centered bars whose height varies uniformly up and down (see Fig.~\ref{fig:colored_barcodes_PrimarySchool_HighSchool}) --- we call it \textit{``center-based ordering''}.
In both cases, the bars are arranged in such a way as to reduce the space they occupy in the interface and the lengths of the vertical flows.

\subsection{System prototype}
\label{sec:systemPrototype}

We now describe the interface and interactions that compose the prototype of our ZigzagNetVis system (see a screenshot in Fig.~\ref{fig:teaser}), a web-based visual analytics tool that incorporates all steps of our workflow and was used by our user study participants (Sec.~\ref{sec:userStudy}).

When first loaded, the system automatically opens a menu through which it is possible to input a network and node categorical metadata (optional). The system then suggests temporal resolutions for the inputted network following the procedure described in Sec.~\ref{subsec:resolution_suggestion_method}. 
To help users choose among the suggested resolutions, they can ask for quantitative network measures (or features). For each resolution, the system will display values for burstiness~\cite{samplingLuis}, average lifetime of edges~\cite{samplingLuis}, normalized stability~\cite{stability} and the inverse of the normalized fidelity --- the original fidelity~\cite{stability} gives us a distance measurement and we use the similarity counterpart.
In our case, a higher value indicates greater faithfulness of the network under the selected resolution to the original network ($r = 1$).
After choosing a resolution, users can filter out bars (i.e., connected components) with less than $x$ node or with duration less than $y$ timestamps, $x$ and $y$ being user-defined. We provide a visual comparison of different filtering parameters in our supp. material (Sec.~C.1).

Once the network, temporal resolution, and the other parameter values are chosen, the system exhibits its first and main view (Fig.~\ref{fig:teaser}(A)), which contains the colored barcode and appears with maximized height and width, i.e., also occupying the screen space on Fig.~\ref{fig:teaser}(C-E). This view adopts as default the bottom-based component ordering, but the user is free to change it at any time (Fig.~\ref{fig:teaser}(H)).
Besides zoom in/out and pan, users can select specific connected components or bars representing nodes that share the same label (Fig.~\ref{fig:teaser}(G)). In this way, it is possible to analyze their behavior at particular timestamps (tasks T1, T3) and evolution throughout time (task T2). Nodes sharing the same label can be selected in the layout by hovering over the label of interest in the color legend or the bar with the color associated with that label. Likewise, a connected component can be selected by hovering over any of its bars (Fig.~\ref{fig:teaser}(A)). It is also possible to persist the current selection (left-click) and select multiple labels (CTRL + left-click).

\begin{figure}[t]
\centering
\includegraphics[width=\linewidth]{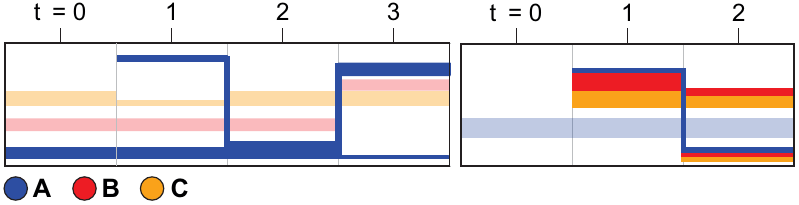}
\caption{Toy example showing the behaviors when marking the checkbox \textit{``See flows under interaction''} and hovering over a bar: (left) if selecting by label, the system tracks same label nodes throughout time; (right) if selecting by conn. component, the system shows merges/splits in the selected component.}
\label{fig:didatic_interaction_flow}
\end{figure}

\begin{figure*}[t]
\centering
\includegraphics[width=\linewidth]{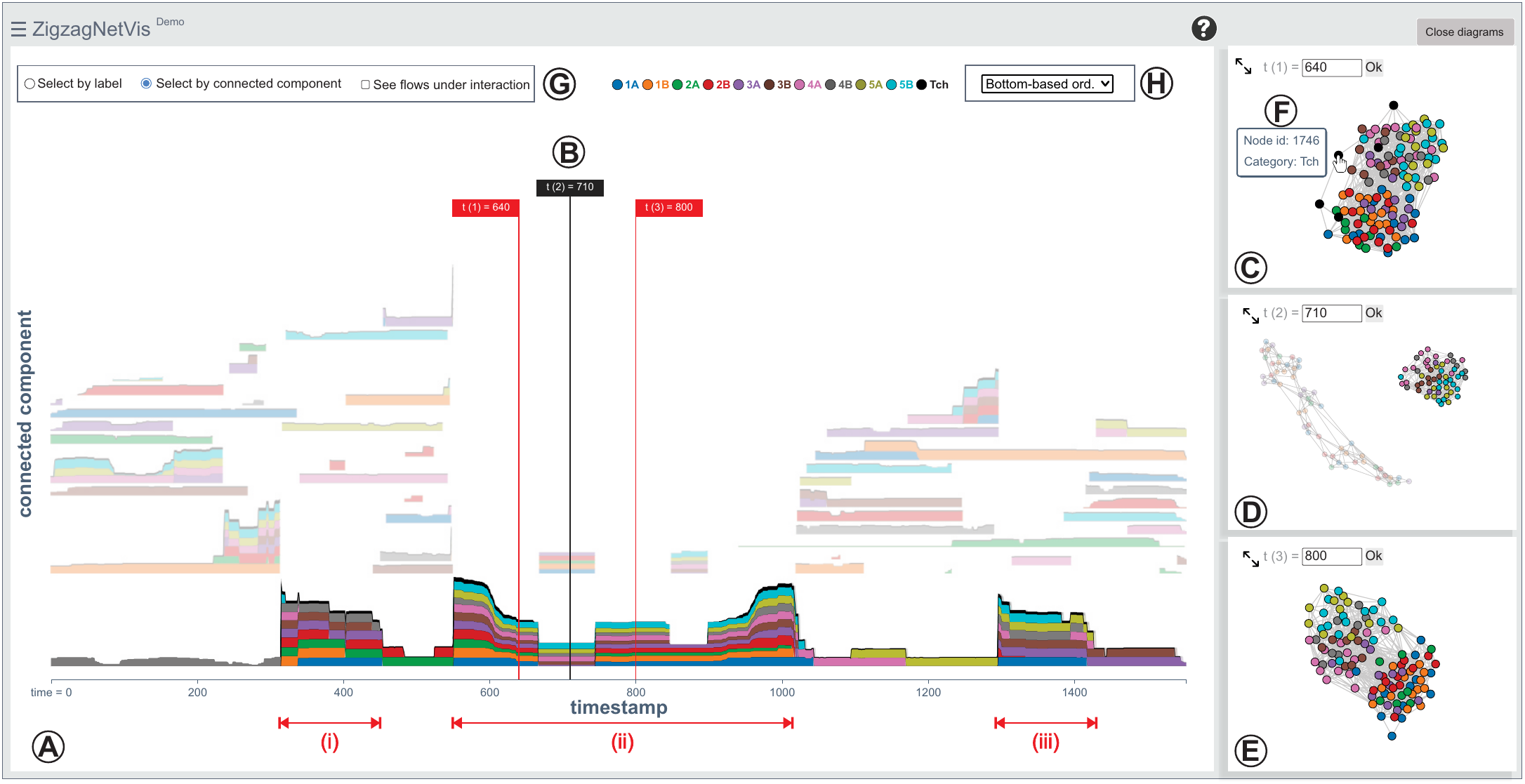}
\caption{
ZigzagNetVis system prototype, an interactive and web-based system with linked views designed to assist the analysis of temporal graphs by highlighting connected components' structure and evolution. (A) Colored barcode with bottom-based ordering that highlights the longest connected component in the graph --- \blue{note that (i), (ii), and (iii) represent time intervals with few connected components compared to others}. (B) Timestamp markers indicating the three timestamps being depicted by (C-E) the three node-link diagrams. (F) Tooltip showing extra information. (G) Users can select groups of nodes by label or by connected component. (H) Users can choose between two available component positioning strategies. }
\label{fig:teaser}
\end{figure*}

Two behaviors are expected when marking the checkbox \textit{``See flows under interaction''} (Fig.~\ref{fig:teaser}(G)) and hovering over any bar of a component: (i) if selecting by label (see Fig.~\ref{fig:teaser}(G) again), the system enables tracking the nodes with that (or those) label(s) throughout the connected components over time, as illustrated in Fig.~\ref{fig:didatic_interaction_flow}(left); (ii) if selecting by connected component, the system shows the events that connect that component to others over time through vertical flows that indicate merges and splits (Fig.~\ref{fig:didatic_interaction_flow}(right)).

After finding a potentially relevant timestamp or interval for analysis, the user double-clicks near it and the system opens three node-link diagrams as presented in Fig.~\ref{fig:teaser}(C-E), one showing the network structure at the timestamp of interest (referred to as $t(2)$, see Fig.~\ref{fig:teaser}(D)) and two others, by default, for $t(2) \mp 10$ timestamps (referred to as $t(1)$ and $t(3)$, see Fig.~\ref{fig:teaser}(C,E)). 
\textit{Timestamp markers} are inserted in the colored barcode to highlight the three timestamps whose node-link diagrams are opened (Fig.~\ref{fig:teaser}(B)).

Users can freely change the three timestamps being analyzed --- note, e.g., the values for $t(1)$, $t(2)$, and $t(3)$ in Fig.~\ref{fig:teaser}. This way, they can analyze the structure of groups of elements at different granularity levels (from the entire network to individual nodes) for any timestamp (task T1), as well as identify and compare structures and temporal behaviors by analyzing multiple node-link diagrams (tasks T2, T3). There are two ways for the user to reach a new timestamp of interest. If the user knows \textit{a priori} which timestamp is relevant for analysis, they can simply type the new timestamp value in the node-link diagram area to update it; the system then repositions the corresponding timestamp marker accordingly. However, if the user is interested in analyzing a timestamp or interval that caught their attention because of an unexpected behavior found on the colored barcode, they can drag and drop one or more timestamp markers to that timestamp or interval; the system then updates the node-link diagram(s) accordingly.

\myParagrapho{Node-link diagram.} Given a selected timestamp of interest $t_k$, our node-link diagram shows all nodes and edges active at $t_k$ using a spring-force node positioning~\cite{ForceDirectedStructural}. Nodes are colored using the same color scale as in the colored barcode. In addition, the system also shows a tooltip with node id and label whenever a node is hovered over, as illustrated in Fig.~\ref{fig:teaser}(F). The user can expand one or more node-link diagrams (button \img{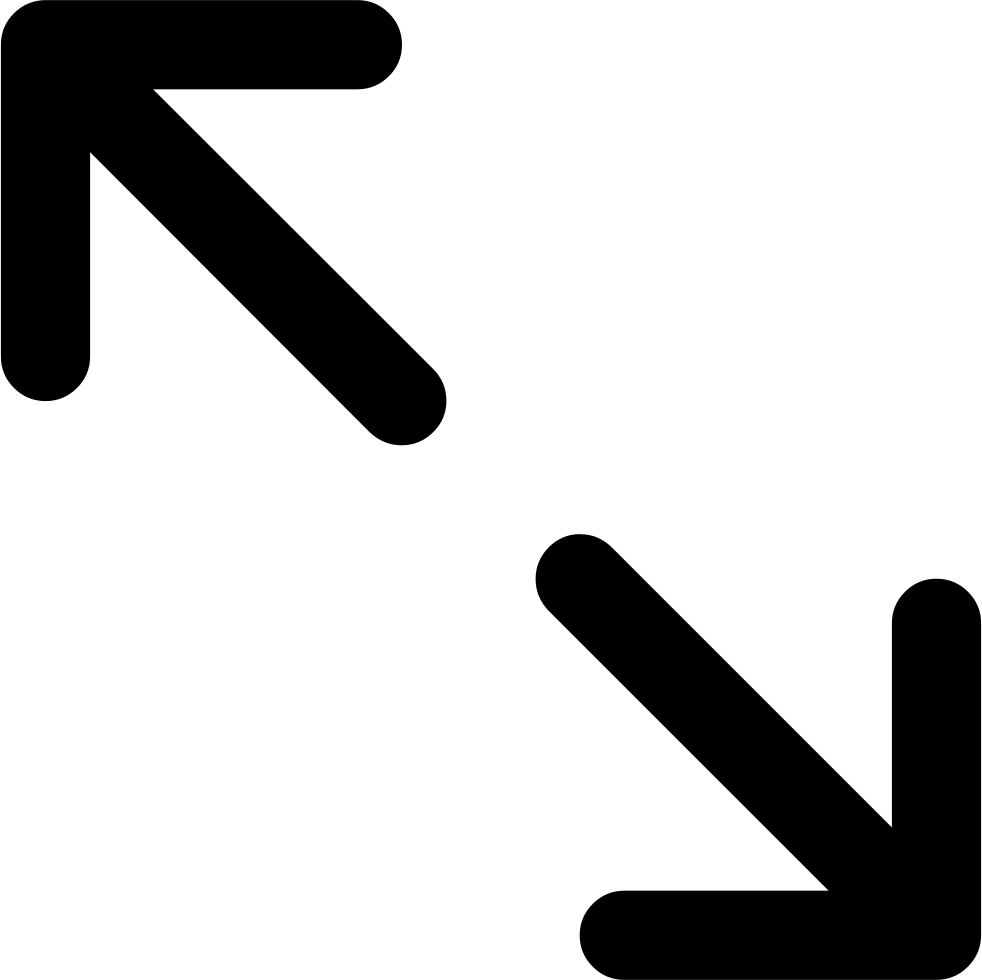}) and drag/drop their maximized versions, e.g., to put them side-by-side and optimize comparisons.
Depending on the type of selection (recall Fig.~\ref{fig:teaser}(G)), a click on a node $x$ in a diagram (expanded or not) selects all nodes that contain the same label as $x$ or all nodes that belong to the same connected component as $x$ (T1).
To help users compare structures and temporal behaviors (T2, T3), all node-link diagrams (expanded and non-expanded) are coordinated with each other and with the colored barcode: groups of nodes selected in one of them are automatically selected in the others (see, e.g., the non-selected connected component in Fig.~\ref{fig:teaser}(A,D), $t(2) = 710$).

\myParagrapho{Design decisions.} 
Besides the decisions made on the colored barcode (recall Sec.~\ref{subsec:ColoredBarcodeDesign}), we also studied \textit{alternative approaches} before choosing static node-link diagrams to explore the network structure at particular times. First, we considered using animations to show the evolution of the network during the time interval selected through the timestamp markers. We gave up this idea because animations have limitations on tasks involving multiple and distant timestamps~\cite{SurveyDynamicVisualization}. After opting for ``static'' visualizations, we considered node-link diagrams and adjacency matrix-based visualizations~\cite{matrix}. We chose the former as it would be easier to identify connected components using the diagram, especially when adopting spring-force node positioning. Finally, we decided to enable the analysis of three timestamps (three node-link diagrams at once) based on the intuitive notion of past, present, and future.

As mentioned, our system prototype associates different colors to nodes (or bars) with different labels when this metadata information is available. Color-blind users can use a color scheme that is safe from color blindness. 
Our prototype also provides a series of features that help colorblind users in their analysis, e.g., by allowing selections and by showing informative tooltips. In the user study, we validated visualizations and color scheme with two self-declared colorblind participants (see Sec.~\ref{ux_results}).

\myParagrapho{Implementation details.} We use a client-server architecture. The server side was implemented in Python and uses popular libraries and frameworks (e.g., \texttt{NetworkX}, \texttt{Flask}, and \texttt{Dionysus2}). We used the \texttt{D3} library in our views.
A demo version of the system, used by our user study participants and already including suggestions, is available at \url{https://github.com/raphaeltinarrage/ZigzagNetVis}.

\myParagrapho{Computational complexity.}
The overall ZigzagNetVis process can be divided into three steps: open the dataset (1), compute the suggestion curve (2), and compute the colored barcode for one resolution (3).
Let $m$ be the number of pairs $(\mathrm{edge}$, $\mathrm{time})$ in the temporal graph, and let $n$ be the number of resolutions tested.
Step 1 consists in reorganizing these pairs in a dictionary, and creating a list of unique edges, resulting in a computational complexity of $O(m)$.
In Step 2, we create $n$ zigzag filtrations, compute their $H_0$-homology barcode, and then compute the consecutive bottleneck distances.
The respective complexities are $O(nm)$, $O(nm\alpha(m))$, and $O(nm^{1.5})$, where $\alpha$ is the inverse Ackermann’s function (approximately constant in practice) \cite{dey2021computing}.
Last, Step 3 consists of one computation of zigzag persistence, which therefore has a computational complexity of $O(n\alpha(n))$.
In general, the complexity of the process is $O(nm^{1.5})$.
We should mention, however, that our personal implementation of the persistence algorithm does not reach the complexity mentioned above and can potentially yield longer execution times.
In our supp. material (Sec.~C.3), we give the running times observed in practice for eight temporal graphs.

\section{Datasets}
\label{datasets}

Our usage scenario and user study explore the first day of data from two real-world and face-to-face temporal graphs collected in educational environments, the Primary School~\cite{primarySchool} and the High School~\cite{highSchool} networks. We have chosen these graphs as they have been extensively analyzed in the context of temporal graph visualization~\cite{3399922,8365984,7192717, LargeNetVis, PONCIANO2021170, 0228728, artigoSara} and because they contain relevant node metadata information.

The first day of the Primary School network~\cite{primarySchool} contains 236 nodes (students and teachers from the first to the fifth grade, each having classes A and B) and 60,623 edges, which represent face-to-face interactions. There are 1,555 timestamps in the original resolution ($r = 1$), each comprising a 20-sec interval. 
Data were collected from 8:45 am to 5:20 pm. There is a lunch break from 12pm to 2pm and two smaller breaks (20-25 min), one in the morning (around 10:30am) and one in the afternoon (around 3:30pm). Each of the 10 school classes has an assigned teacher. For convenience, we will refer to each class using simple terms, for example, 1B to refer to ``first grade, class B''.

The High School network~\cite{highSchool} contains face-to-face interactions between students from nine classes related to different subjects: chemistry and physics (classes PC and PC1), mathematics and physics (classes MP, MP1, and MP2), engineering (class PSI), and biology (classes 2BIO1, 2BIO2, and 2BIO3). The first day contains 312 nodes and 28,780 edges distributed in 899 timestamps (a 20-sec interval each) when adopting the original resolution. 




\section{Usage Scenario}

We demonstrate the suitability of a suggested resolution for analyzing the Primary School and the usefulness of our colored barcode and system to assist in this analysis. 
We also show in Section C.4 of our supp. material that the patterns found using ZigzagNetVis are comparable to those identified using LargeNetVis~\cite{LargeNetVis}, a state-of-the-art approach.
%

\begin{figure}[t]
\includegraphics[width=\linewidth]{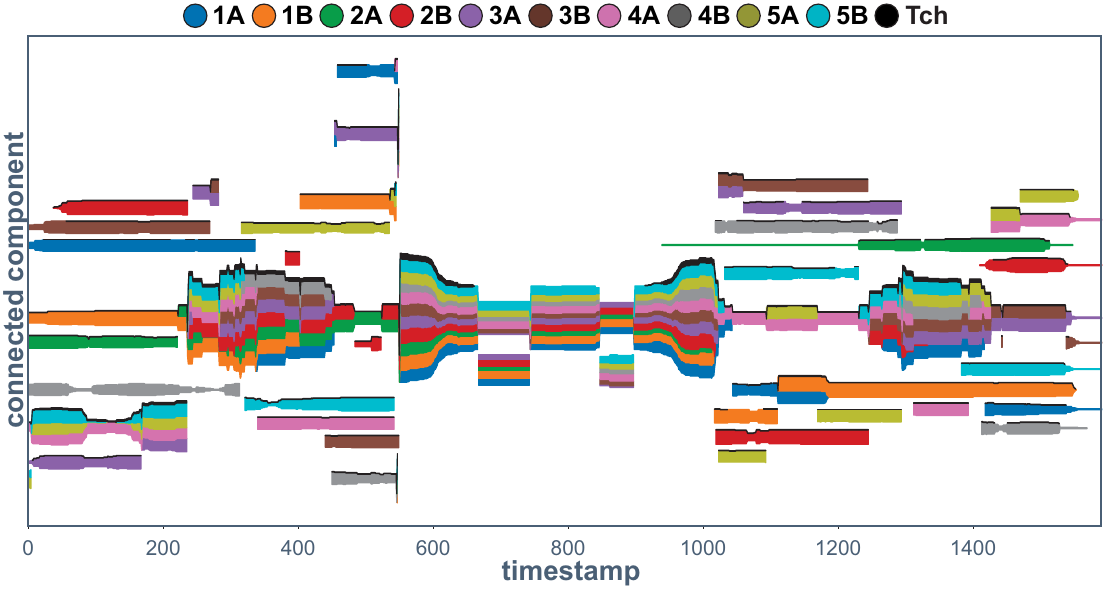}\\
\centering
\caption{Colored barcode with center-based ordering for the Primary School and the suggested resolution $76$.}
\label{fig:colored_barcodes_PrimarySchool_HighSchool}
\end{figure}

ZigzagNetVis suggested the resolutions $r = 8$, $18$, $76$, $154$, and $282$ for the Primary School. Fig.~\ref{fig:teaser}(A) shows our colored barcode for the median resolution $r~=~76$, empirically chosen among the suggested ones due to its interesting patterns.
This visualization was produced after (i) filtering out components with less than 10 nodes and 10 timestamps and (ii) selecting the component with the longest duration. 
Disregarding the component selection (we will discuss it later on), we can already enumerate some patterns and interesting behaviors in the graph data. First, we see that most of the non-selected components (i.e., components with low opacity in Fig.~\ref{fig:teaser}(A)) are composed of students from a single class, along with their teacher (tasks T1, T2). This is expected since these students were having classes in their respective classrooms. This pattern can also be seen in Fig.~\ref{fig:colored_barcodes_PrimarySchool_HighSchool}, which shows the same network and resolution using a different ordering.

There are also time intervals with few connected components compared to others, possibly indicating school breaks (one in the morning, lunch break, and another in the afternoon --- see Fig.~\ref{fig:teaser}(i,ii,iii), respectively) (tasks T2, T3). The first time we have a single component in the graph delineates the beginning of lunch break (see the selected component near timestamp $t = 580$ in Fig.~\ref{fig:teaser}(A)) (task T1). As the students go home for lunch~\cite{primarySchool}, we observe a decrease in the number of nodes in the graph (see just after timestamp $t = 600$) (tasks T2, T3). During lunch, this component is eventually decomposed into two parts, as illustrated in Fig.~\ref{fig:teaser}(A,D)($t = 710$), one containing students from classes 1A, 1B, 2A, 2B, 3A and the other containing a few other students from 3A and 3B, 4A, 4B, 5A, 5B (task T1). This division is explained by the location of the students that stay at the school: some children stay in the cafeteria while others stay at the courtyard~\cite{primarySchool}; these groups encounter each other when they switch places, leading to a single component again (see Fig.~\ref{fig:teaser}(A,E)($t = 800$)). Note also the absence of teachers during the lunch break: they are present at first (see Fig.~\ref{fig:teaser}(A,C,F)($t = 640$)), but they leave (there are no teachers in $t = 710$ and $t = 800$, for example) and come back near the end of the lunch break (task T3), when we start seeing many connected components in the graph again (task T2).

\section{User Study}
\label{sec:userStudy}




\subsection{Participants and Experiment setup}

The experiment recruited 27 participants, including undergraduates (9), Master's students (6), Ph.D.s candidates (7), postdocs (2) and professors (1). According to their self-reports, 4 participants had advanced knowledge in graphs, 4 in visualization, 2 in TDA, and 3 in Informatics in Education. 
%
%
We conducted the experiment using the think-aloud protocol, a common technique to obtain a more accurate perception of the participants' thoughts~\cite{Carpendale2008}. 
To avoid participants being influenced by our presence and not mentioning negative aspects, we explicitly asked them to highlight our approach's limitations. Before the experiment, we conducted a pilot study with two participants not included in the final analysis. 


\subsection{Questionnaire}

First, the participants were presented with a 7-minute video tutorial that introduced the concepts of graph, temporal resolution, and connected components, and explained the proposed layout and system functionalities. The questionnaire was divided into four main sections: (i) background and experience; (ii) a hands-on experience with defined tasks; (iii) nine questions that address the Primary and High School networks; and (iv) Likert-scale questions to collect the participants' feedback. 



The questions were designed to evaluate layout perception, test functionalities, find patterns, and freely explore the given networks. First, we assessed comprehension of the basic functionalities through hands-on experience, where we asked the participants to open the Primary School network using the default configuration. Then, we asked them to verbally describe the definitions of some concepts necessary to understand the experiment (e.g., connected components and temporal resolution) and to follow a set of 12 simple tasks (ST1-ST12) to check if they were familiarized with the system's functionalities (e.g., shortcuts and interaction features). They were also asked to validate our visualization by exploring the Primary School network with resolutions $r = 76$ (SQ1-SQ3) and $r = 154$ (SQ4-SQ6), and the High School under $r = 46$ (SQ7-SQ9). Due to time limitations, we focused on analyzing only these three resolutions, all suggested by ZigzagNetVis using sliding-window timeslicing. In addition, our focus on school networks aimed to provide participants with familiar contexts for understanding nodes and edges, which have the same meaning on both networks. 

The SQ1-SQ9 questions were open questions in which we guided the participants to identify specific patterns (SQ1-SQ3 and SQ7-SQ8), asked them to compare the results of two resolutions (SQ4, SQ5), and encouraged them to explore the system freely (SQ6, SQ9). Finally, we evaluated the participants' preferences for ZigzagNetVis using a series of Likert-Scale questions (LQ1-LQ10) and asked them to describe the positive and negative aspects of the system. The complete description of the questions and expected answers are available in the supp. material (Sec.~D.1).

After preliminary tests, we fixed both filters for bars in 10 (recall Sec.~\ref{sec:systemPrototype}) to avoid receiving too many different results, which would hinder the analysis of the collected data.

\subsection{Results}
\label{ux_results}



\myParagrapho{Hypotheses on data analysis.} All participants answered at least one of the points that we expected for each open question (SQ1-SQ5, SQ7, SQ8). Also, during the experiment, we encouraged participants to raise hypotheses that could justify specific patterns considering a school environment. For instance, in question SQ1 (Primary School), we asked them to evaluate the relationship between students and teachers from classes 4A, 5A, and 5B (which form a connected component at some point). 
All participants mentioned that class 4A was far from the others. Furthermore, 62\% of the participants identified that the two subgroups (4A, 5A-5B) were linked by an edge that involved a teacher; 37\% noticed that this edge actually involves two teachers. 
The hypotheses put forward to explain the strong interaction between classes 5A and 5B mentioned that, since both classes belong to the fifth grade (30\% remembered this information), it could be due to interdisciplinary events such as laboratory activity (22\%) or group studies (14.81\%).

\begin{figure}[t]
\includegraphics[width=\linewidth]{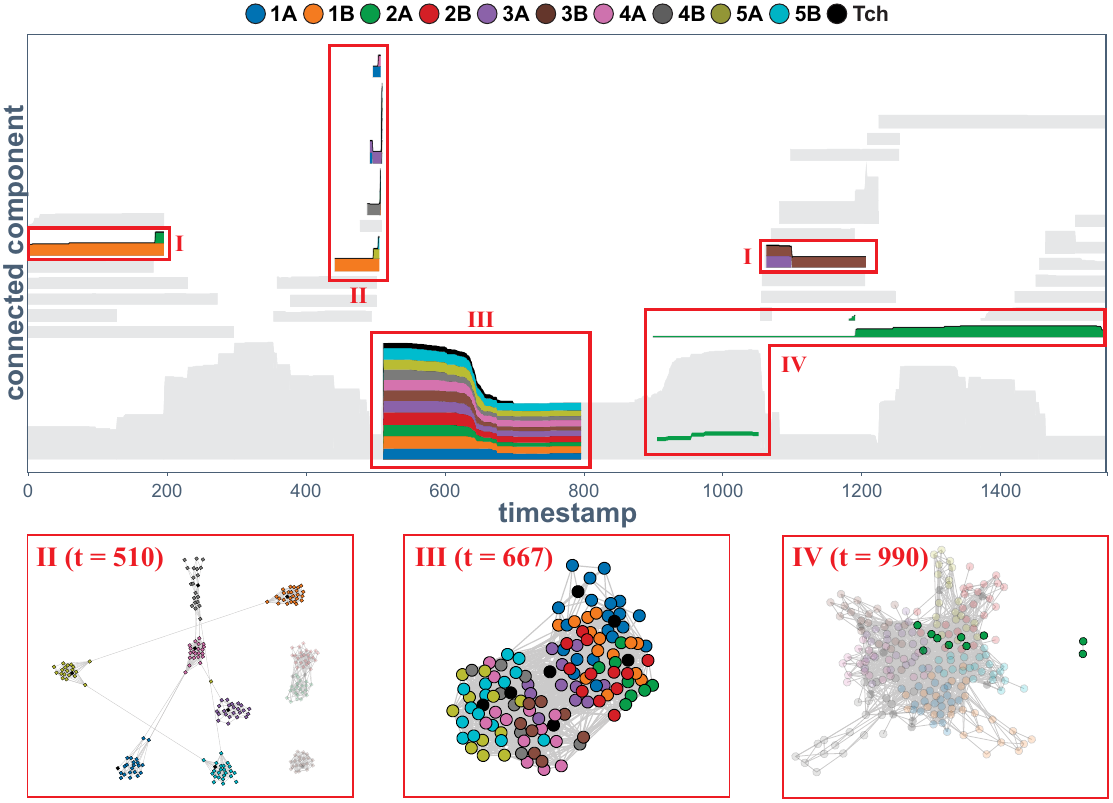}
\caption{Four patterns (I-IV) mentioned via SQ6 (Primary School,~$r=154$). The colored barcode adopts bottom-based ordering.
}
\label{ux_primary_77}
\end{figure}

\myParagrapho{Exploratory analyses.} We proposed questions where the participants could freely explore the system and identify patterns not described by other questions (SQ6, SQ9). More than 85\% of the participants mentioned new patterns or anomalies in their exploratory analyses of the Primary School (SQ6), and 74\% found new ones in the High School network (SQ9). Among the patterns and anomalies found, the most cited for the Primary School using $r=154$ were (Fig.~\ref{ux_primary_77}): (I) merges and splits between related students; (II) peaks of interaction in a short time period; (III) a single connected component containing all students and teachers (even though the teachers leave the network at some point); and (IV) same-class students divided into two connected components. Although this question was not designed to compare patterns identified in different resolutions, most participants tried to compare patterns visible with $r=154$ with those from $r=76$. For instance, Fig.~\ref{fig:teaser}(A,D) illustrates that there are two connected components around timestamp 710 when using $r=76$, which is hidden in the higher resolution (see Fig.~\ref{ux_primary_77}(III)). About that, a participant mentioned that \textit{``you can clearly see how patterns vary according to the selected resolution when analyzing the primary school''}.

\begin{figure}[t]
\includegraphics[width=\linewidth]{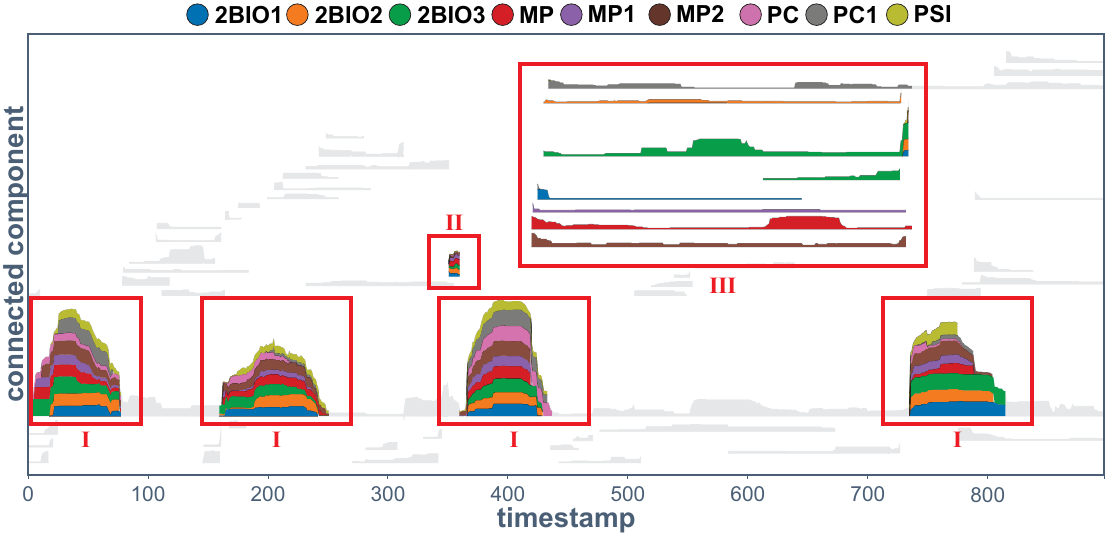}
\caption{Three patterns (I-III) mentioned via SQ9 (High School,~$r=46$). The colored barcode adopts bottom-based ordering.
}
\label{ux_highschool_23}
\end{figure}

The participants identified three common patterns in the second exploratory question (SQ9, High School). The first refers to peaks of activity in the same connected component over time (see Fig.~\ref{ux_highschool_23}(I)). In the High School, there are also intervals where all students merge into a single and highly connected component. 
The participants could see that these intervals correspond to break periods, lunch break, or group activities. 
The second pattern is related to a small connected component just before a large peak (Fig.~\ref{ux_highschool_23}(II)). Based on the node-link diagram, there were just a few connections between the students, which represented the beginning of a group activity or a break. Finally, the third pattern refers to connected components with varying lengths over time but composed of single classes (Fig.~\ref{ux_highschool_23}(III)). According to the participants, they allow one to see the class hours, but, contrary to the primary school, where the number of students per component is quite stable over time during classes (see Fig.~\ref{fig:teaser}), this network presents classes with non-uniform activity over time.


\myParagrapho{Interactive features.} We also validated the functionalities mainly used to answer representative questions. 
In summary, the participants preferred to move timestamp markers rather than type new timestamp values for the exploratory tasks. On average, the feature used mainly in the node-link diagrams was zoom (41.48\%), which is justified by the small size of nodes and edges initially applied. Not least, the similar rate of usage involving selection by label (44.44\%) and by component (41.48\%) indicates that both were appreciated. Please refer to the supp. material (Sec.~D.2) for details.

\begin{figure}[t]
\includegraphics[width=\linewidth]{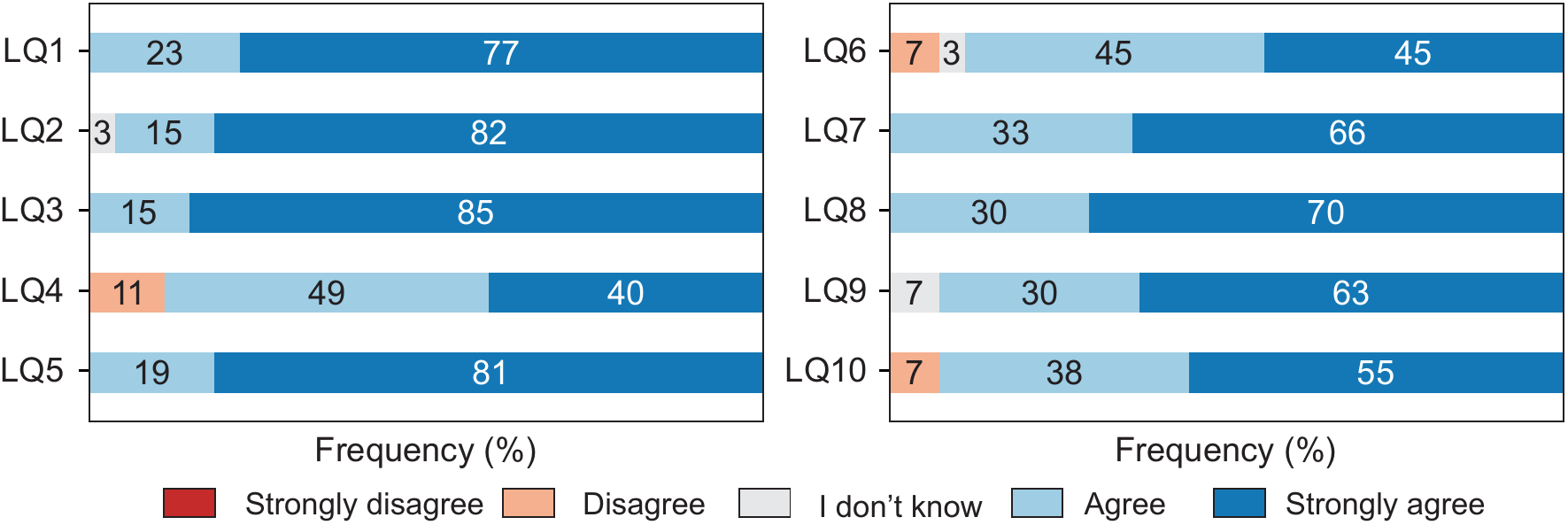}
\caption{Participants’ feedback using a Likert scale.
}
\label{likert_ux}
\end{figure}

\myParagrapho{Likert-scale questions and participants' feedback.} Fig.~\ref{likert_ux} shows the participants' assessments of the colored barcode's (LQ1) and node-link diagrams' (LQ2) quality and usefulness, their coordination and interaction (LQ3), and the system's intuitiveness and ease of use (LQ4), usefulness (LQ5), and response time (LQ6). There were also questions related to specific tasks, such as understanding the temporal evolution (LQ7), comparing structure at different times (LQ8) or at node level (LQ9), and analyzing the network under different resolutions (LQ10).

First, considering the negative evaluations, three participants mentioned that the system was not intuitive (LQ4) because it lacked a ``help'' button summarizing the main functionalities. Regarding response time (LQ6), two users complained about loading time, although the system's interactions worked satisfactorily. One of the experts added that \textit{``I can't say about speed, for the tested datasets I agree but generally I don't know, it depends on the network size''}. At last, about the analysis under different resolutions (LQ10), two participants considered that the comparison was difficult since it depended on the memory load of the user.


Besides the negative evaluations, ZigzagNetVis achieved a 95\% of acceptance rate for the raised criteria (LQ1-LQ10), considering the average agreement (29\%) and strong agreement (66\%) rates.
Several participants raised positive points about the system and colored barcode, claiming that \textit{``The proposed system is simpler and more efficient in analyzing temporal networks than the other tools I know''}, and \textit{``the colored barcode is great (pretty and very interesting), both for the color distinction and the subtlety of increases and decreases in a bar over time''}. For another participant, \textit{``It is the union of both views (colored barcode and node-links) that is most useful. Each alone would not allow us to understand well what is happening''}. At last, one expert complemented that \textit{``the barcodes are very good for quickly visualizing long interactions, while the node-link diagrams allow you to understand to what degree these interactions are happening''}.

It should be noted that we tested the system with two colorblind participants, who validated that there were enough features (such as tooltips, different color scales, and interactive color legend) to perform all tasks without hindering the analyses. Finally, some participants suggested improvements already incorporated, such as the selection of multiple labels, improvements in readability (e.g., better contrast in menus) and the mentioned ``help'' button.

\section{Analysis of the suggested resolutions}
\label{sec:analysis_suggested_resolutions}

This section is devoted to the analysis of the resolutions suggested by ZigzagNetVis' methodology.
We aim to demonstrating empirically that these suggestions are relevant.
As mentioned in Sec.~\ref{sec:related_work}, few studies have tackled the problem of choosing a resolution.
In particular, no reference data sets are available.
As a means of comparison, we will use, for each of the temporal graphs considered in this paper, the resolutions used in the literature --- that we stress are mainly chosen ``by hand".
However, a direct comparison cannot be made.
Indeed, as pointed out in Sec.~\ref{subsec:description_method}, one cannot define ``optimal" resolutions, but rather \textit{meaningful ranges} of resolutions.
Consequently, to assess the quality of ZigzagNetVis' suggested resolutions, one must analyze them qualitatively by understanding the behaviors of the corresponding dynamic graphs and comparing them with the literature.
This study will be conducted here, in particular using the visual tool provided by the bottleneck distance.
In our supp. material (Section C.2.3), we extend this study by comparing our suggestion curves with commonly used feature of temporal graphs, revealing when they coincide and when they complement each other.

\begin{figure*}[t]
\includegraphics[width=\linewidth]{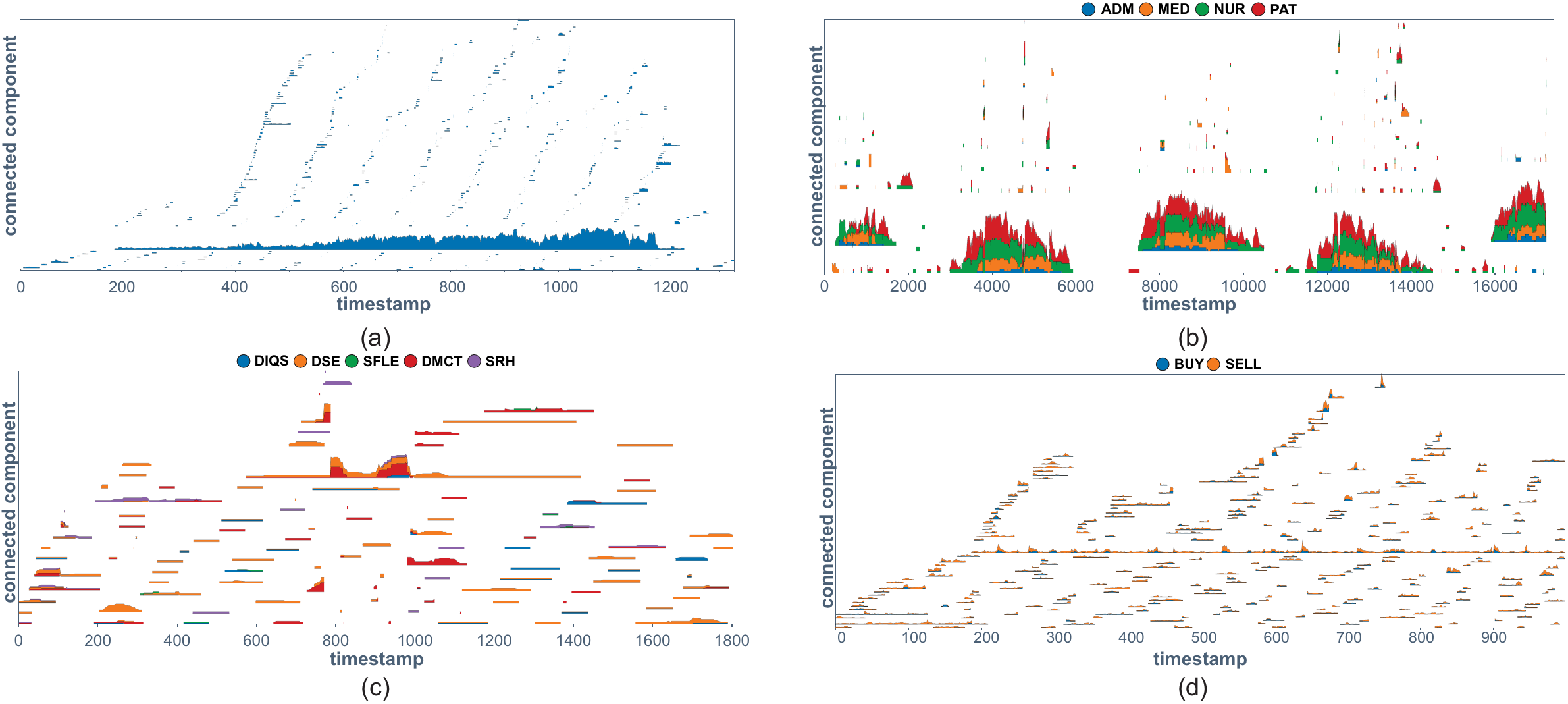}
\centering
\caption{
Colored barcodes with bottom-based ordering for four networks from Tab.~\ref{tab:recommended_resolutions}. (a) Enron with $r = 6$. (b) Hospital with $r = 74$. (c) InVS with $r = 66$. (d) Sexual with $r = 6$. All resolutions adopted were suggested by our method (see Tab.~\ref{tab:recommended_resolutions}). There is no component filtering except for the Sexual network (f), whose colored barcode shows only components with at least 10 node members and a duration of at least 10 timestamps.
}
\label{fig:barcode_other_networks}
\end{figure*}



\myParagrapho{Visualization of suggested values.}
Tab.~\ref{tab:recommended_resolutions} presents the resolutions suggested by our approach, using sliding-window timeslicing and considering eight different graphs of varying characteristics and sizes.
The corresponding normalized suggestion curves are shown in Fig.~6 of our supp. material.
To construct the curves for these eight temporal graphs, we used, respectively, a maximal time of $2000$, $2000$, $2000$, $20000$, $1300$, $1400$, $3000$, and $1000$, and resolutions up to a quarter of these values. Note that the maximal time for Primary and High Schools covers the first day, as in Sec.~\ref{datasets}.

Fig.~\ref{fig:barcode_other_networks} shows the colored barcodes for four networks in Tab.~\ref{tab:recommended_resolutions} to emphasize the usefulness of both the resolution suggestion method and this visualization in assisting the analysis of real-world networks. The colored barcodes exhibit the entire graphs, except for the InVS network (Fig.~\ref{fig:barcode_other_networks}(c)), which shows only the first day of data to better present the visual pattern we want to discuss.

Even though the Enron network (Fig.~\ref{fig:barcode_other_networks}(a)) does not provide node metadata, it is easy to identify global patterns that do not rely on such information, for example, the gradual increase in the number of connections and node activity over time~\cite{enron}. The increasing size of the main connected component, followed by an abrupt decrease near the end of the network, is related to important events in the context of these network data, including the CEO resignation and bankruptcy.
Temporal patterns related to circadian rhythms can also be identified in face-to-face networks, as shown in Fig.~\ref{fig:barcode_other_networks}(b) for the Hospital network~\cite{Hospital}. We can easily identify intervals with bursts of events (five days) followed by intervals with few or no interaction (four nights).

Incorporating node metadata greatly improves network analysis by allowing us to observe local patterns in the data. In the InVS network~\cite{InVS}, for example, most connected components contain only nodes that share the same label (in this case, employees of the same department), as illustrated in Fig.~\ref{fig:barcode_other_networks}(c). That makes sense in the context of this network, as most of the employees are of type ``residents'', i.e., they interact mainly with others in their own department. 
This is a pattern we do not observe in the Sexual network~\cite{sexual_contacts} (Fig.~\ref{fig:barcode_other_networks}(d)). Since it is a bipartite graph, all connected components will have at least one node from each label, i.e., a buyer and a seller. Note that the Sexual network is much larger than the others we have considered. Its original form (resolution 1) contains 12,157 nodes, 34,060 edges, and 1,000 timestamps, each representing a 1-day interval~\cite{sexual_contacts}.

\begin{table}[t]
\centering
\caption{Suggested resolutions for eight distinct networks.}
\label{tab:recommended_resolutions}
\resizebox{0.49\textwidth}{!}{%
\begin{tabular}{lll}
\hline
\textbf{Network} & \textbf{Suggested resolutions} & \textbf{Used in the literature} \\ \hline
Primary School~\cite{primarySchool} & 8, 18, 76, 154, 282  & 10~\cite{0228728}, 25~\cite{PONCIANO2021170}  \\
High School~\cite{highSchool} & 8, 12, 46, 92, 104 & 18~\cite{7192717}, 180~\cite{8365984}, 45~\cite{3399922} \\
Hospital~\cite{Hospital} & 14, 26, 32, 74, 352 & \{9, 45, 60, 90\}~\cite{CNO, wim}, 69~\cite{KAIS}\\
InVS~\cite{InVS} & 66, 148, 158, 164, 202 & --- \\ 
Museum~\cite{museum_conference} & 6, 12, 36, 52, 320 & 1~\cite{combinacao} \\
Enron~\cite{enron} & 6, 12, 24, 36, 68 & 1~\cite{DyNetVis, 6596125, fish2017supervised}, 2~\cite{PONCIANO2021170}, 5~\cite{KAIS}, \\
& & \{1, 7, 15, 30, 90, 180\}~\cite{orman2021finding}, \\ & & \{1, 5, 12\}~\cite{sulo2010meaningful} \\
Conference~\cite{museum_conference} & 12, 22, 30, 42, 224 & 30~\cite{fish2017supervised} \\
Sexual~\cite{sexual_contacts} & 6, 160, 186, 226, 240 & 1~\cite{combinacao} \\
\hline
\end{tabular}%
}
\end{table}



\myParagrapho{Resolutions used in the literature.}
In general, studies that analyze temporal graphs use resolution directly or indirectly. Some focus on comparing different resolutions~\cite{orman2021finding, PONCIANO2021170, sulo2010meaningful}, while others select arbitrary resolutions according to the analysis needs~\cite{3399922, KAIS, 6596125}. For instance, some works prioritize high resolution values for global pattern identification~\cite{3399922, 8365984}, while others focus on small ones and local patterns~\cite{6596125, fish2017supervised, combinacao}. Note that ZigzagNetVis suggests resolutions suitable for both types of analysis (Tab.~\ref{tab:recommended_resolutions}).

Tab.~\ref{tab:recommended_resolutions} summarizes our suggested resolutions and others used in literature for eight popular graphs. For the well-known Enron network~\cite{enron}, while some studies use the original resolution $r=1$ as an arbitrary value to perform local analyses~\cite{DyNetVis, 6596125, fish2017supervised}, others compare resolutions coming from a small set of arbitrary values~\cite{orman2021finding, sulo2010meaningful}. For example, Sulo et al.~\cite{sulo2010meaningful} analyze this network under resolutions $r = 1$, $r = 5$, and $r = 12$, highlighting the different patterns each resolution allows one to identify. According to the authors, the pattern ``CEO resignation'' is easily identified when adopting resolutions between 4 and 7~\cite{sulo2010meaningful}. Note that ZigzagNetVis suggested resolutions $r = 6$ (therefore included in the mentioned ``good-quality'' range) and $r = 12$ (a resolution also used by the authors). The suggestion of a resolution that matches exactly the one used by previous studies also occurred with the Conference network ($r = 30$, as depicted in Tab.~\ref{tab:recommended_resolutions}).

As another example, some studies mention the same circadian rhythm pattern discussed in Fig.~\ref{fig:barcode_other_networks}(b) for the Hospital network, i.e., days with bursts of activity and idle nights~\cite{KAIS, CNO, wim}. ZigzagNetVis and these studies allow one to identify this pattern, even though they use different but close resolution values. Our method also suggests a resolution many times greater than those used in the literature for this network ($r = 352$). This is probably the resolution in which the idle intervals are lost.
In general, our approach suggests resolutions that are close to those used by the related literature. In addition, it can also suggest other resolution values that potentially lead to unexplored visual patterns.





\myParagrapho{Resolution comparison and explainability.}
As discussed in Sec.\ref{subsec:intro_zigzag}, the bottleneck distance offers a clear interpretability.
Namely, the distance $d_\mathrm{B}(\mathcal{B},\mathcal{B}')$ between two barcodes is always caused by a pair of bars or a bar alone, that is, such that the cost of this pair, or of this bar alone, is equal to the distance.
Consequently, highlighting these bars allows us to observe precisely \textit{where} the barcodes differ the most.
This is particularly useful for understanding the suggested resolutions of ZigzagNetVis.

\begin{figure}[t]
\includegraphics[width=1\linewidth]{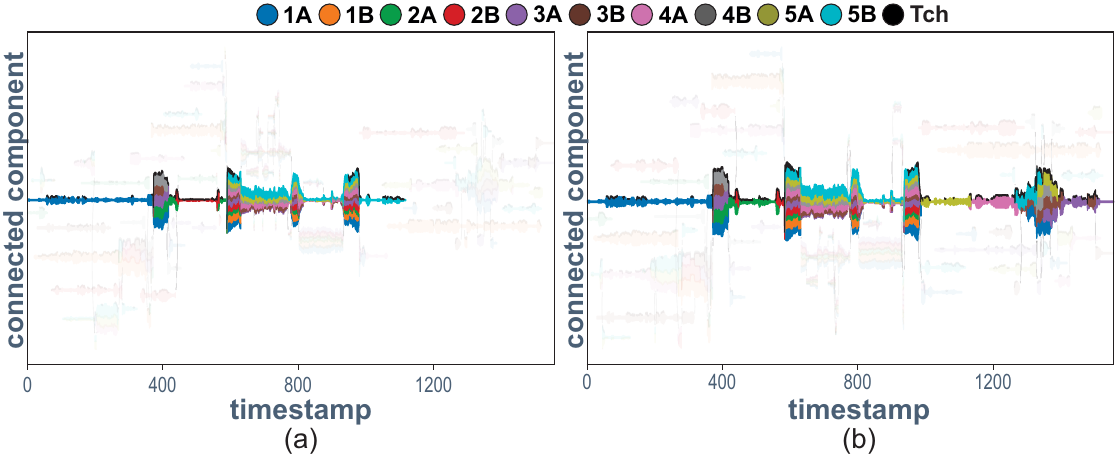}\\
\caption{Visualization of the bottleneck distance for the first day of the Primary School. (a-b) $r = 6$ and $r = 8$ showing only bars with height larger than 50. Highlighted components represent the bars that differ the most between these two resolutions, according to the bottleneck distance.} 
\label{fig:comparison_resolution_primarySchool}
\end{figure}

Taking into account the first day of the Primary School network, the algorithm suggests resolution $r = 8$ (see Tab.~\ref{tab:recommended_resolutions}).
The resolution just before this one is $r = 6$, since sliding-window timeslicing only accepts even values of resolutions.
In order to visualize what structural change has occurred between resolutions 6 and 8, we show in Fig.~\ref{fig:comparison_resolution_primarySchool}(a,b) the corresponding colored barcodes, while highlighting the pair of bars that provoked the bottleneck distance.
As we can see, when going from $r = 6$ to $r = 8$, a large bar is formed, which lasts throughout the observation period. Please refer to the supp. material (Sec.~C.2.2) for other networks.


\myParagrapho{Classification of structural changes.}
A manual analysis of the resolutions suggested by ZigzagNetVis compels us to classify the structural changes into three categories.
The first category contains the initial resolutions.
We have observed, in the suggestion curves, the phenomenon of a chaotic start, followed by a relatively flat phase.
These resolutions correspond to critical points indicating the formation of the first persisting connected component.
A second type of easily identifiable structural change is that of the connection between days of the temporal graph. 
At the critical resolution connecting two consecutive days, assuming no activity is recorded during the night, the suggestion curve shows a significant peak.
The last group of resolutions generally contains those that cause a persistent connected component to merge with a larger one.

This classification allows, at least heuristically, to divide the range of resolutions into three intervals: a chaotic start, followed by a range where the resolution curve only exhibits relevant peaks, and at last a few values caused by the merging of the days.
This observation can be used when the user, through manual inspection, seeks relevant resolutions to study.
We stress that the last case is not observed in the figures for the Primary and High School networks, since we selected resolutions smaller than the length of a night.

\section{Discussion and Limitations}

\myParagrapho{\blue{Timeslicing.}}
\blue{ZigzagNetVis is not designed for graphs with continuous real-valued timestamps as timeslicing approaches fail to represent these graphs faithfully~\cite{LargeNetVis}. Considering graphs with discrete times, we have described two uniform timeslicing approaches that may be used with ZigzagNetVis: partition and sliding-window-based. Regardless of the chosen approach, the suggested resolution is a global and static value that is used to represent the entire graph. In future work, we intend to investigate whether non-uniform timeslicing would lead to better results.}

\myParagrapho{Visual scalability.}
Our colored barcode is better suited for small to mid-size graphs, in terms of the number of timestamps or connected components (even though too few and large components also hinder the analysis). Although we provide filters and interactions that help with large networks, we intend to improve our visual scalability to better meet this type of network. \blue{Specifically, we plan to extend the representation to deal with more timestamps and components, e.g., by collapsing/expanding based on the graph dynamics. We also intend to incorporate sampling strategies and more sophisticated filters, e.g., based on structural properties such as the strength of the connected components or edge weights (explicit or inferred).}

\myParagrapho{Resolution comparison.} 
\blue{Some participants would also like to simultaneously compare suggested resolutions with each other and with non-suggested ones. Although one could open the system many times or perform a side-by-side comparison using multiple instances of the system, we believe that incorporating such a capability into our system would enhance the identification of patterns coming from different resolutions, help users to understand and follow changes that regions of interest suffer when varying resolutions (recall Sec.~\ref{sec:analysis_suggested_resolutions}), and also increase the user's confidence in the suggestions or reveal room for improvement in the suggestion procedure, e.g., by incorporating user feedback.}


\myparagraph{Zigzag persistent homology.}
Through the lens of homology, all connected components are treated identically, regardless of the number of nodes they contain.
Consequently, in extreme cases, a structural change in the temporal network can be provoked by a single node.
Since this situation might not be convenient for the analysis of large networks, where relevant features are commonly understood as those involving many nodes, we intend to design and adopt a variation of the bottleneck distance that would take into account the number of nodes. 
%
Besides, our work focused on homology $H_0$.
The inclusion of higher topological features, such as in \cite{myers2023temporal}, may contain further relevant information, that we intend to add in future works.
%
Not least, we adopted in this work a simple peak detection via their prominence, which is well established and easy to interpret.
In a follow-up study, we intend to test more sophisticated approaches, for example, peak detection via Z-scores, or TDA-inspired techniques based on peaks' persistence.

\myParagrapho{Running time.}
\blue{
Tab. 1 in our supp. material shows that, in practice, the most time-consuming step of our algorithm is the computation of $m$ persistence diagrams, $m$ being the number of resolutions tested.
To reduce this cost, we could take advantage of the fact that two consecutive resolutions should yield barcodes close to each other; an idea known as \textit{updating barcodes} \cite{dey2021updating}.
Although we have not investigated this aspect further, since the running times obtained empirically were satisfactory, such a technique could open the door to larger-scale graphs.
}

\myParagrapho{Visual improvements and new features.}
Based on feedback from reviewers and participants, we've added new features to the system prototype: a center-based component positioning, merge/split visual representation, and a table with quantitative measurements for suggested resolutions. While participants did not test these features, they do not directly affect the results outlined in this paper.

\section{Conclusion}
\label{conclusion}

This paper presented ZigzagNetVis, a methodology that suggests potentially relevant temporal resolutions for graph analysis using zigzag PH, a well-established technique from TDA, and the tool system that implements it. Our methodology can be summarized as follows. First, we build persistence barcodes for candidate resolutions. Then, we compute the bottleneck distance between pairs of barcodes and build a suggestion curve based on the distance values. Finally, we suggest resolutions based on the curve's peaks.
ZigzagNetVis also incorporates a timeline-based visualization inspired by the persistence barcodes of TDA. Our visualization assists researchers and practitioners in exploring temporal graphs by highlighting the connected components' structure and evolution. We validated ZigzagNetVis and our web-based and interactive system prototype through a usage scenario and a user study with 27 participants, who assessed its usefulness and effectiveness.

\ifCLASSOPTIONcompsoc
  \section*{Acknowledgments}
\else
  \section*{Acknowledgment}
\fi

This work was supported by grants \#2023/18026-8, \#2021/07012-0, \#2020/10049-0, \#2020/07200-9, \#2022/13190-1, \#2016/17078-0 from S\~ao Paulo Research Foundation (FAPESP), by grant \#E-26/204.593/2024 from Carlos Chagas Filho Foundation for Research Support of Rio de Janeiro State (FAPERJ), by Fundação Getulio Vargas (FGV), by grant \#311144/2022-5 from Conselho Nacional de Desenvolvimento Cientifico e Tecnologico (CNPq). The Article Processing Charge (APC) for the publication of this work was funded by  Coordena\c{c}\~ao de Aperfei\c{c}oamento de Pessoal de Nivel Superior - CAPES (ROR identifier: 00x0ma614). For open access purposes, the authors have attributed a Creative Commons CC BY license to any accepted version of the article.



\bibliographystyle{abbrv-doi}

\bibliography{ms}


\begin{IEEEbiography}[{\includegraphics[width=1in,height=1.25in,clip,keepaspectratio]{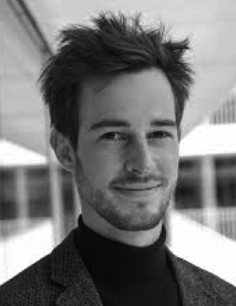}}]{Raphaël Tinarrage} 
holds a PhD in pure and applied mathematics from the Université Paris-Saclay (2020). 
He passed the agrégation (national teaching qualification) while studying at the École normale supérieure Paris-Saclay.
He is currently a research fellow at the Institute of Science and Technology Austria, and previously at the Fundação Getúlio Vargas EMAp. His work focuses on Topological Data Analysis, in its theoretical developments and practical applications.

\end{IEEEbiography}

\begin{IEEEbiography}[{\includegraphics[width=1in,height=1.25in,clip,keepaspectratio]{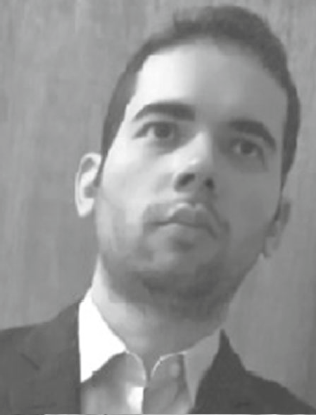}}]{Jean R. Ponciano} is an Assistant Professor with the Mathematics and Computer Science Institute - University of S\~ao Paulo, Brazil. His research interests include information visualization, visual analytics, network science, and data streams. He frequently serves as a program committee member for relevant conferences, including IEEE Vis, EuroVis, and ASONAM, and as an external reviewer for other relevant venues, such as IEEE TVCG, Comp. Graph. Forum, Computers \& Graphics.
\end{IEEEbiography}

\begin{IEEEbiography}[{\includegraphics[width=1in,height=1.25in,clip,keepaspectratio]{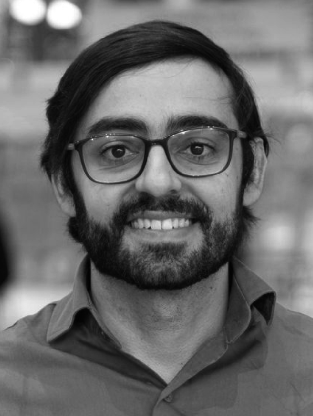}}]{Claudio D. G. Linhares}
is a Senior Lecturer at Linnaeus University, at the Department of Computer Science and Media Technology Faculty of Technology, in Växjö, Sweden. My research interests include information visualization, network visualization, human-computer interaction, visual analytics, and human-in-the-loop AI, with applications especially in forestry and healthcare.
\end{IEEEbiography}

\begin{IEEEbiography}[{\includegraphics[width=1in,height=1.25in,clip,keepaspectratio]{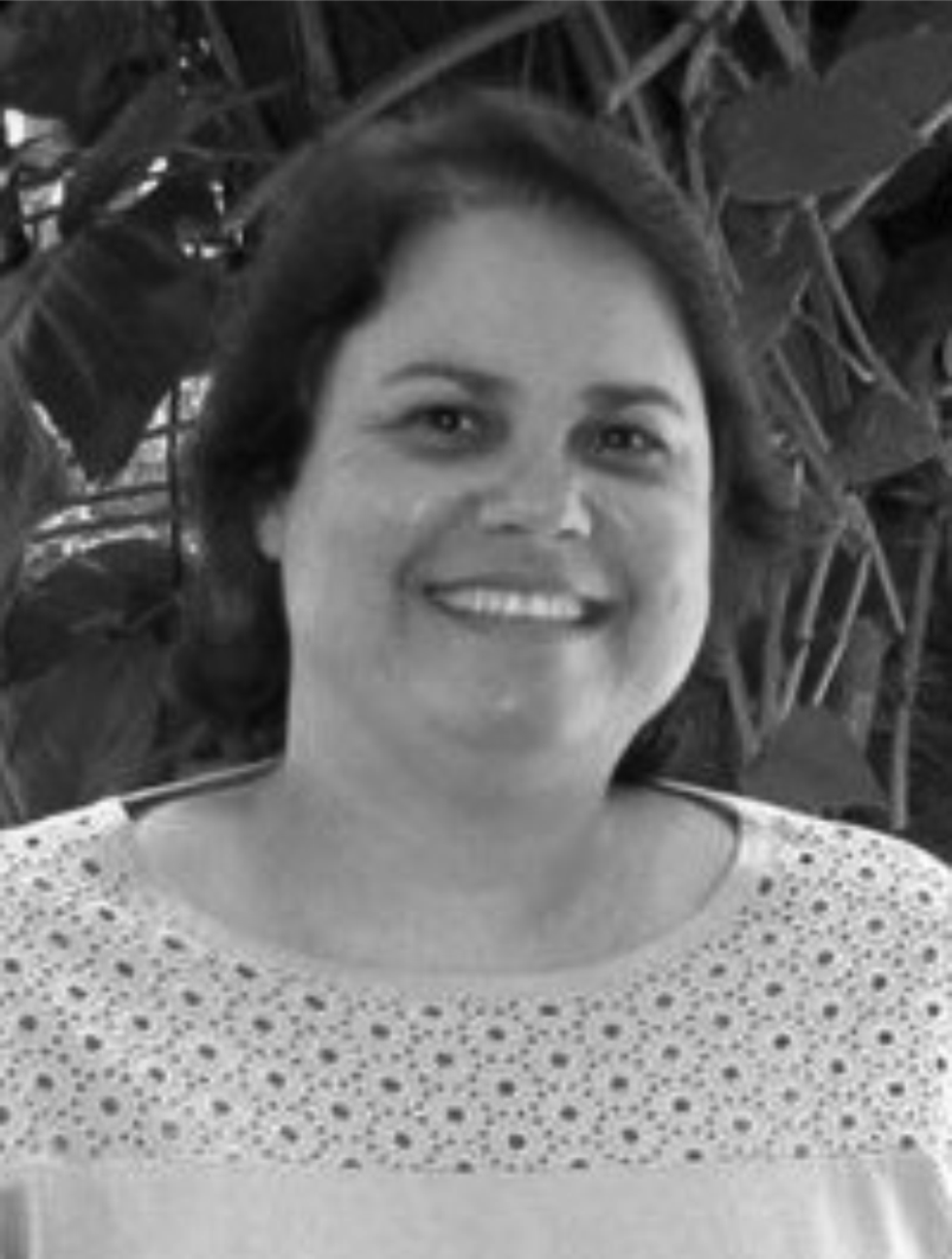}}]{Agma~J.~M.~Traina} 
received the BSc and MSc degrees in Computer Science and PhD in Computational Applied Physics all from the University of São Paulo, Brazil in 1983, 1987 and 1991 respectively. She also spent a sabbatical leave as a visiting researcher at the Computer Science Department of the Carnegie Mellon University (1998-2000) working on Multimedia Databases. She is a full professor with the Mathematics and Computer Science Institute - University of São Paulo, since 2008.
Her research interests include indexing and retrieval of complex data by content, similarity queries, data visualization, visual data mining, as well as image and video processing. Agma has been working in the integration of the results of her research lines with applications to medicine, aimed at the development of applied computational systems. She is a member of the Brazilian Computer Society, ACM and IEEE Computer Society.
\end{IEEEbiography}

\begin{IEEEbiography}[{\includegraphics[width=1in,height=1.25in,clip,keepaspectratio]{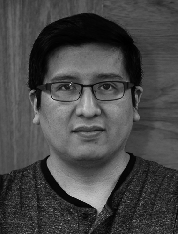}}]{Jorge~Poco} 
is an Associate Professor at the School of Applied Mathematics, Fundação Getúlio Vargas (FGV), Brazil. He earned his Ph.D. in Computer Science from New York University, an M.Sc. in Computer Science from the University of São Paulo, Brazil. 
His research focuses on data visualization, visual analytics, machine learning, and data science. He has actively contributed to program committees for IEEE SciVis, IEEE InfoVis, VAST, and EuroVis.
\end{IEEEbiography}

\end{document}


\maketitle

\appendix

\section{Definition of bottleneck distance}

The most common distance between barcodes $\mathcal{B}$ and $\mathcal{B}'$ is the \emph{bottleneck distance}, defined as follows. 
First, we say that a subset $C \subset \mathcal{B} \times \mathcal{B}'$ is a \emph{partial correspondence} if for any bar $b \in \mathcal{B}$ (resp. $b' \in \mathcal{B}'$) there exists at most one bar $b' \in \mathcal{B}'$ (resp $b \in \mathcal{B}$) such that $(b,b')\in C$.
Next, the cost of a matched pair of bars $(b,b') \in C$ is defined as the maximum difference between their endpoints. A bar $b \in \mathcal{B}$ (resp. $b' \in \mathcal{B}'$) for which there is no $b' \in \mathcal{B}'$ (resp $b \in \mathcal{B}$) such that $(b,b')\in C$ is said unmatched, and its cost is defined as its half-length.
The cost of the partial correspondence $C$ is defined as the maximal costs of its matched and unmatched bars.
Last, the bottleneck distance, denoted $d_\mathrm{B}(\mathcal{B},\mathcal{B}')$, is defined as the minimal cost of a partial correspondence:
\vspace{-0.15cm}
\begin{equation*}
d_\mathrm{B}(\mathcal{B},\mathcal{B}')
= \inf_C \max \bigg\{
\sup_{(b,b')\in C} \|b-b'\|_\infty,
\sup_{(b,\cdot)\notin C} \|b\|_\infty,
\sup_{(\cdot,b')\notin C} \|b'\|_\infty
\bigg\}.
\end{equation*}

It is interesting to note that, by definition, this distance is either equal to $\|b-b'\|$ for a pair of bars $(b,b')\in\mathcal{B}\times\mathcal{B}'$, or to $\|b\|$ for a single bar $b\in\mathcal{B}\cup\mathcal{B}'$.
That is to say, the bottleneck distance is caused either by a pair of bars, or by a bar alone. Identifying this cause allows us to derive an explainability pipeline, in Sec.~9 of the main document.

\section{Methodology details}
\label{subsec:instability}


When studying temporal graphs, two common uniform timeslicing methods are employed.
The first one, partition timeslicing, consists of choosing a multiple $\alpha r_0$ of the initial resolution and subdividing the time by stacking intervals of length $\alpha r_0$. It is employed, for instance, in LargeNetVis~\cite{LargeNetVis}. The other one, sliding-window timeslicing, is obtained by allocating to each edge an activation window of semilength $\alpha r_0$, as used in \cite{myers2023temporal}.
In practice, we observed that our resolution suggestion method, described in the main document (Sec.~4.2), gives better results when considering sliding-window timeslicing.
The partition timeslicing suffers from \emph{instability}, which we will exemplify here.

\subsection{Instability of partition timeslicing}

\begin{figure}[t]
\includegraphics[width=\linewidth]{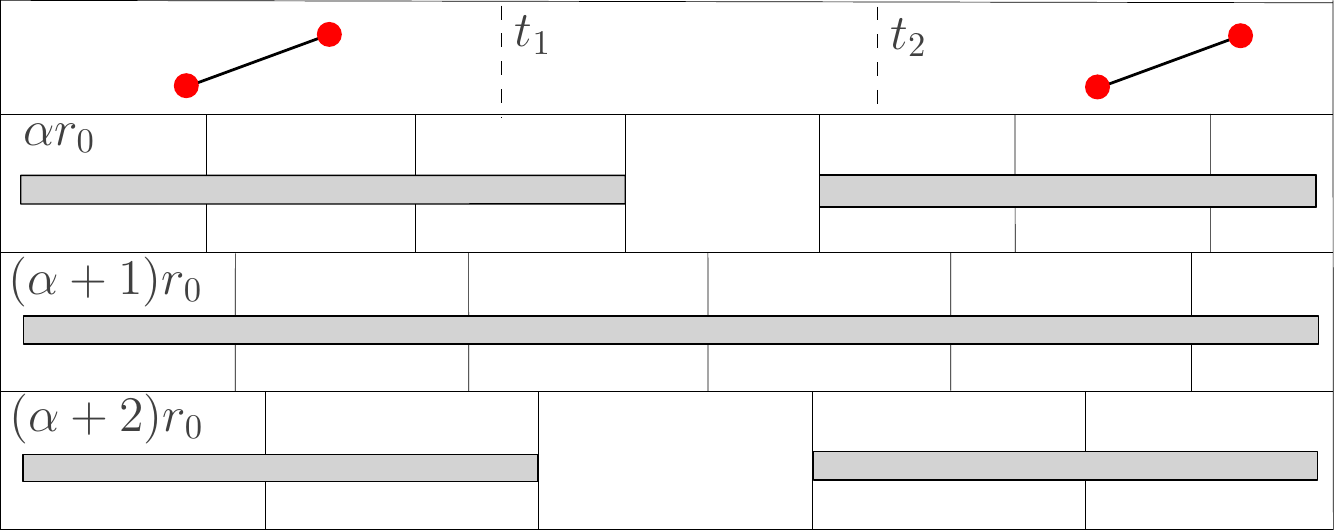}
\vspace{-0.3cm}
\caption{From top to bottom: a temporal graph, and the barcodes obtained for three consecutive resolutions, using the partition timeslicing. The two bars of the barcode get glued and cut again, showing the instability of this timeslicing method.}
\label{fig:instability}
\end{figure}

Consider a temporal graph that is made of only two nodes. Let the time interval $[0,T]$ be subdivided as $[0,t_1]\cup[t_1,t_2]\cup[t_2,T]$, and suppose that the edge is active only on $[0,t_1]$ and $[t_2,T]$.
Let $r=\alpha r_0$ be a multiple of the initial resolution such that $r < t_2-t_1 < 2 r$.
By applying partition timeslicing, a whole interval $[kr, (k+1)r]$ may be included in $[t_1,t_2]$. Since no edge is active in this interval, we obtain a graph $G_k$ that is empty. Hence the barcode consists of two bars, the barcode being empty on the interval $[t_1,t_2]$.
However, if by chance no $[kr, (k+1)r]$ is included in $[t_1,t_2]$, then the barcode will consist of only one bar.
The situation is depicted in Fig.~\ref{fig:instability}.

\begin{figure}[t]
\includegraphics[width=\linewidth]{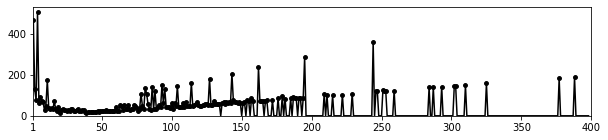}
\vspace{-0.4cm}
\caption{Normalized suggestion curve for partition timeslicing in the Primary School network. 
We observed that the peaks witness an alternation of the barcode between having a large bar and two smaller bars. The $x$ and $y$ axes represent the resolution values and the consecutive bottleneck distance, respectively.}
\label{fig:instability2}
\vspace{-0.3cm}
\end{figure}

In practice, we have observed that increasing the resolution slightly allows us to go from one situation to the other and vice versa. 
A concrete example for the Primary School network is given in Fig.~\ref{fig:instability2}.
We observed that most of the peaks correspond to the same bar of the barcode, which gets cut and merged again.
In this case, our resolution selection method will detect all these critical resolution values.
To analyze the graph, we would rather have detected this change in behavior only once.

The sliding-window timeslicing, in contrast, does not exhibit this problem. Indeed, the activation windows' length on each edge grows monotonically. Therefore, for our main paper's analyses, experiments, and illustrations, we choose to use the sliding window method. The only exception is when we compare our ZigzagNetVis approach with LargeNetVis~\cite{LargeNetVis} (see Sec.~\ref{comparison_largenetvis} in this document). This is because, as mentioned earlier, LargeNetVis employs partition timeslicing.

\subsection{Visual comparison of timeslicing methods}
\label{visual_comparison_timeslicing_methods}

\begin{figure}[h]
\includegraphics[width=\linewidth]
{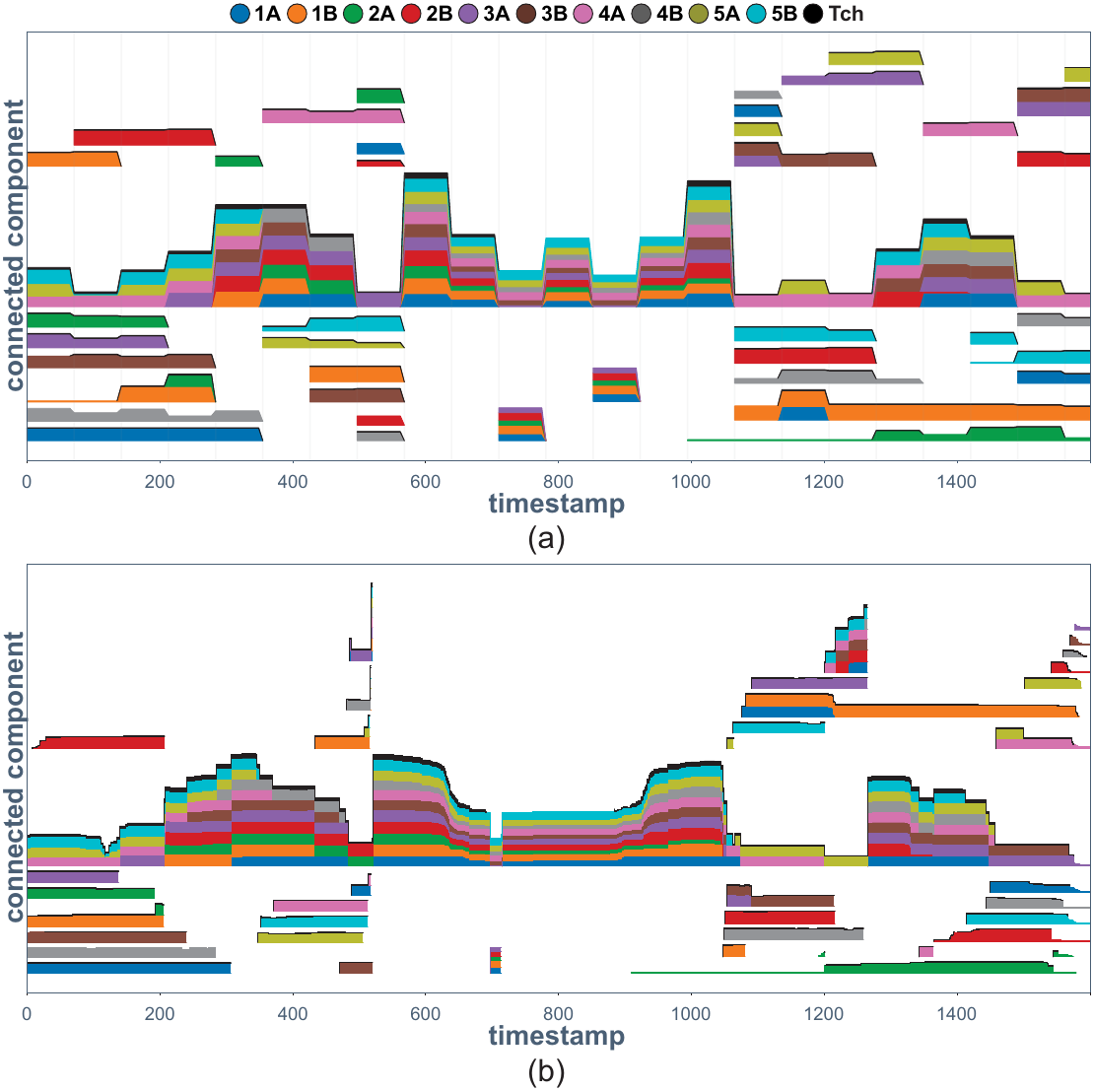}
\vspace{-0.5cm}
\caption{Comparing colored barcodes using the first day of the Primary School network for (a) partition and (b) sliding window timeslicing. Both cases use the suggested resolution $r = 154$ and filter out components with less than 10 node members and 10 timestamps of duration.}
\label{partition_vs_sliding}
\vspace{-0.2cm}
\end{figure}

Given the first day of the Primary School network (recall Sec.~6 from the main document), Fig.~\ref{partition_vs_sliding} illustrates our colored barcode created using partition (Fig.~\ref{partition_vs_sliding}(a)) and sliding window timeslicing (Fig.~\ref{partition_vs_sliding}(b)). While the partition presents a smoother representation, the sliding window faithfully represents the activity variation over time. In the case of the partition, the number of connected components is computed only at each partition, equivalent to layouts representing the activity in grouped timestamps or timeslices, such as the LargeNetVis's Global View~\cite{LargeNetVis}. In contrast, the number and evolution of the connected components are highly dynamic when using sliding windows due to the quick identification of temporal events such as grows, merges, splits, and disappearances.

\section{Additional experiments}

\subsection{Visual comparison of filtering options}

\begin{figure}[t]
\includegraphics[width=\linewidth]{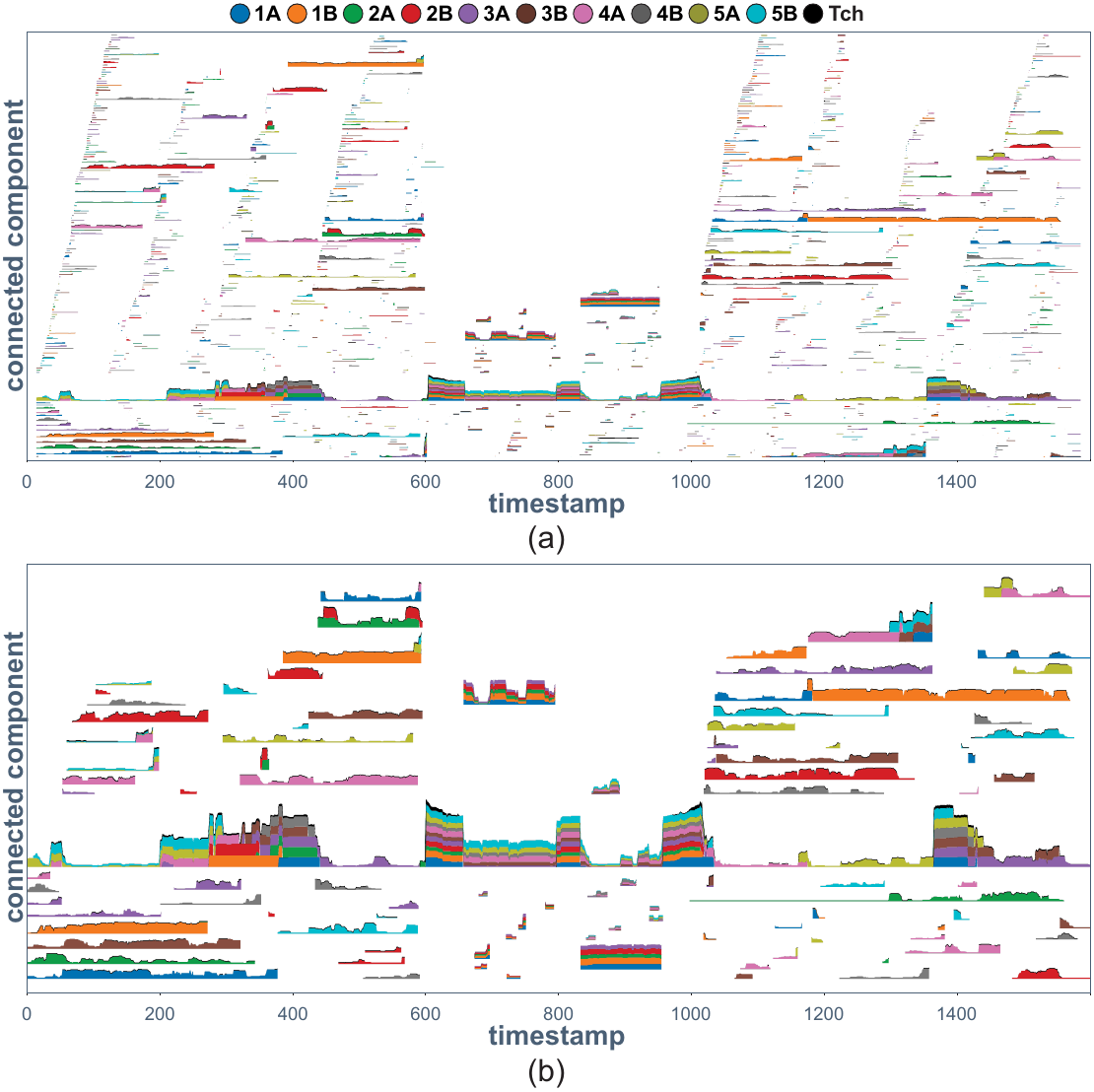}
\vspace{-0.5cm}
\caption{Comparing the colored barcodes for the Primary School network using sliding window: (a) without filtering and (b) after filtering out components with less than 10 nodes and 10 timestamps of duration. Adopted resolution: $r = 18$.}
\label{filter_length_height}
\vspace{-0.2cm}
\end{figure}

Fig.~\ref{filter_length_height} compares the colored barcode without filtering connected components (Fig.~\ref{filter_length_height}(a)) with the one after filtering out components with less than 10 nodes and 10 timestamps of duration (Fig.~\ref{filter_length_height}(b)). The dataset used is the first day of the Primary School with resolution $r = 6$. Although the colored barcode without filtering faithfully represents all active connected components in every timestamp, this layout leads to a high level of visual clutter, hindering the analyses. Focusing on large connected components leads to important regions of interest and improves the scalability of the approach. The minimum number of nodes and duration a component must have are currently user-defined thresholds; we believe further investigation may lead to a method that automatically suggests suitable values.


\subsection{Suggested resolutions}

\subsubsection{Resolution curves}

As mentioned in Sec.~9 of the main document, Fig~\ref{fig:suggestion_curves} depicts the normalized suggestion curves for the networks Primary School, High School, Hospital, InVS, Museum, Enron, Conference, and Sexual.

\subsubsection{Comparison of resolutions}

\blue{As also studied in Sec.~9 of the main document, Fig~\ref{fig:comparison_resolutions} shows the bars that differ the most when considering two selected resolutions, according to the bottleneck distance. The figure considers the networks Hospital, InVS, Museum, Enron, Conference, and Sexual. To see the bars that differ the most in the Primary School, please refer to Fig.~16 of the main document.}


\subsubsection{Comparison with other features}
\label{subsec:analysis_resolutions_quantitative}

We now compare our novel method with existing techniques.
In the literature, features of temporal graphs are of two sorts: either they are features of (non-temporal) graphs, adapted to the temporal case by taking their mean of their list over all the snapshots, or they directly depend on the temporal structure \cite{sizemore2018dynamic,orman2021finding}.

\myParagrapho{Geometric and topological features of snapshots.}
Let $G$ be a temporal graph to which we apply a sliding-window timeslicing of resolution $r$.
Given a snapshot $G_t$, we consider:
\begin{itemize}
\itemsep0cm
\item its number of nodes and edges, denoted $N(t)$ and $E(t)$,
\item its density $D(t) = 2E(t) / (V(t)(V(t)-1))$,
\item its number of connected components $CC(t)$,
\item the mean degree $MD(t)$ of its nodes,
\item its transitivity $T(t)$, defined the ratio between the number of triangles and triads (i.e., pairs of edges sharing a vertex).
\end{itemize}
Taking the mean value of such a feature over all times $t$ yields a feature of the temporal graph $G$.
We denote them respectively $N(r)$, $E(r)$, $D(r)$, $CC(r)$, $MD(r)$ and $T(r)$, making explicit the dependence on the resolution.

\begin{figure}[t]
\begin{center}    
\includegraphics[width=1\linewidth]{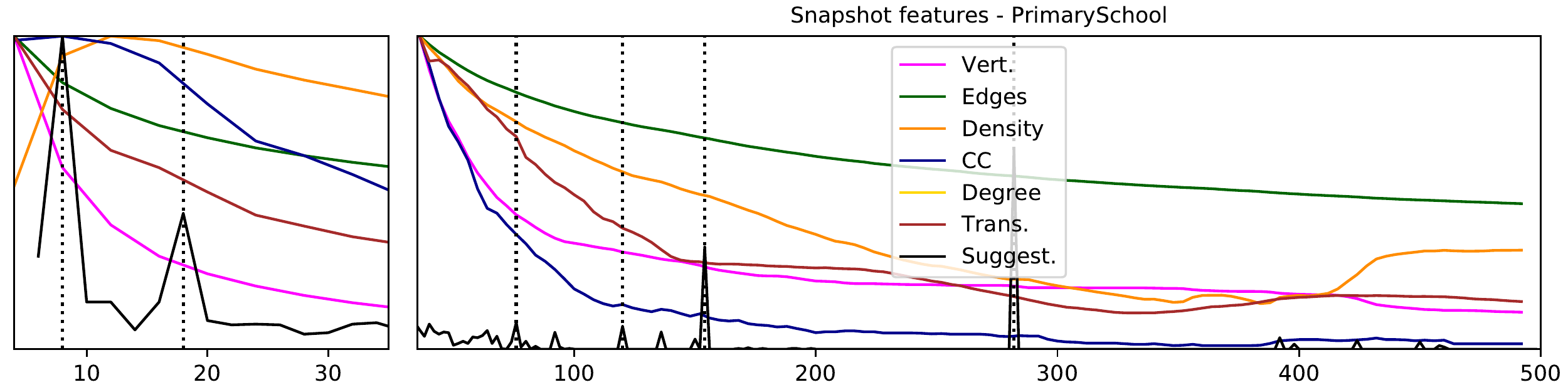}
\\
\includegraphics[width=1\linewidth]{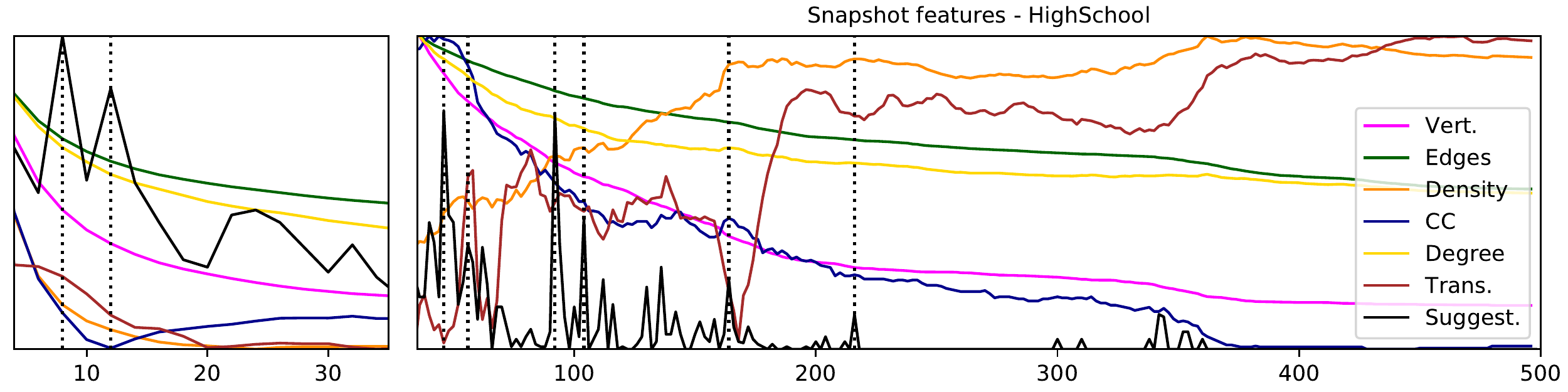}
\end{center}
\vspace{-0.5cm}
\caption{Suggestion curve and derivative of the mean snapshot features for the networks Primary School (top) and High School (bottom). 
The $x$ axis represents the resolution values.
The curves are normalized, and a few interesting values are highlighted with a dashed line.}
\label{fig:snapshot_features}
\vspace{-0.3cm}
\end{figure}

Note that, when increasing the resolution $r$, the features $N(r)$, $E(r)$ and $D(r)$ increase.
That is, they are non-decreasing functions.
In general, we except that abrupt changes in these values reflect the fact that the temporal graph exhibits a new behavior.
To visualize such changes, one can plot these curves or, more efficiently, their derivative.
These curves are represented in Fig.~\ref{fig:snapshot_features} for the Primary and High School networks, and in Fig.~\ref{fig:features_mean} for the other graphs.
Since we are only interested in the peaks or qualitative behaviors of these curves, and not their absolute values, we normalize them so that their maxima equals one.
In order to ease the reading, the $x$-axis is divided in two windows, and the curves are normalized both times.

A manual inspection of these curves allows us to compare them with our suggestion curve.
For instance, in the Primary School Network, one sees that the first peak, at $r=8$, corresponds to the global maximum of the derivative of the number of connected components.
Besides, the peak at resolution 154 seems to appear simultaneously as the transitivity curve shows a constant derivative.

Similar observations can be made on the High School network.
The peaks of our suggestion curve found at resolutions 46, 204, and 216 correspond, respectively, to global maxima of the derivative of the number of connected components, transitivity, and density.
Besides, the peaks at resolutions 12 and 56 correspond, respectively, to a global minimum of the derivative of the number of connected components and a significant local maximum of the transitivity.

These observations suggest that our curve captures information coming from various features of graphs.
However, some peaks remain unexplained, and hence we will study other features in the paragraphs below.
It must also be noted that certain features' peaks do not correspond to a peak of the suggestion curve.
This may be caused by the fact that the suggestion curve, based on the homology group $H_0$, is blind to certain purely geometric properties of graphs, and works only in terms of connected components.

\begin{figure}[t]
\begin{center}    
\includegraphics[width=1\linewidth]{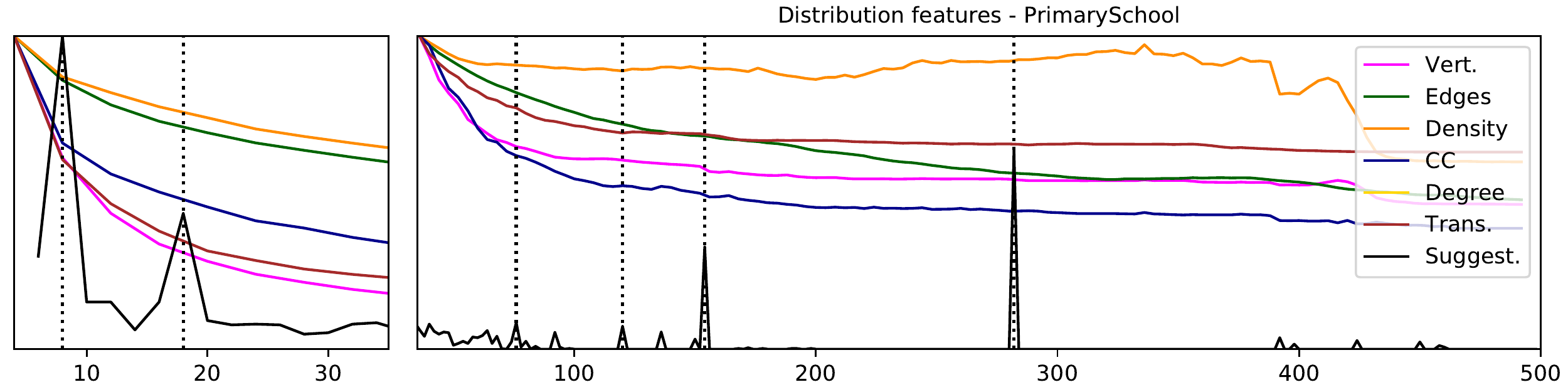}
\\
\includegraphics[width=1\linewidth]{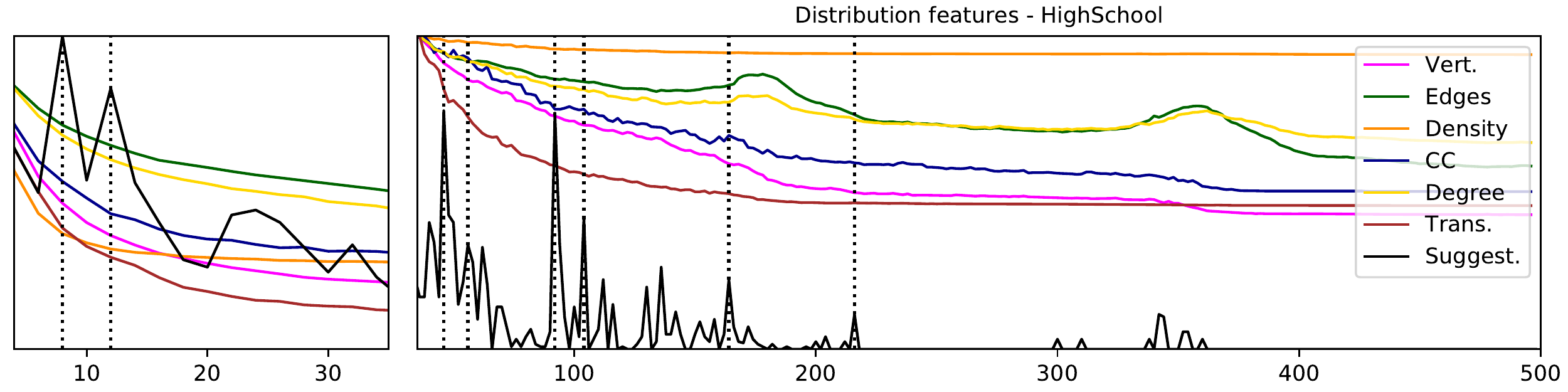}
\end{center}
\vspace{-0.5cm}
\caption{Consecutive distances between the distribution of the snapshot features for the networks Primary School (top) and High School (bottom). 
The $x$ axis represents the resolution values.
The curves are normalized, and a few interesting values are highlighted with a dashed line.}
\label{fig:distribution_features}
\vspace{-0.3cm}
\end{figure}

\myParagrapho{Distribution of features of snapshots.}
Instead of taking the average value of a feature over all the snapshots, we can compare their distribution over time.
To do so, we consider the entire curves 
\begin{center}    
$f_{N,r}\colon t\mapsto N(t)$, 
~~~~$f_{E,r}\colon t\mapsto E(t)$, 
~~~~etc.
\end{center}
Given two consecutive resolutions $r$ and $r+2$, we compare these curves via their $\ell^2$-norm 
\begin{center}    
$\|f_{N,r}-f_{N,r+2}\|_2$, 
~~~~$\|f_{E,r}-f_{E,r+2}\|_2$, 
~~~~etc.
\end{center}
These are functions of $r$, on which we expect to observe abrupt changes in the graph's behavior.
These curves are represented in Fig.~\ref{fig:distribution_features} for the Primary and High School networks and in Fig.~\ref{fig:features_distrib} for the other graphs.

As before, one draws correspondences between the suggestion curve and these features.
For example, on the Primary School network, the first suggested resolution, $r=8$, happens precisely during an abrupt change in the derivative of all the curves.
Besides, the two peaks at resolutions 120 and 154 delimit the only interval where the curve of transitivity increases and then decreases.
This last correspondence has already been observed in the last paragraph while considering the mean transitivity of the temporal graph.

In a similar fashion, on the High School network, one observes that the selected resolution 46 corresponds to an abrupt change in the curve built from transitivity.
In a few words, comparing the distribution of the snapshot features offers compatible but also complementary information to the average values alone.

\begin{figure}[t]
\begin{center}    
\includegraphics[width=1\linewidth]{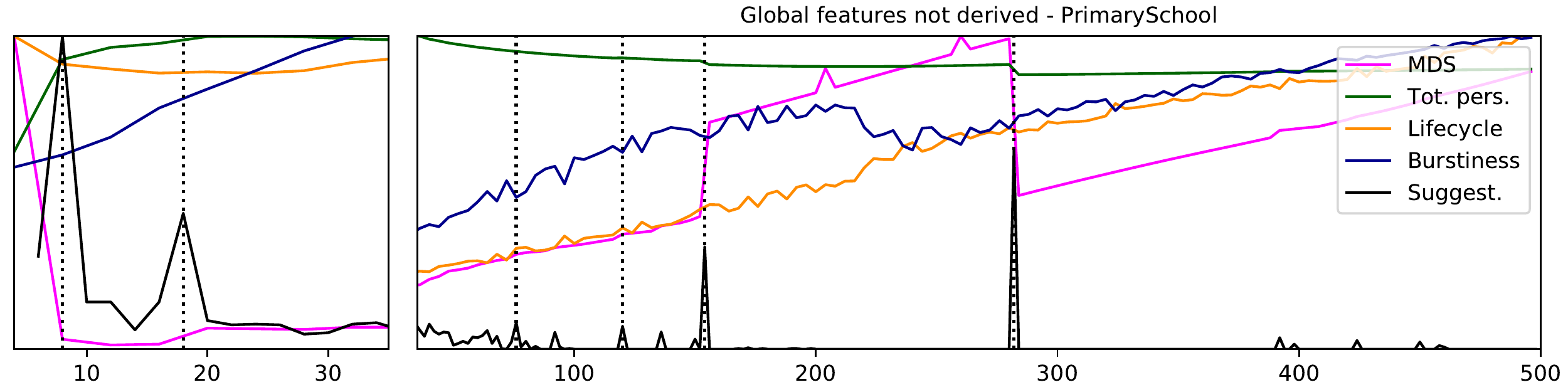}
\\
\includegraphics[width=1\linewidth]{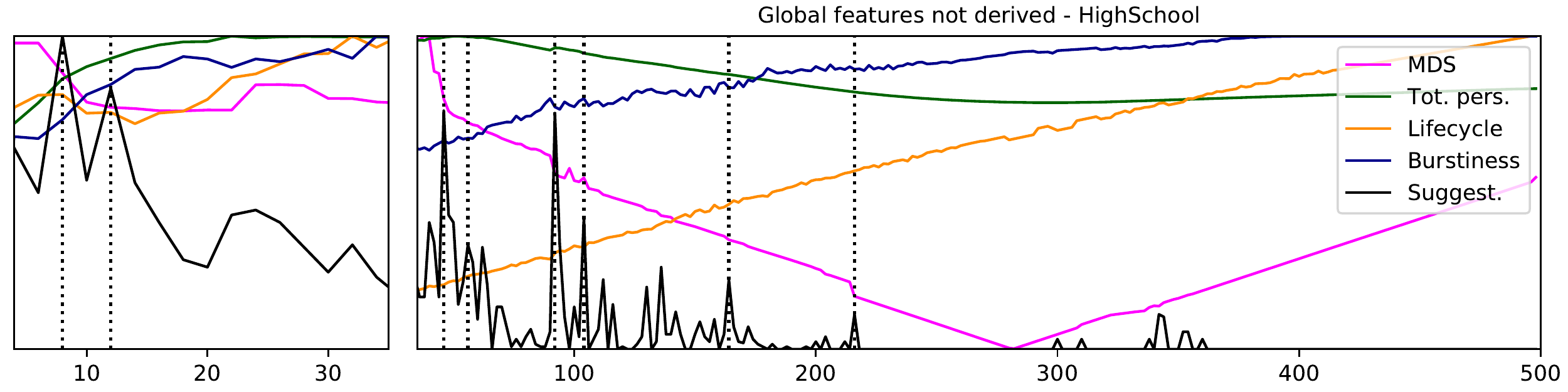}
\end{center}
\vspace{-0.5cm}
\caption{Global features as functions of the resolution parameter, for the networks Primary School (top) and High School (bottom).
The $x$ axis represents the resolution values.
The curves are normalized, and a few interesting values are highlighted with a dashed line.}
\label{fig:global_features}
\vspace{-0.3cm}
\end{figure}

\begin{figure*}[!ht]
\begin{center}    
\begin{minipage}[t]{.49\linewidth}
\includegraphics[width=1\linewidth]{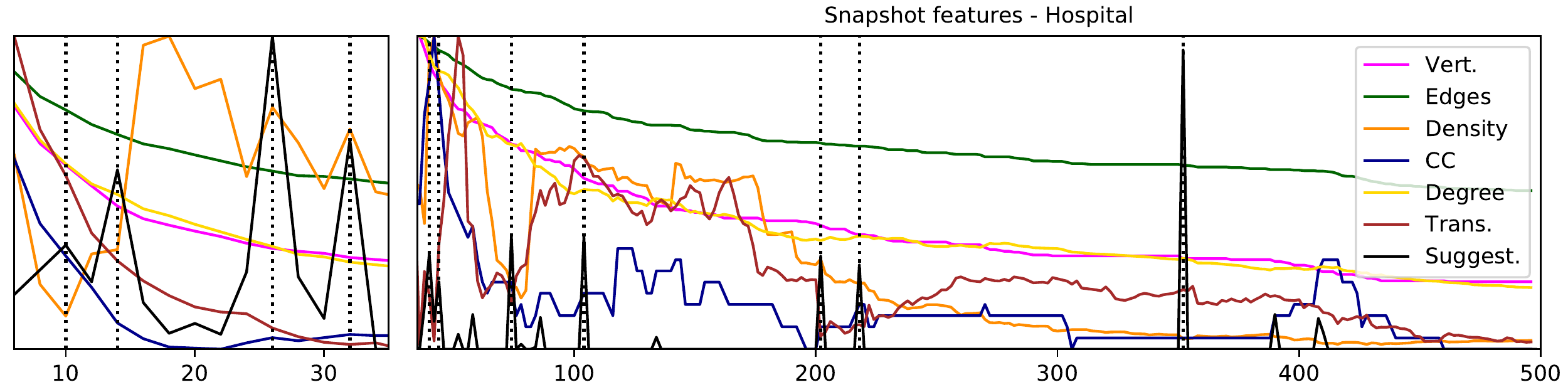}
\end{minipage}
\begin{minipage}[t]{.49\linewidth}
\includegraphics[width=1\linewidth]{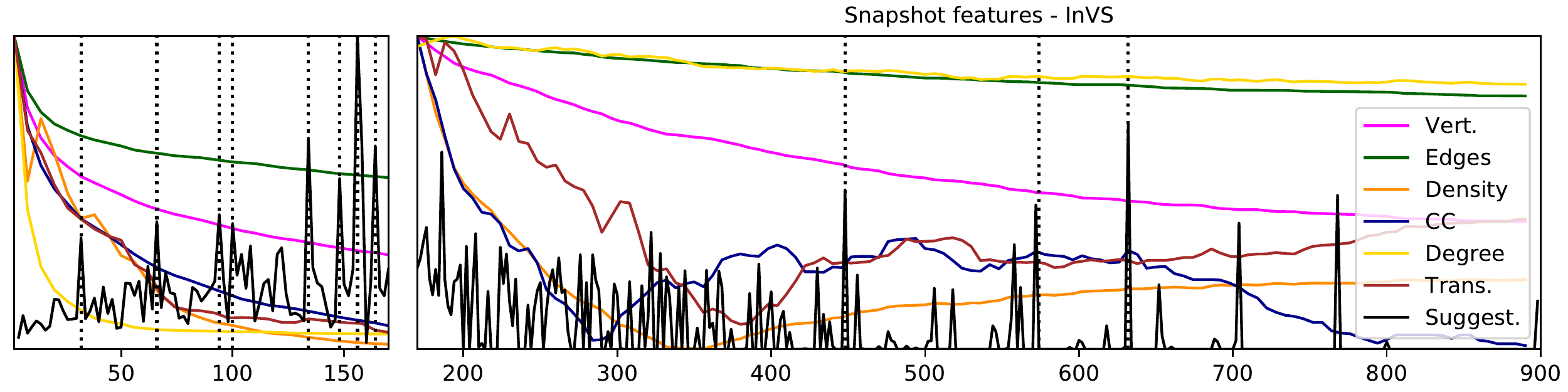}
\end{minipage}
\begin{minipage}[t]{.49\linewidth}
\includegraphics[width=1\linewidth]{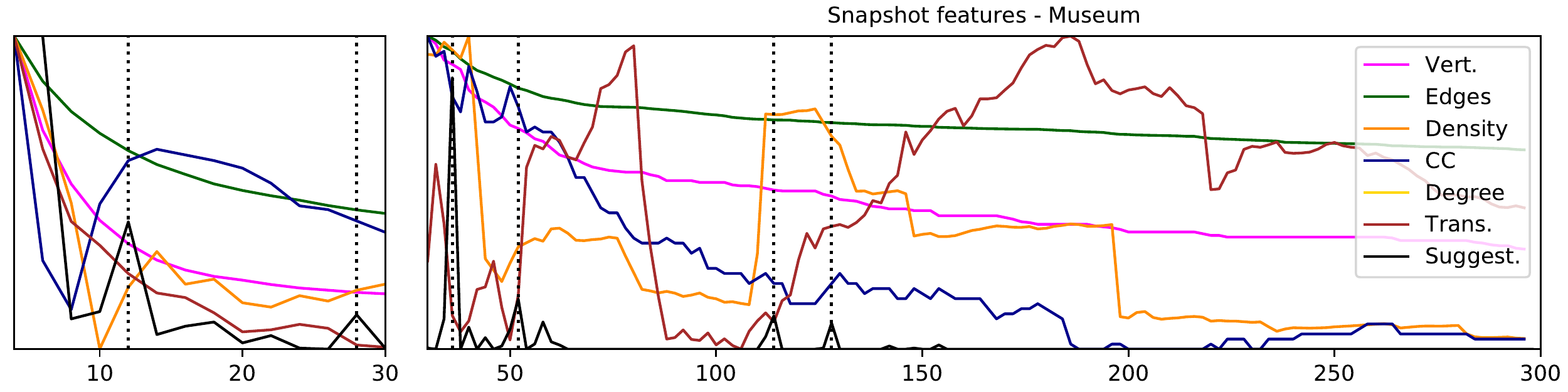}
\end{minipage}
\begin{minipage}[t]{.49\linewidth}
\includegraphics[width=1\linewidth]{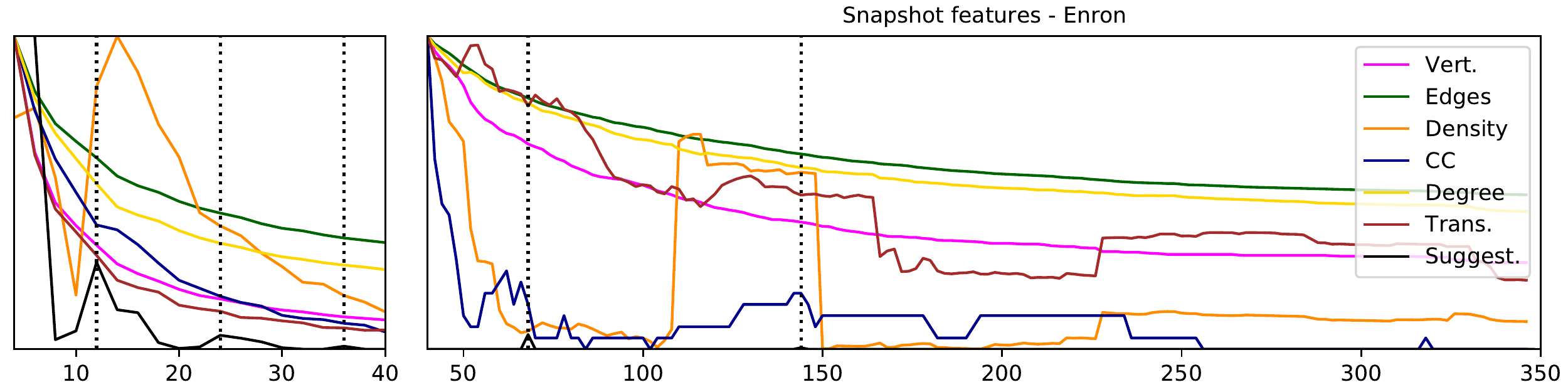}
\end{minipage}
\begin{minipage}[t]{.49\linewidth}
\includegraphics[width=1\linewidth]{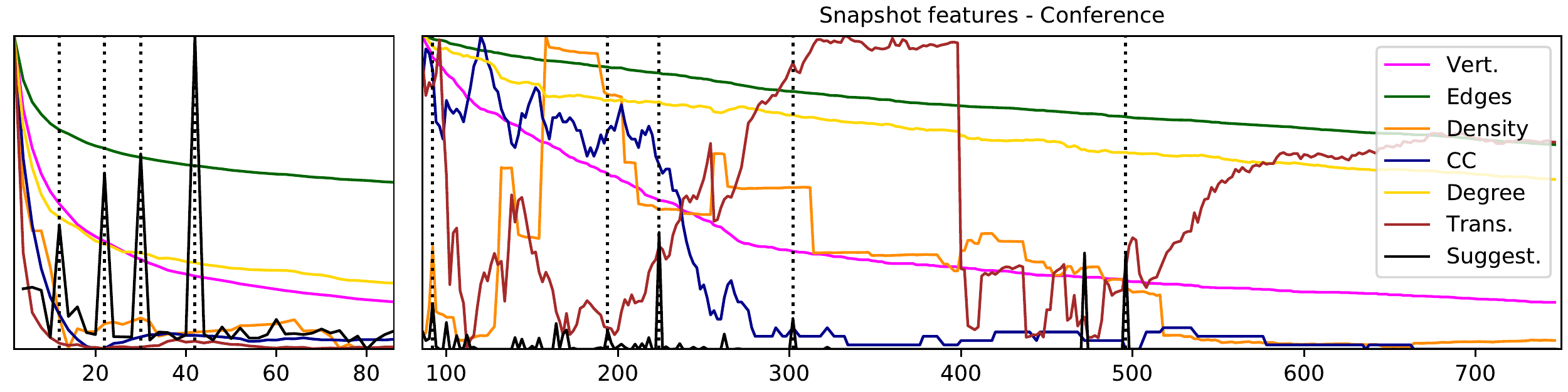}
\end{minipage}
\begin{minipage}[t]{.49\linewidth}
\includegraphics[width=1\linewidth]{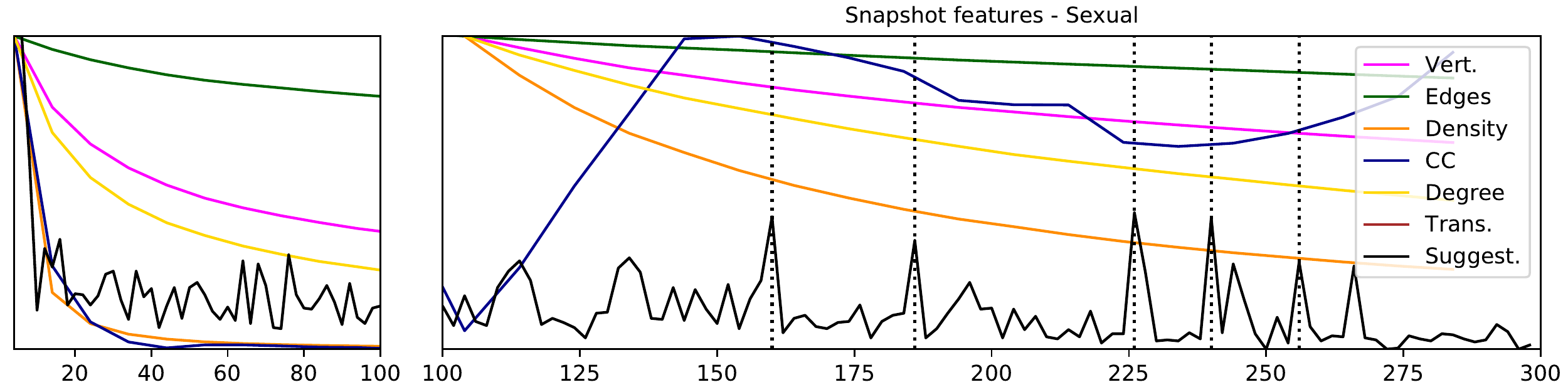}
\end{minipage}
\end{center}
\vspace{-0.7cm}
\caption{\blue{Suggestion curve and derivative of the mean snapshot features
for the networks Hospital, InVS, Museum, Enron, Conference, and Sexual, from left-to-right then top-to-bottom.
The x-axis represents the resolution values. The curves are normalized, and a few interesting
values are highlighted with a dashed line.}}
\label{fig:features_mean}
\end{figure*}

\begin{figure*}[!ht]
\begin{center}    
\begin{minipage}[t]{.49\linewidth}
\includegraphics[width=1\linewidth]{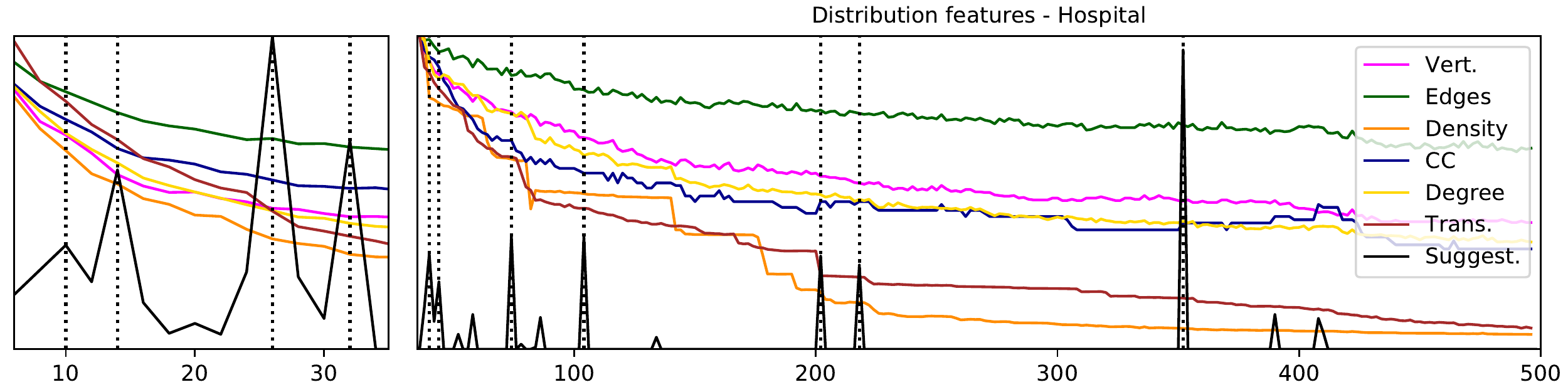}
\end{minipage}
\begin{minipage}[t]{.49\linewidth}
\includegraphics[width=1\linewidth]{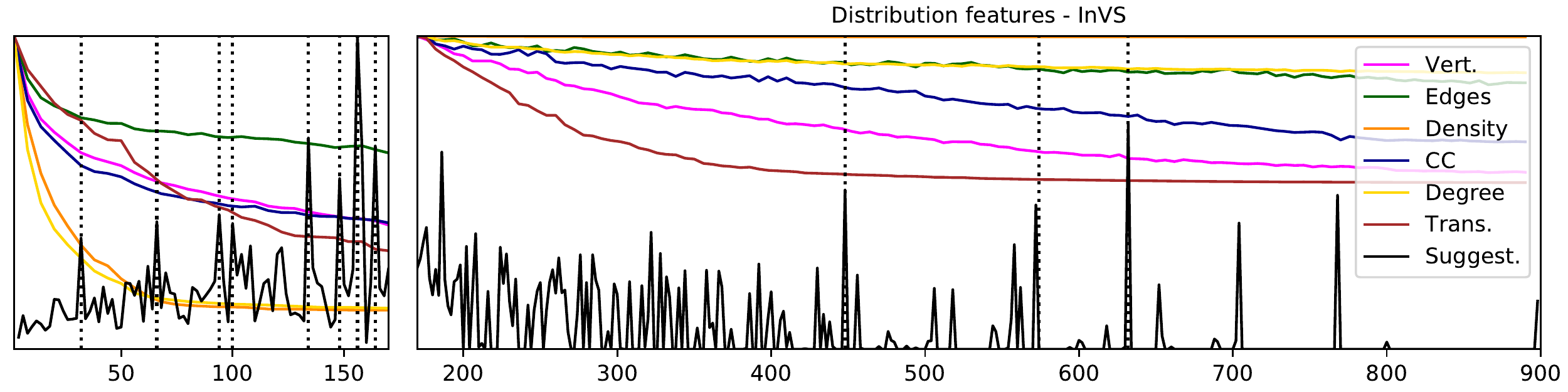}
\end{minipage}
\begin{minipage}[t]{.49\linewidth}
\includegraphics[width=1\linewidth]{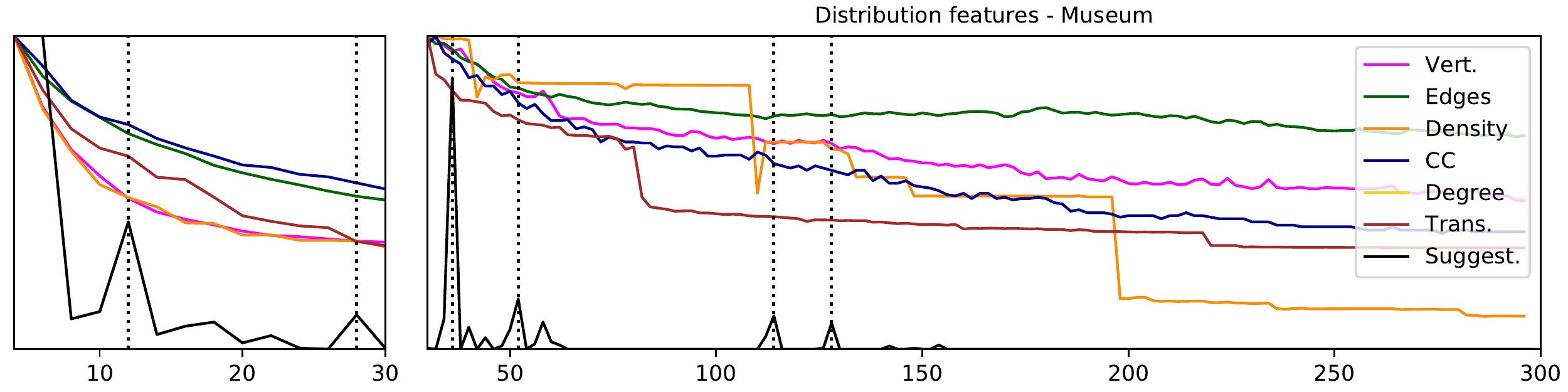}
\end{minipage}
\begin{minipage}[t]{.49\linewidth}
\includegraphics[width=1\linewidth]{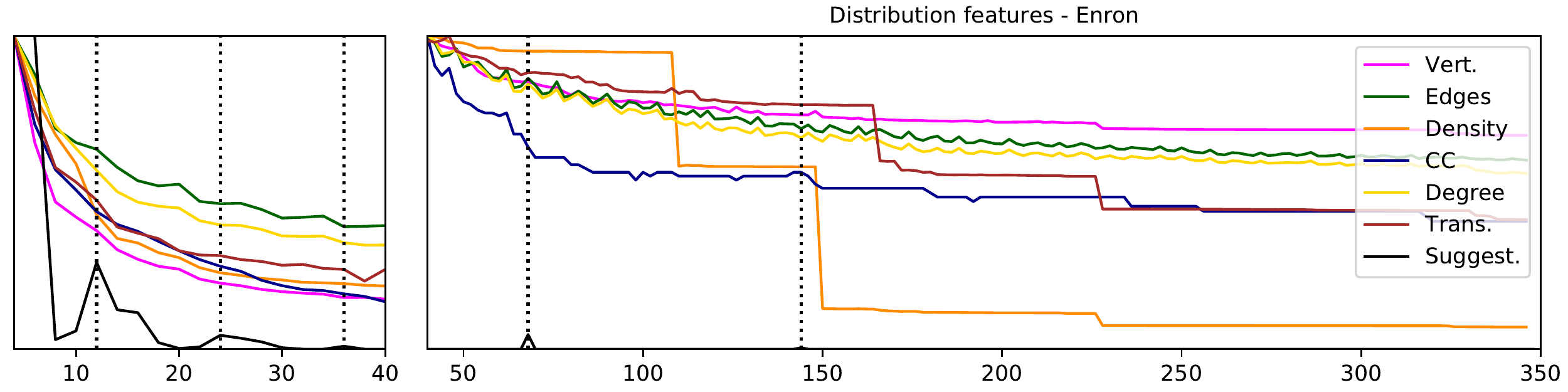}
\end{minipage}
\begin{minipage}[t]{.49\linewidth}
\includegraphics[width=1\linewidth]{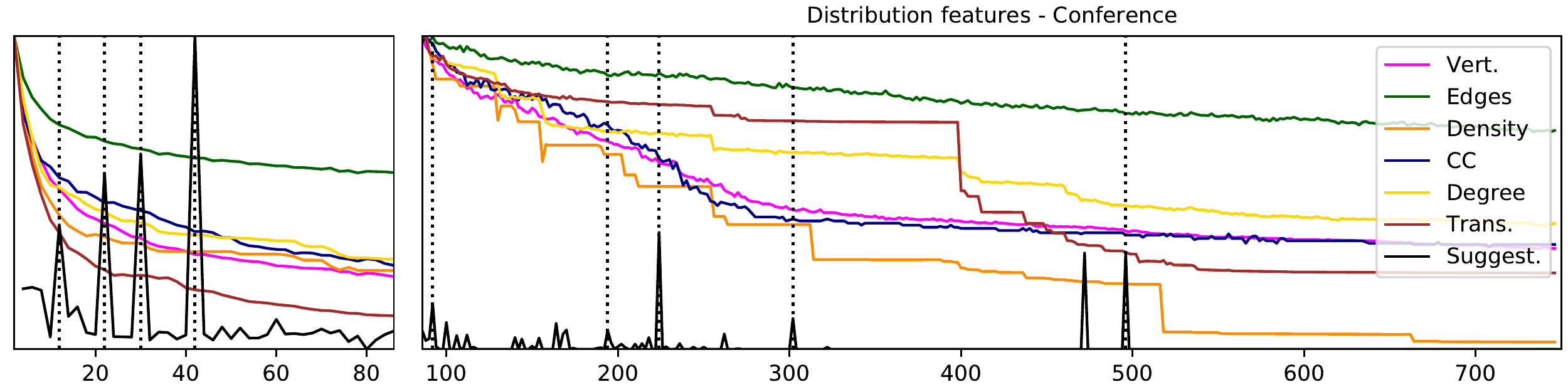}
\end{minipage}
\begin{minipage}[t]{.49\linewidth}
\includegraphics[width=1\linewidth]{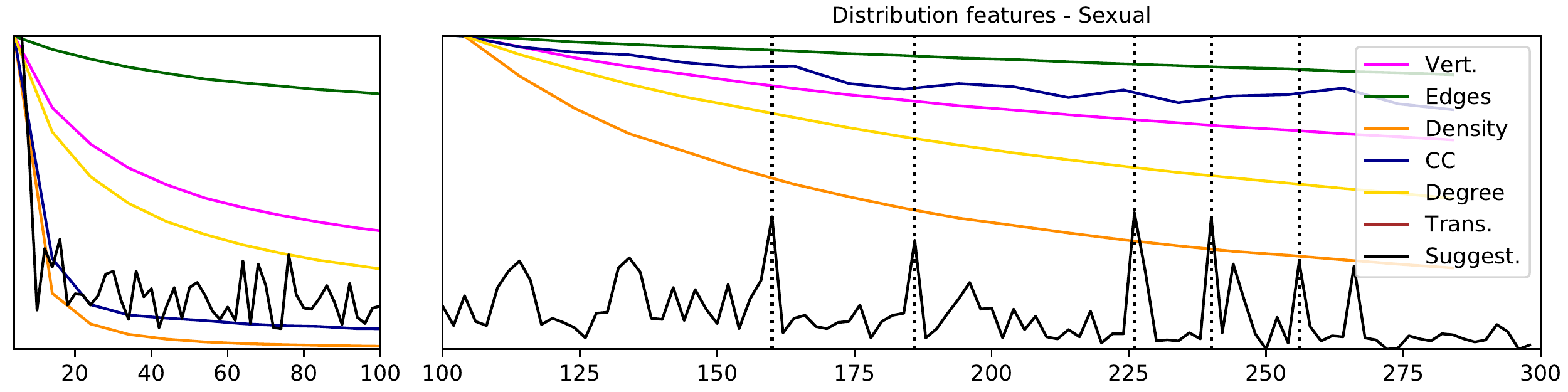}
\end{minipage}
\end{center}
\vspace{-0.7cm}
\caption{\blue{Consecutive distances between the distribution of the snapshot
features, for the networks Hospital, InVS, Museum, Enron, Conference, and Sexual, from left-to-right then top-to-bottom.
The x-axis represent the resolution values. The curves are normalized, and a few interesting values are highlighted with a dashed line.}}
\label{fig:features_distrib}
\end{figure*}

\begin{figure*}[!ht]
\begin{center}    
\begin{minipage}[t]{.49\linewidth}
\includegraphics[width=1\linewidth]{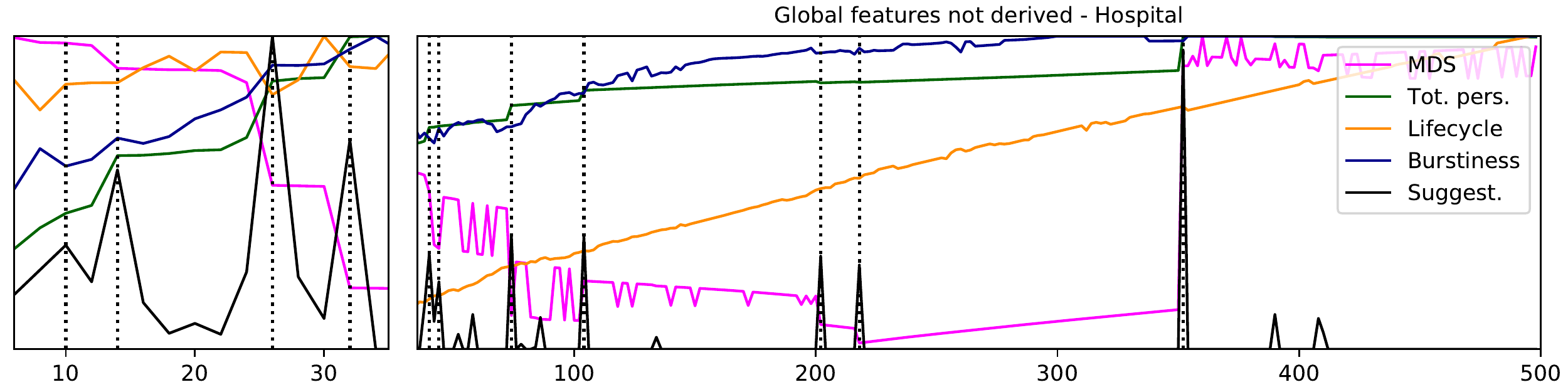}
\end{minipage}
\begin{minipage}[t]{.49\linewidth}
\includegraphics[width=1\linewidth]{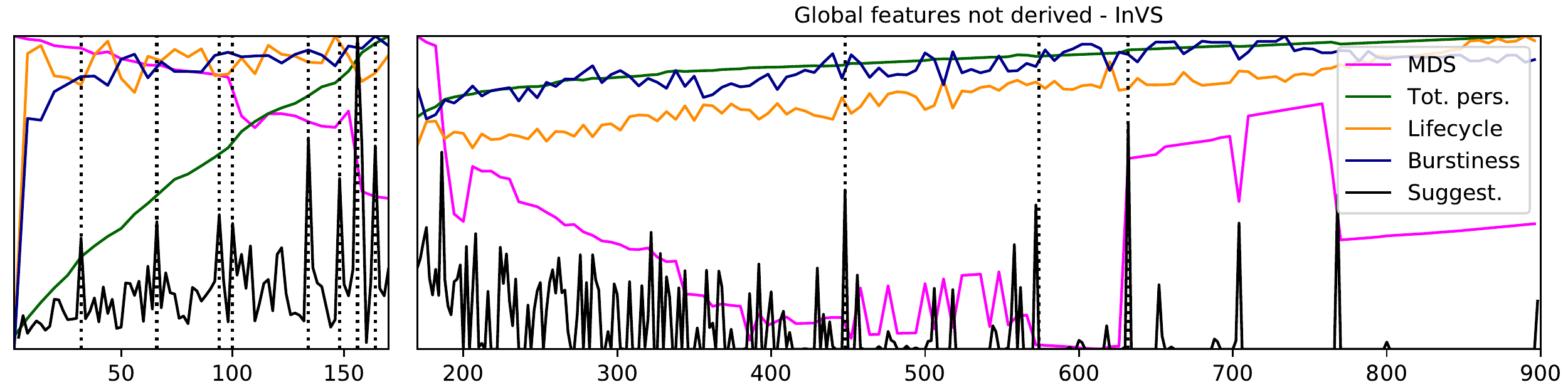}
\end{minipage}
\begin{minipage}[t]{.49\linewidth}
\includegraphics[width=1\linewidth]{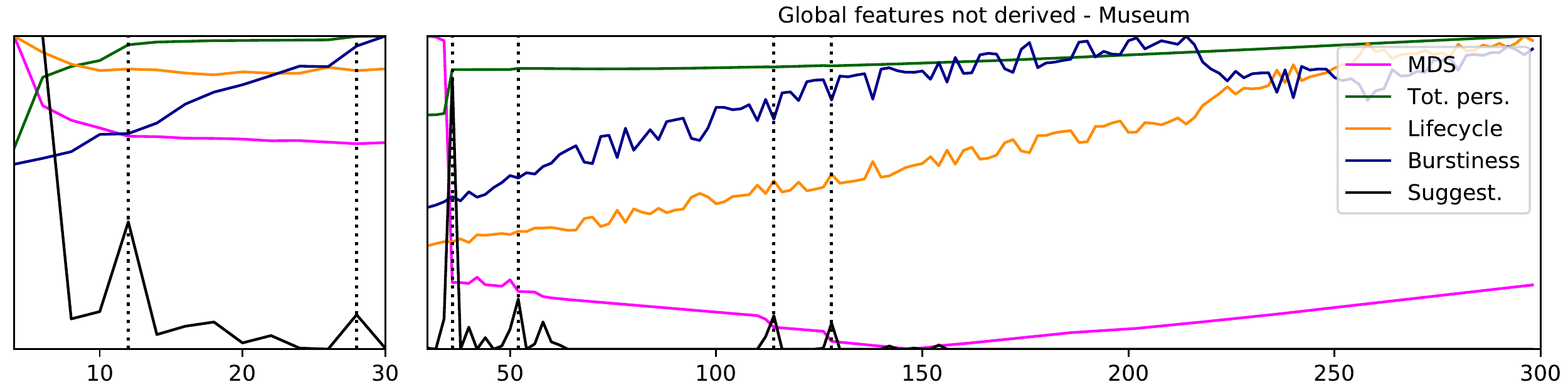}
\end{minipage}
\begin{minipage}[t]{.49\linewidth}
\includegraphics[width=1\linewidth]{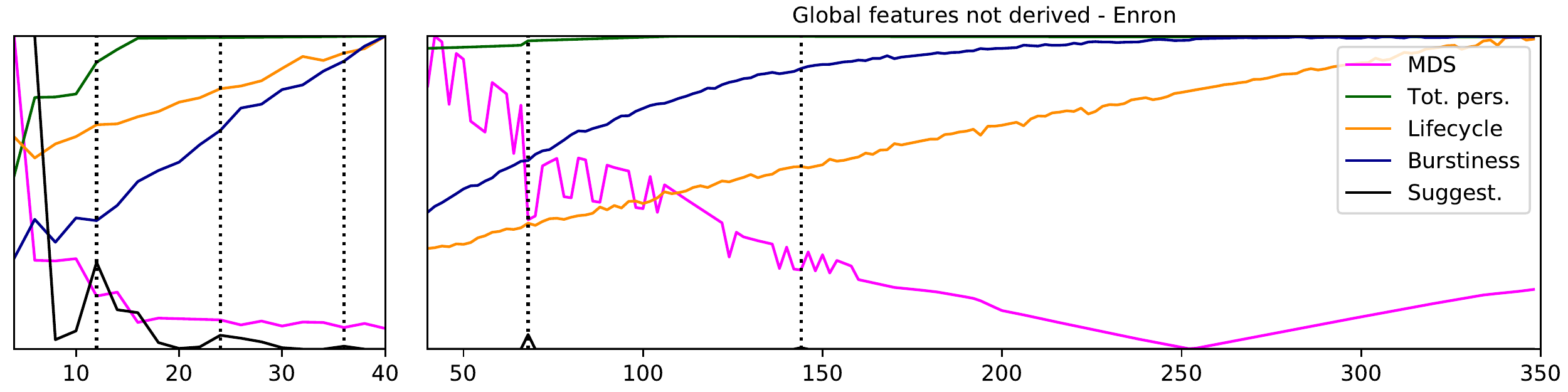}
\end{minipage}
\begin{minipage}[t]{.49\linewidth}
\includegraphics[width=1\linewidth]{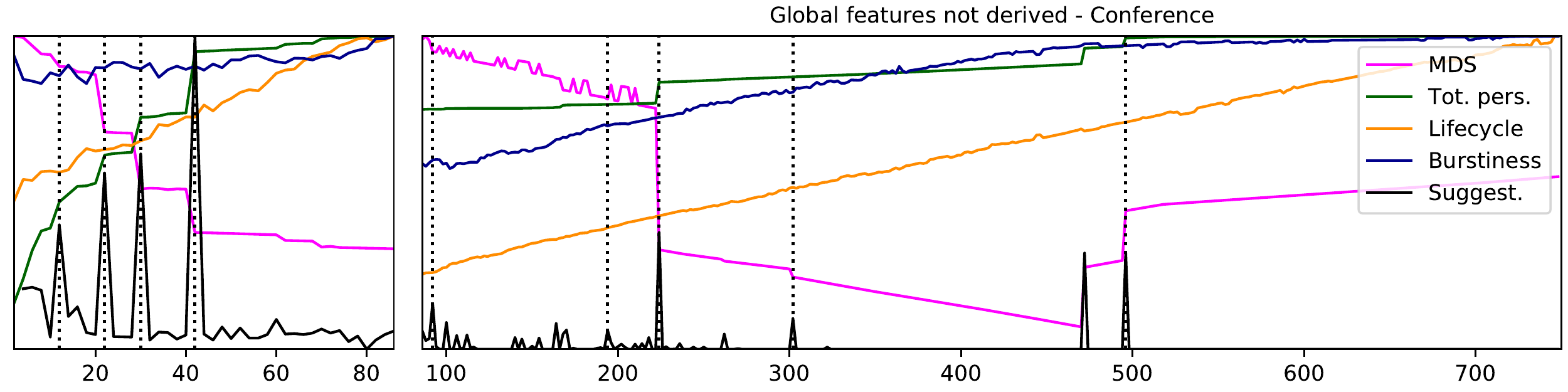}
\end{minipage}
\begin{minipage}[t]{.49\linewidth}
\includegraphics[width=1\linewidth]{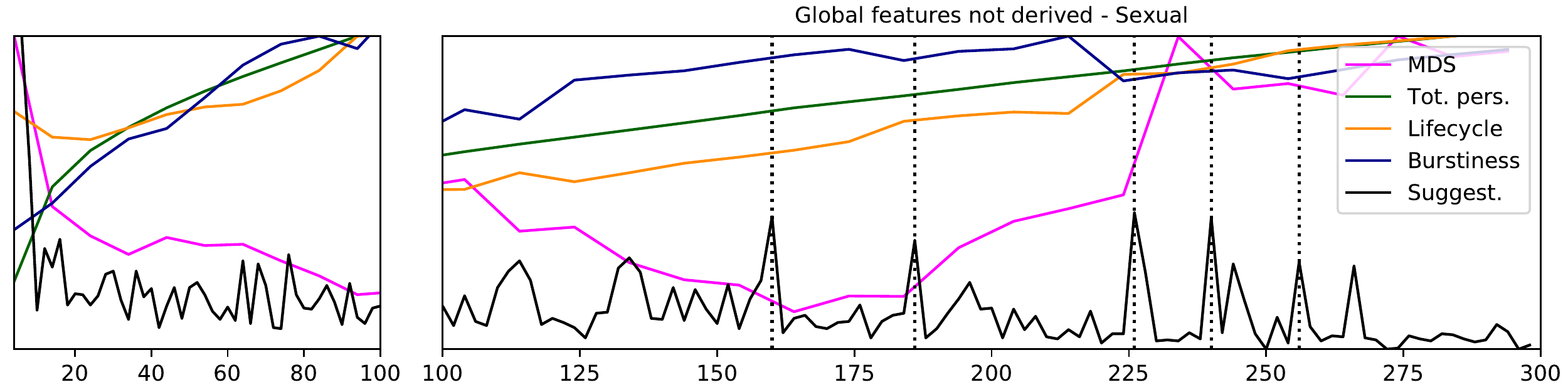}
\end{minipage}
\end{center}
\vspace{-0.7cm}
\caption{\blue{Global features for the networks Hospital, InVS, Museum, Enron, Conference, and Sexual, from left-to-right then top-to-bottom.
The x-axis represents the resolution values. The curves are normalized, and a few interesting
values are highlighted with a dashed line.}}
\label{fig:features_global}
\end{figure*}

\myParagrapho{Global features.}
Lastly, we consider features of dynamic graphs that do not come from features of snapshots. 
Given a temporal graph $G$, timesliced at a resolution $r$, we consider:
\begin{itemize}
\itemsep0cm
\item its burstiness $B(r)$ and average lifecycle $LC(r)$, defined in~\cite{samplingLuis}.
\item its stability $S(r)$ and fidelity $F(r)$, defined in \cite{stability}.
\end{itemize}
Moreover, we will also consider the \textit{total persistence} $TP(r)$ of its corresponding zigzag persistence module, defined as the quadratic mean of the length of its bars.
Finally, we will employ the curve $MDS(r)$ defined in \cite{8365984}. 
It consists of the multidimensional scaling (MDS) in dimension 1, whose input is the set of bottleneck distances between the persistence barcodes of the temporal graph for all the resolutions considered.
These curves are represented in Fig.~\ref{fig:global_features}  for the Primary and High School networks and in Fig.~\ref{fig:features_global} for the other graphs.

On all the graphs, one observed a high correlation between our suggestion curve and the curves of MDS and total persistence.
This is expected since all these features are related to the persistence barcodes of the zigzag modules.
We also observe that the peak at $r=18$ of our suggestion curve for the Primary School corresponds to a global minimum of the lifecycle. The same occurs with $r=12$ for the High School network.

Regarding stability and fidelity, we observe that the curves $r\mapsto S(r)$ and $r\mapsto F(r)$ are convex and no abrupt change can be observed.
Instead, we consider a relevant feature that can be defined from them: the point of intersection between the normalized stability and the inverse of the normalized fidelity reveals the resolution value that balances the best between these antagonists' values.
The resolutions suggested by this strategy are $r = 24$ and $r = 12$ for the Primary and High School, respectively. Note that ZigzagNetVis also suggests $r = 12$ for the High School.


In conclusion of this section, the peaks of the suggestion curve can, most of the time, be mapped to peaks or bumps of other features in the literature.
We stress that our analysis does not reveal the exact nature of this connection; we simply showed how the suggestion curve can be understood as related to other features.


\subsubsection{Comparison with Wasserstein distance}
\begin{figure}[!t]
\includegraphics[width=.99\linewidth]
{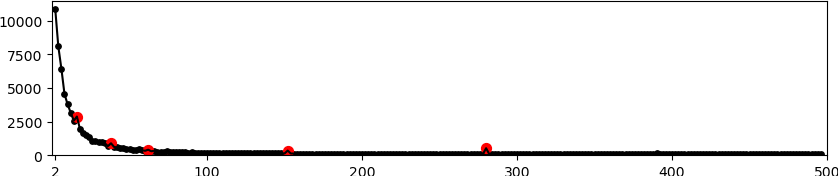}\\
\includegraphics[width=.99\linewidth]
{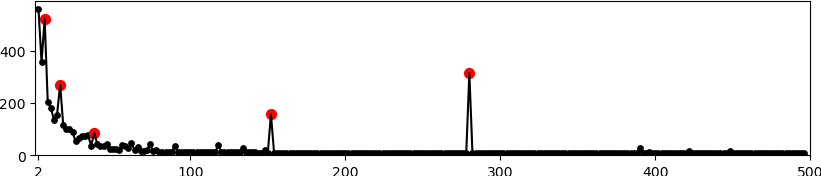}\\
\includegraphics[width=.99\linewidth]
{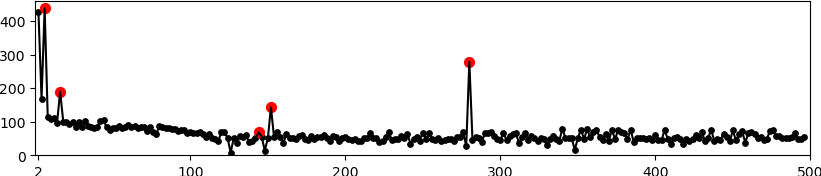}
\vspace{-0.2cm}
\caption{
Normalized suggestion curves with sliding-window timeslicing and Wasserstein distance, for the network Primary School. 
The distance is computed respectively with order 1, 2 and 10.
The $x$ and $y$ axes represent the resolution values and the consecutive Wasserstein distances.}
\label{fig:suggestion_Wasserstein_distance}
\end{figure}

\begin{figure*}[!t]
\begin{center}    
 \begin{minipage}[t]{.49\linewidth}
 \includegraphics[width=1\linewidth]{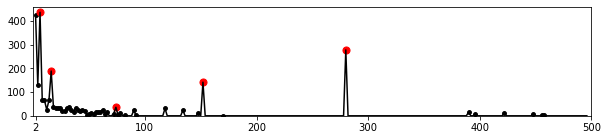}
 \end{minipage}
 \begin{minipage}[t]{.49\linewidth}
 \includegraphics[width=1\linewidth]{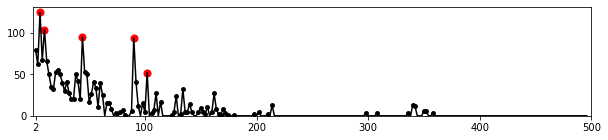}
 \end{minipage}
\begin{minipage}[h]{.49\linewidth}
\includegraphics[width=1\linewidth]{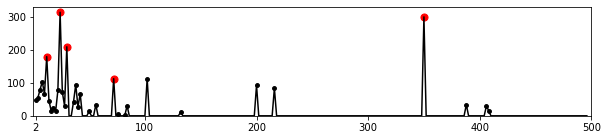}
\end{minipage}
\begin{minipage}[h]{.49\linewidth}
\includegraphics[width=1\linewidth]{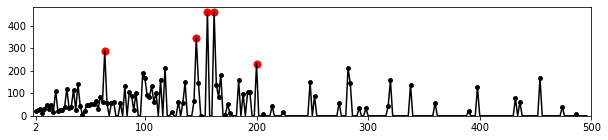}
\end{minipage}
\begin{minipage}[h]{.49\linewidth}
\includegraphics[width=1\linewidth]{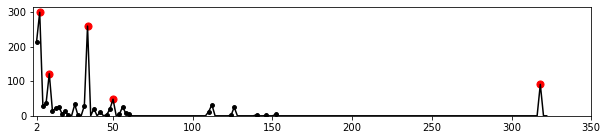}
\end{minipage}
\begin{minipage}[h]{.49\linewidth}
\includegraphics[width=1\linewidth]{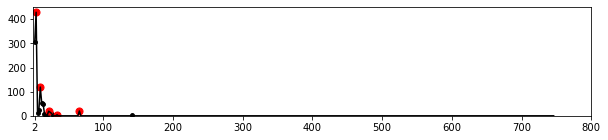}
\end{minipage}
\begin{minipage}[h]{.49\linewidth}
\includegraphics[width=1\linewidth]{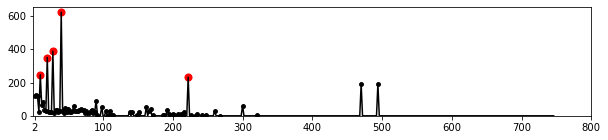}
\end{minipage}
\begin{minipage}[h]{.49\linewidth}
\includegraphics[width=1\linewidth]{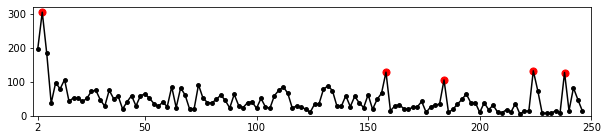}
\end{minipage}
\end{center}
\caption{Normalized suggestion curves, using sliding-window timeslicing, for the networks Primary School, High School, Hospital, InVS, Museum, Enron, Conference, and Sexual, from left-to-right then top-to-bottom. The $x$ and $y$ axes represent the resolution values and the consecutive bottleneck distance, respectively.}
\label{fig:suggestion_curves}
\end{figure*}

\begin{figure*}[!ht]
\begin{center}    
\begin{minipage}[t]{.49\linewidth}
\includegraphics[width=1\linewidth]{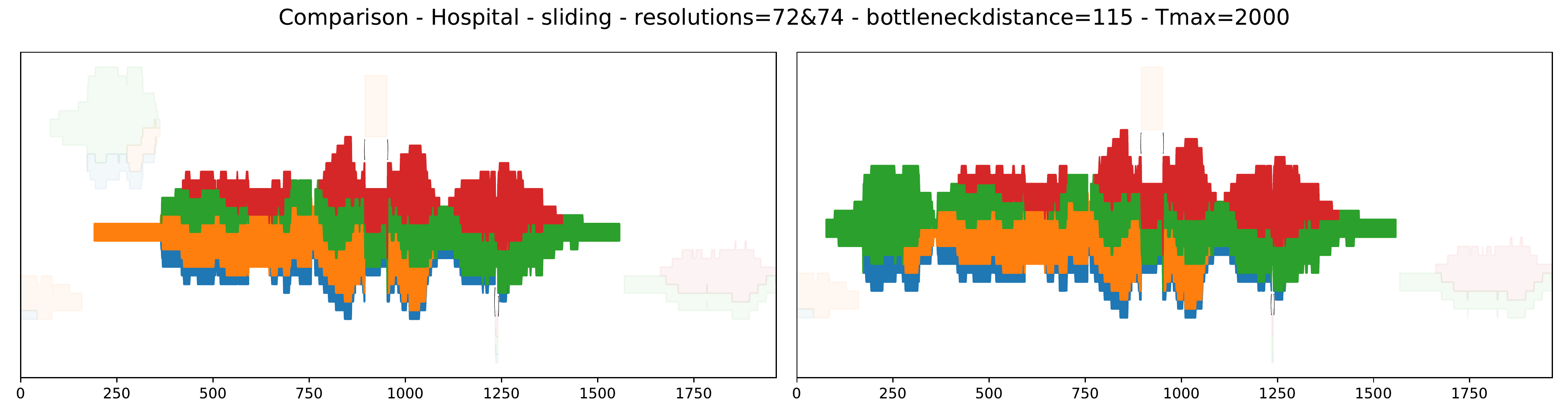}
\end{minipage}
\begin{minipage}[t]{.49\linewidth}
\includegraphics[width=1\linewidth]{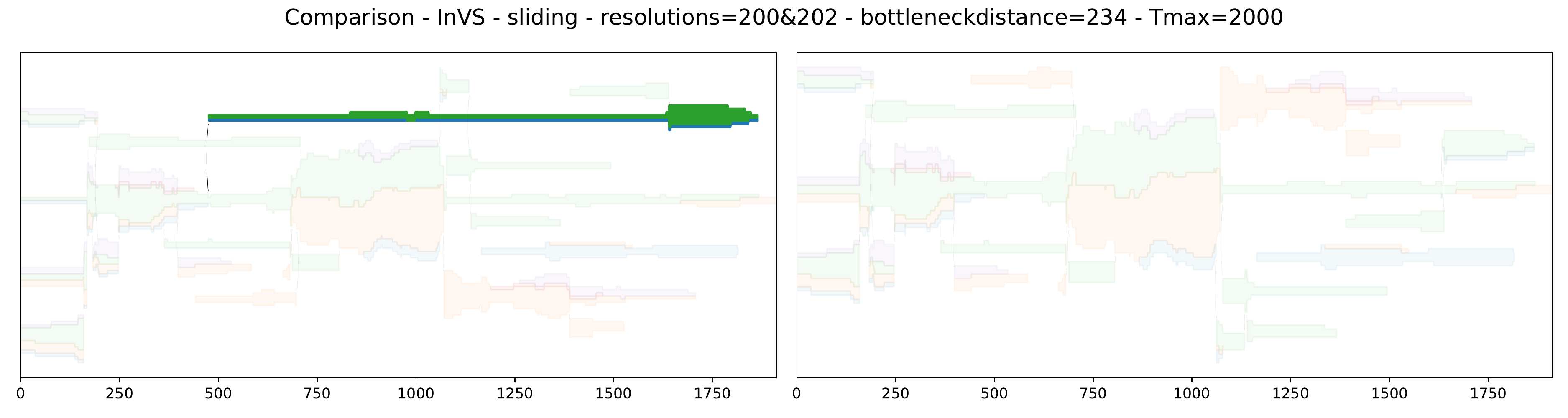}
\end{minipage}
\begin{minipage}[t]{.49\linewidth}
\includegraphics[width=1\linewidth]{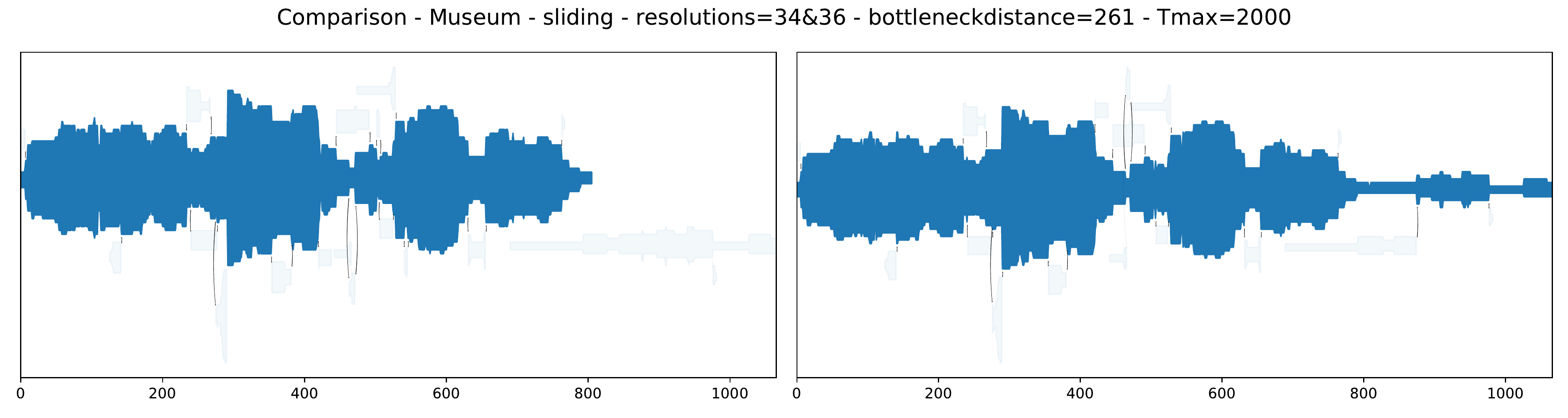}
\end{minipage}
\begin{minipage}[t]{.49\linewidth}
\includegraphics[width=1\linewidth]{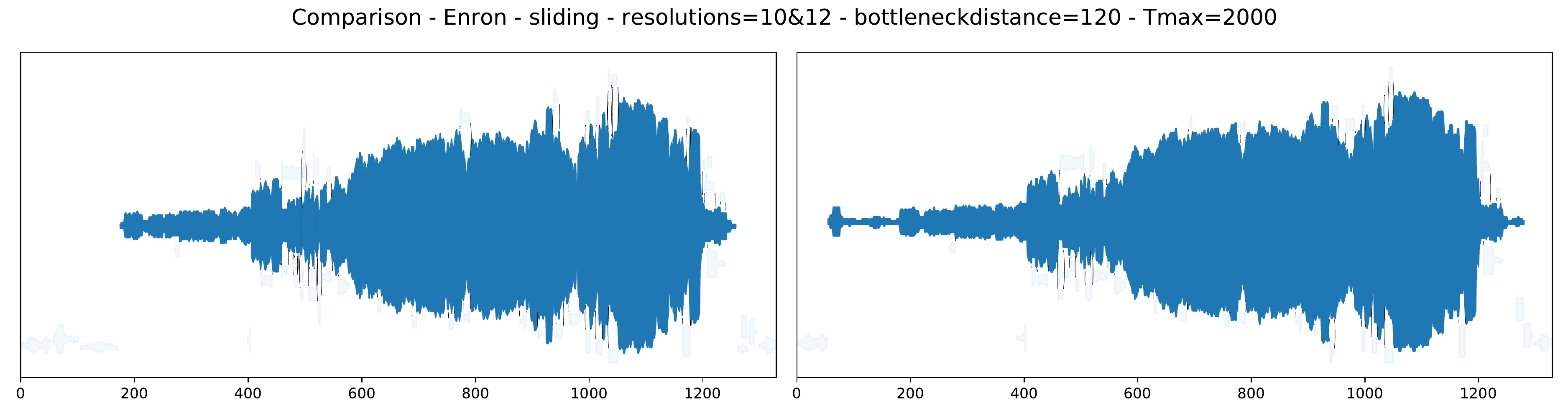}
\end{minipage}
\begin{minipage}[t]{.49\linewidth}
\includegraphics[width=1\linewidth]{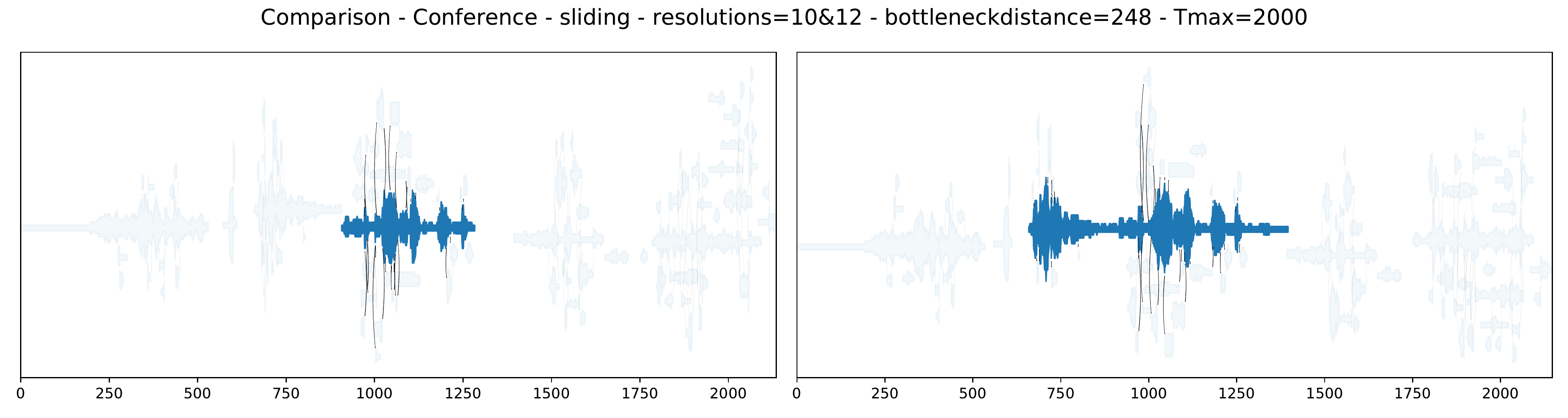}
\end{minipage}
\begin{minipage}[t]{.49\linewidth}
\includegraphics[width=1\linewidth]{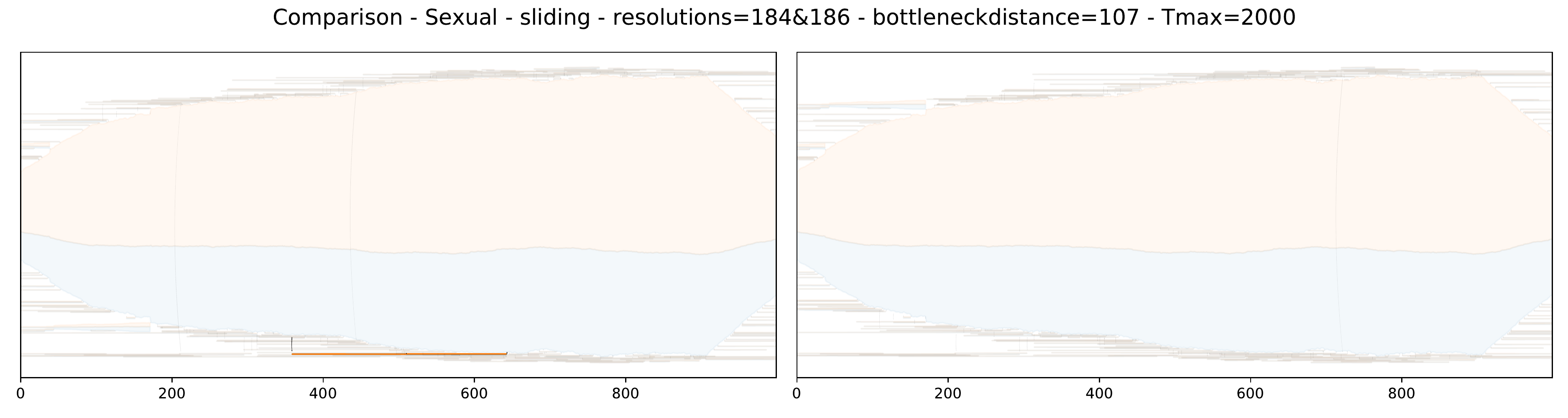}
\end{minipage}
\end{center}
\vspace{-0.3cm}
\caption{\blue{Visualization of the bottleneck distance for the networks Hospital, InVS, Museum, Enron, Conference, and Sexual, from left-to-right then top-to-bottom. The $x$ and $y$ axes represent the resolution values and the consecutive bottleneck distance, respectively.
Highlighted connected components represent the bars that differ the most when considering the two selected resolutions, according to the bottleneck distance.}}
\label{fig:comparison_resolutions}
\end{figure*}

\textcolor{black}{
As discussed in the Sec.~4.2 of our article, in the context of PH, a concurrent distance to the bottleneck is the Wasserstein distance.
We represent in Fig.~\ref{fig:suggestion_Wasserstein_distance} the suggestion curves obtained from the Wasserstein distance, on the Primary School dataset, for orders 1, 2 and 10.
The five suggested resolutions are respectively $\{18, 40, 64, 154, 282\}$, $\{8, 18, 40, 154, 282\}$ and $\{8, 18, 146, 154, 282\}$.
This is to be compared with the resolutions $\{8, 18, 76, 154, 282\}$ obtained with the bottleneck distance (Fig.~\ref{fig:suggestion_curves}(Primary School)).
Although yielding similar resolutions, the peaks of the suggestion curves obtained with the Wasserstein distance appear smaller or less isolated than that of bottleneck distance.
}

\subsection{Running times}

Tab.~\ref{tab:running_times} depicts our average running time of 10 executions for every procedure, i.e., the average running time needed to \textbf{open the dataset (1)}, \textbf{compute the suggestion curve (2)}, and \textbf{compute the colored barcode for a given resolution (3)}. The experiments were performed on a personal computer with Intel(R) Core(TM) i5-8350U x 8 CPU @ 3.60GHz, 16 GB RAM, and Ubuntu 22.10.
The table considers the eight networks and maximal time values from the main document, Sec.~9.

\begin{table}[!h]
\centering
\caption{Running times in seconds for eight distinct networks. }
\label{tab:running_times}
\resizebox{0.35\textwidth}{!}{%
\begin{tabular}{llll}
\hline
\textbf{Network} & \textbf{Step 1} & \textbf{Step 2} & \textbf{Step 3} \\ \hline
Primary School~\cite{primarySchool}&31&241&7\\
High School~\cite{highSchool}      &53&79&4\\
Hospital~\cite{Hospital}           &9&5&1\\
InVS~\cite{InVS}                   &3&5&1\\
Museum~\cite{museum_conference}      &3&12&1\\
Enron~\cite{enron}      &10&78&5\\
Conference~\cite{museum_conference}      &6&26&1\\
Sexual~\cite{sexual_contacts}      &10&571&600\\
\hline
\end{tabular}%
}
\end{table}

\subsection{Comparison with LargeNetVis}
\label{comparison_largenetvis}

To further validate the colored barcode, we performed a direct comparison with LargeNetVis~\cite{LargeNetVis}, an established approach to visualize large temporal networks. To be coherent with the partition timeslicing used by LargeNetVis, we decided to use partition timeslicing in ZigzagNetVis as well (see Sec.~\ref{visual_comparison_timeslicing_methods} for a visual comparison between partition and sliding timeslicing). We also forced the number of timeslices in LargeNetVis to be equal to the number of partitions in ZigzagNetVis for a fair comparison.

\begin{figure}[!t]
\includegraphics[width=\linewidth]{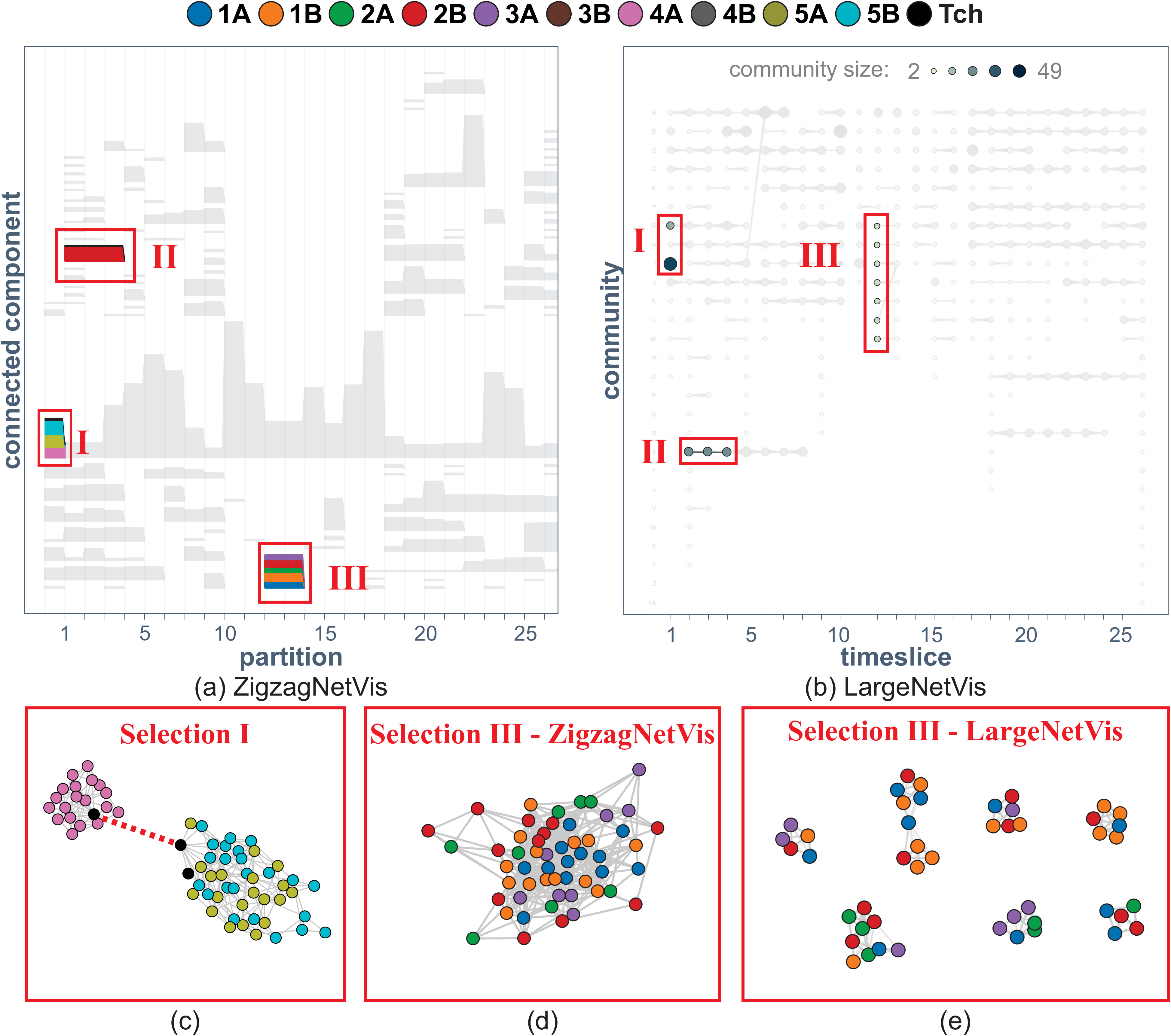}
\caption{Comparison between (a) ZigzagNetVis with bottom-based ordering and (b) LargeNetVis highlighting three distinct patterns (I-III) in the Primary School network.
}
\vspace{-0.2cm}
\label{barcode_largenetvis}
\end{figure}

Using the Primary School network, Fig.~\ref{barcode_largenetvis} shows a comparison between our colored barcode and LargeNetVis' Global View, also showing node-link diagrams that support the comparison. The first highlighted pattern refers to a single connected component on ZigzagNetVis containing students and teachers from three classes (4A, 5A, 5B) (Fig.~\ref{barcode_largenetvis}(a,I)) and the two equivalent communities from LargeNetVis (Fig.~\ref{barcode_largenetvis}(b,I)). When analyzing the corresponding node-link diagrams (Fig.~\ref{barcode_largenetvis}(c)), we see that these two communities form a single connected component thanks to a single edge (dotted in red) linking two teachers (task T1). Also, we see that students from class 4A interact with each other but not with the other two classes (5A and 5B); on the other hand, students from 5A interact with students from 5B and vice-versa (T1).
This finding is supported by the fact that students in the same class interact more often with themselves than with students in other classes and that the same goes for same-grade students.
Note also that LargeNetVis only allows us to identify the school classes that are found in a community through the node-link diagram. However, this information is immediate with ZigzagNetVis' colored barcode.

Regarding the second pattern (Fig.~\ref{barcode_largenetvis}(a-b,II)), both layouts were able to faithfully represent the continuous level of interactions involving students in class 2B and their teacher (task T2). Once again, note that the colored barcode already shows the school class involved in the interactions. Regarding the third pattern, the colored barcode highlights a component that contains students from five different classes interacting with each other during lunch break (Fig.~\ref{barcode_largenetvis}(a,III)). When analyzing the node-link diagram (Fig.~\ref{barcode_largenetvis}(d)), we see several interactions between these students during that period, i.e., the component is strongly connected (task T1). In LargeNetVis, due to the nature of the community detection algorithm, this strongly connected component was divided into seven small communities, which impaired the finding of this strongly connected group (see Fig.~\ref{barcode_largenetvis}(b,III) and Fig.~\ref{barcode_largenetvis}(e)).

Each layout has advantages and disadvantages depending on the user task. We aimed to demonstrate that our approach compares to well-validated visualizations, producing equally relevant results. 

\section{User study}

\subsection{Complete questionnaire}

The questionnaire used in the user study was originally written in Brazilian Portuguese, in which all participants were fluent. The questions were translated into English in this document.

\myParagrapho{Background and experience}

\begin{itemize}
    \item What is your age group? Choose one option: (i) Between 18 and 24 years old; (ii) Between 25 and 34 years old; (iii) Between 35 and 44 years old; (iv) Between 45 and 64 years old;  (v) More than 65 years old; 
    \item Are you aware of any visual difficulties you may have?
    \item What area is your education in (e.g., computer science, statistics)?
    \item What is your most relevant academic title/function? Choose one option: (i) I'm an undergraduate student; (ii) I'm pursuing my master's degree; (iii) I'm a Ph.D. student/candidate; (iv) I'm a postdoctoral researcher; (v) I'm a professor.
    \item What is your degree of prior knowledge in the Information Visualization field? Choose one option: None, Basic, Intermediate, and Advanced knowledge.
    \item What is your degree of prior knowledge in the Network Science field? Choose one option: None, Basic, Intermediate, and Advanced knowledge.
    \item What is your degree of prior knowledge in the Topological Data Analysis field? Choose one option: None, Basic, Intermediate, and Advanced knowledge.
    \item What is your degree of prior knowledge in the Informatics in Education field? Choose one option: None, Basic, Intermediate, and Advanced knowledge.
    \item Briefly explain your experience with the above fields (Visualization, Network Science, Topological Data Analysis, and Informatics in Education).
\end{itemize}

\myParagrapho{Hands-on experience (ST1-ST12)}


Given the Primary School network with default filters, perform the following tasks:

\begin{description}
\item[ST1] Select the students from class 3B on the colored barcode.  \underline{Action:} Click on the desired barcode using the mouse's left button.

\item[ST2] Select the students from class 1B using the color legend. \underline{Action:} Click on the desired color legend label using the mouse's left button.

\item[ST3] Remove all selection. \underline{Action:} Click on a blank screen space using the mouse's right button.

\item[ST4] Zoom in and zoom out on the colored barcode. \underline{Action:} Scroll the mouse wheel up (in) and down (out).

\item[ST5] Go back to the default zoom level. \underline{Action:} Click on a blank screen space using the mouse's right button.

\item[ST6] Select the connected component that initiates with students from classes 4A, 5A, 5B, and Teachers, and describe patterns and behaviors perceived after visually analyzing that connected component. \underline{Action:} Find the connected component that contains the students 4A, 5A, 5B, and Teacher and select it using the mouse's left button.

\item[ST7] Select the timestamps 200, 300, and 400 by moving the timestamp markers on the colored barcode. \underline{Action:} Double-click on the colored barcode to show the three timestamp markers and then drag and drop each one to the desired position.

\item[ST8] Select any connected component in any node-link diagram and describe what occurs to it in the other two timestamps. \underline{Action:} First, select the flag ``Select by connected components'' and then click on a node in the node-link diagram using the mouse left button to select a specific component.

\item[ST9] Remove all selections again and perform zoom-in and zoom-out in any node-link diagram. \underline{Action:} Click in the node-link diagram using the mouse's right button to reset the selection and use the mouse wheel to perform zoom-in (up) and zoom-out (down).

\item[ST10] Change the timestamp of the second node-link diagram (i.e., the one in the middle) from 300 to 350 by typing the new value. \underline{Action:} Type 350 on the textbox positioned in the middle node-link diagram and click OK or press Enter to change the selected timestamp.

\item[ST11] Expand two node-link diagrams and position them side-by-side. \underline{Action:} Click on the expand button (\img{maximize.png}) positioned on the top-left portion of the node-link diagram, and drag and drop the opened window to position it.

\item[ST12] Define temporal resolution and give an example. \underline{Action:} No action is required in the system.

\end{description}

\myParagrapho{Six questions for the Primary School network (SQ1-SQ6)}

Given the Primary School network  with resolution $r = 76$ (SQ1-SQ3), answer:

\begin{description}
\item[SQ1] Consider the bars and node-link diagrams that refer to the students and teachers from the first and large connected component (i.e., the one containing students from classes 4A, 5A, 5B, and their teachers). Around timestamp 50, how is the relationship between students and teachers of different classes? 

\underline{Expected answer:} There is a strong relationship between the students from class 5A and 5B and their respective teachers, forming a cluster. Also, there is a strong relationship between the students of class 4A and their teacher, forming a second cluster. These two clusters connect to each other through a single edge involving two teachers, leading to one connected component. 

\item[SQ2] Select three timestamps to compare: 640, 710, and 800. What can you tell about the relationship between students and teachers when comparing those three timestamps?

\underline{Expected answer:} There was a strong relationship between all students and teachers in the first timestamp (640). However, in the next timestamp (710), the students were divided into two groups, consisting of younger students (1A, 1B, 2A, 2B, and partially 3A) and older students (partially 3A, 3B, 4A, 4B, 5A, 5B), but with no interaction from the teachers. At the last timestamp (800), the students merged again and were highly connected, but again without the presence of their teachers.

\item[SQ3] Tell the classes that, at the beginning of their activities, were increasing in size over time. Look only at the beginning of the network (from the beginning until around timestamp 100).

\underline{Expected answer:} Almost every class that is divided into a single connected component has this behavior of starting with just a few students and increasing this size over time.

\end{description}

Given the Primary School network with resolution $r = 154$ (SQ4-SQ6), answer:

\begin{description}
\item[SQ4] Tell the classes that, at the beginning of their activities, were increasing in size over time. Look only at the beginning of the network (from the beginning until around timestamp 100). Notice that this question was already answered for the same network but using a different resolution.

\underline{Expected answer:} In the case of this resolution, almost no class has this pattern visible except for class 2B.

\item[SQ5] Considering the last question, according to your perception, did both resolutions lead to the same answer? In a negative case, why do you think the answers differed for the two distinct resolutions?

\underline{Expected answer:} No, both resolutions had very different answers. In the case of the smaller resolution ($r = 76$), many components followed this behavior (increasing size over time). On the other hand, this pattern almost disappeared when using $r = 154$. 

\item[SQ6] Based on the currently selected network and resolution, freely explore the system and try to find new patterns or anomalies. If there are any, mention findings you consider relevant and tell us which part of the visualization helped you find them.

\underline{Expected answer:} No expected answer since the participants were free to explore the system and find new patterns or anomalies.

\end{description}

\myParagrapho{Three questions for the High School network (SQ7-SQ9)}

Given the High School network with $resolution=46$, answer:

\begin{description}
\item[SQ7] Can you identify two distinct connected components containing  students from the same class at the same timestamp? If so, cite a class and at which timestamp this behavior occurs.

\underline{Expected answer:} Multiple cases highlight this pattern, such as  2BIO3 at any timestamp between 614 and 725, or 2BIO2 at any timestamp between 586 and 725.

\item[SQ8] Analyze the relationship of students from the class MP2 after timestamp 810.

\underline{Expected answer:} After timestamp 810, until the end of the network, there exist only three students from class MP2 with connections.

\item[SQ9] Based on the currently selected network and resolution, freely explore the system and try to find new patterns or anomalies. If there are any, mention findings you consider relevant and tell us which part of the visualization helped you find them.

\underline{Expected answer:} No expected answer since the participants were free to explore the system and find new patterns or anomalies.

\end{description}

\myParagrapho{Questions to compare ZigzagNetVis with other techniques}

\begin{itemize}

\item Have you tried to do analyses similar to those described in this study? Options: Yes or no.

\item (Optional if the previous answer is yes) What systems/techniques do you commonly use? Do you prefer ZigzagNetVis or the other systems/techniques you know? Why?

\end{itemize}

\myParagrapho{Likert-scale questions (LQ1–LQ10)}

We used two 5-point Likert-scale questionnaires to assess the participants' preferences about ZigzagNetVis and the provided visual components  (LQ1 -- LQ6) and specific tasks (LQ7 -- LQ10). For each question below, the participants should choose between (i) Strongly disagree; (ii) Disagree; (iii) I don't know; (iv) Agree; (v) Strongly agree.

\begin{description}

\item[LQ1] The colored barcode is useful and helps when analyzing the networks.

\item[LQ2] The node-link diagrams are useful and help when analyzing the networks.

\item[LQ3] The interaction and coordination between the views are useful and help when analyzing the networks.

\item[LQ4] ZigzagNetVis is intuitive and easy to use.

\item[LQ5] ZigzagNetVis is useful.

\item[LQ6] ZigzagNetVis is fast (i.e., the provided interactions work in a satisfactory time).

\end{description}

\begin{description}

\item[LQ7] It is easy to understand the temporal evolution of connected components and particular classes, in terms of what happens with groups that interact with one another over time, when using ZigzagNetVis.

\item[LQ8] It is easy to compare the network structure at different timestamps when using ZigzagNetVis.

\item[LQ9] It is easy to analyze the network structure at a node level when using ZigzagNetVis.

\item[LQ10] It is easy to analyze the network under different resolutions when using ZigzagNetVis.

\end{description}

\myParagrapho{Questions to collect the participants’ feedback}

\begin{itemize}
\item What are the most useful visual aids offered by the ZigzagNetVis system? Why?

\item What are the most useful visual aids offered by the ZigzagNetVis system? Why?
    
\item What other visual aids could be helpful if incorporated into the ZigzagNetVis system?
    
\item Do you have any final comments?

\end{itemize}

\subsection{Interactive features - complete analysis}

Since we recorded the participants' screens, we validated the functionalities mainly used for some questions. Note that since the questions for the same network are sequential (e.g., SQ1-SQ3), in some cases, the participant did not need to interact to find new patterns; the interaction from the previous question might have helped them to answer the current one. 
We consider ten possible interactions categorized into general, colored barcode, and node-link diagram (Tab.~\ref{tab:table_interactions}): users can select connected components (III) or nodes with the same label (II) --- in this case by clicking on the color legend (I) or the colored barcode (IV); users navigate throughout time by moving the timestamp markers (V) or by typing the new value on a node-link diagram (VII); they can pan and zoom in/out on the colored barcode (VI) and diagrams (X); and they can expand the diagrams (VIII) and use the informative tooltip (IX).
 Tab.~\ref{tab:table_interactions} presents the percentage of participants who used these interactive features for questions SQ1-SQ3, SQ6, and SQ9. We chose these questions (the first three of the primary school and the two exploratory ones) as we believe they are sufficient to understand the users' behaviors on defined and exploratory tasks.

Participants interacted with the system differently. For instance, in the already mentioned SQ1, all participants made selections using connected components (justified by the question description) and zoomed in the node-link diagram to understand the relationship between classes. Since they were asked to select ``around timestamp 50'', some participants chose to move the timestamp markers, others typed the timestamp value on the node-link diagram, and others did both (Tab.~\ref{tab:table_interactions}(SQ1,~V~and~VII)). For the exploratory questions, moving the timestamp markers was the best option for most participants (Tab.~\ref{tab:table_interactions}(SQ6~and SQ9,~V)). Overall, they interacted more often with the colored barcode than with the diagrams. 
On average, the feature mainly used in the node-link diagrams was zoom (41.48\%), which is justified by the small size of nodes and edges initially applied. Not least, the similar rate of usage involving selection by label (44.44\%) and by component (41.48\%) indicates that both were appreciated.


\begin{table}[t]
\centering
\caption{Percentage of participants who used each of the ten possible interactive features to answer five questions. Features used by more than 50\% of participants are highlighted in bold.}
\label{tab:table_interactions}
\resizebox{0.49\textwidth}{!}{%
\begin{tabular}{c|ccc|ccc|cccc}
             & \multicolumn{3}{c|}{\textbf{General}}       & \multicolumn{3}{c|}{\textbf{Colored barcode}}        & \multicolumn{4}{c}{\textbf{Node-link diagrams}}                \\ \cline{2-11} 
             & \textbf{I} & \textbf{II}    & \textbf{III} & \textbf{IV} & \textbf{V}     & \textbf{VI}    & \textbf{VII}   & \textbf{VIII} & \textbf{IX} & \textbf{X}     \\ \hline
\textbf{SQ1} & 22.22      & 37.03          & \textbf{100} & 25.92       & \textbf{59.25} & 14.81          & \textbf{59.25} & 25.92         & 29.62       & \textbf{81.48} \\
\textbf{SQ2} & 25.92      & 29.62          & 40.74        & 22.22       & 22.22          & 25.92          & \textbf{85.18} & 18.51         & 25.92       & 40.74          \\
\textbf{SQ3} & 37.03      & \textbf{70.37} & 18.51        & 48.14       & 44.44          & \textbf{51.85} & 25.92          & 7.4           & 3.7         & 18.51          \\
\textbf{SQ6} & 7.4        & \textbf{55.55} & 22.22        & 44.44       & \textbf{66.66} & 33.33          & 3.7            & 14.81         & 7.4         & 37.03          \\
\textbf{SQ9} & 18.51      & 29.62          & 25.92        & 18.51       & \textbf{51.85}          & 25.92          & 3.7            & 11.11         & 3.7         & 29.62          \\ \hline
\textbf{Avg} & 22.22      & 44.44          & 41.48        & 31.85       & 48.15          & 30.37          & 35.55          & 15.55         & 14.07       & 41.48          \\ \hline
\end{tabular}%
}
\end{table}

\bibliographystyle{abbrv-doi}

\bibliography{ms}